%% file: main.tex
\documentclass[acmsmall]{acmart}
\usepackage{xurl}
\usepackage{xspace}
\usepackage{float}
\usepackage{tikz}
\usetikzlibrary{quantikz2}
\usetikzlibrary{positioning,arrows.meta,shadows}
\usepackage{enumitem}

\usepackage{booktabs}
\usepackage{caption}
\usepackage[justification=centering]{subcaption}
\usepackage{multirow}
 \usepackage{amsthm}

\DeclareUrlCommand\code{\urlstyle{tt}}

\theoremstyle{definition}
\newtheorem{example}{Example}[section]

\usepackage{outlines}

\usepackage{siunitx}
    
\usepackage[linesnumbered]{algorithm2e}

 \definecolor{mygray}{gray}{0.92}
 \definecolor{belyellow}{RGB}{181, 156, 86}
 \definecolor{commentgreen}{rgb}{0.0, 0.5, 0.0}

\SetCommentSty{mycommfont}

\usepackage[toc,page]{appendix}

\usepackage{adjustbox}
\usepackage{blindtext}

\usepackage{ifsym}

\usepackage{braket}
\usepackage{mathtools}

\usepackage[utf8]{inputenc}
\usepackage[english]{babel}

\theoremstyle{definition}

\newcommand{\revisionA}[1]{\textcolor{black}{#1}}
\newcommand{\revisionB}[1]{\textcolor{black}{#1}}
\newcommand{\revisionC}[1]{\textcolor{black}{#1}}
\newcommand{\revisionM}[1]{\textcolor{black}{#1}}

\newcounter{RihanNOC}
\stepcounter{RihanNOC}

\newcounter{FlorisNOC}
\stepcounter{FlorisNOC}

\newcounter{AaNOC}

\newcounter{AdNOC}

\newcommand{\para}[1]{\noindent\textbf{#1.}}

\newcommand{\sys}{\texttt{InferQ}\xspace}

\usepackage{tabularx}
\usepackage{array}
\usepackage[table,dvipsnames,svgnames]{xcolor}
\newcolumntype{Y}{>{\raggedright\arraybackslash}X}

\newcolumntype{P}[1]{>{\raggedright\arraybackslash}p{#1}}

\newcolumntype{C}{>{\hsize=0.90\hsize\raggedright\arraybackslash}X}
\newcolumntype{J}{>{\hsize=1.10\hsize\raggedright\arraybackslash}X}

\newcommand{\vis}[1]{\begin{tabular}[t]{@{}l@{}}#1\end{tabular}}

\newcommand{\qarrow}{\hspace{0.20em}\(\rightarrow\)\hspace{0.20em}}

\newcommand{\candsep}{,\allowbreak\ }

\newcommand{\qfamily}[2]{%
  \tikz[baseline=(X.base)]\node[
    draw=none,
    rounded corners=2.0pt,
    fill=#2,
    text=white,
    inner xsep=2.4pt,
    inner ysep=1.2pt,
    font=\scriptsize\sffamily\bfseries,
    drop shadow={opacity=0.14,shadow xshift=0.35pt,shadow yshift=-0.35pt}
  ] (X) {#1};%
}

\definecolor{qcStart}{HTML}{334155}
\definecolor{qcEnd}{HTML}{7F1D1D}
\definecolor{qcStatePrep}{HTML}{1D4ED8}
\definecolor{qcOracleA}{HTML}{B45309}
\definecolor{qcHamSim}{HTML}{047857}
\definecolor{qcModExp}{HTML}{6D28D9}
\definecolor{qcGrover}{HTML}{BE123C}
\definecolor{qcQPE}{HTML}{0E7490}
\definecolor{qcAmpEst}{HTML}{A21CAF}
\definecolor{qcVar}{HTML}{3F6212}
\definecolor{qcRandom}{HTML}{475569}
\definecolor{qcMeasure}{HTML}{111827}
\definecolor{qcAmpAmpl}{HTML}{C2410C}
\definecolor{qcQFT}{HTML}{4338CA}

\newcommand{\famSTART}{\qfamily{START}{qcStart}}
\newcommand{\famEND}{\qfamily{END}{qcEnd}}

\newcommand{\famStatePrep}{\qfamily{StatePrep}{qcStatePrep}}
\newcommand{\famOracleA}{\qfamily{Oracle/A}{qcOracleA}}
\newcommand{\famHamSim}{\qfamily{HamSim}{qcHamSim}}
\newcommand{\famModularExp}{\qfamily{ModularExp}{qcModExp}}
\newcommand{\famGroverIter}{\qfamily{GroverIter}{qcGrover}}
\newcommand{\famQPE}{\qfamily{QPE}{qcQPE}}
\newcommand{\famAmpEst}{\qfamily{AmpEst}{qcAmpEst}}
\newcommand{\famVariational}{\qfamily{Variational}{qcVar}}
\newcommand{\famRandom}{\qfamily{Random}{qcRandom}}
\newcommand{\famMeasure}{\qfamily{Measure}{qcMeasure}}
 
\newcommand{\famAmpAmpl}{\qfamily{AmpAmpl}{qcAmpAmpl}}
 \newcommand{\famQFT}{\qfamily{QFT}{qcQFT}}

\DeclareUrlCommand{\feat}{\urlstyle{tt}\small}
\newcommand{\Ablock}{\qfamily{$A$}{qcOracleA}}

\AtBeginDocument{%
  }

\setcopyright{cc}
\setcctype{by-nc-nd}
\acmJournal{PACMMOD}
\acmYear{2026} \acmVolume{4} \acmNumber{4 (SIGMOD)} \acmArticle{278}
\acmMonth{9} \acmDOI{10.1145/3837116}

\ccsdesc[500]{Information systems~Data management systems}
\ccsdesc[300]{Applied computing~Physics}

\keywords{Quantum circuit simulation, benchmarking, relational database management systems, SQL workload generation, learned simulator selection}

\begin{document}
\setlength{\pdfpagewidth}{6.75in}
\setlength{\pdfpageheight}{10in}

\received{17 January 2026}
\received[revised]{19 May 2026}
\received[accepted]{12 June 2026}

\title{InferQ: A Database-Oriented Benchmark for Quantum Circuits Simulation}

\author{Andrei Ilinescu}
\authornote{Both authors contributed equally to this research.}
\email{a.ilinescu@student.tudelft.nl}
\affiliation{%
  \institution{Delft University of Technology}
  \city{Delft}
  \country{The Netherlands}
}

\author{Aadi Patwardhan}
\authornotemark[1]
\email{a.patwardhan@student.tudelft.nl}
\affiliation{%
  \institution{Delft University of Technology}
  \city{Delft}
  \country{The Netherlands}
}

\author{Rihan Hai}
\email{r.hai@tudelft.nl}
\affiliation{%
  \institution{Delft University of Technology}
  \city{Delft}
  \country{The Netherlands}
}

\begin{abstract}
Recent work suggests that relational database management systems (RDBMSs) can execute quantum circuit simulation by compiling the simulation into SQL workloads (primarily join-and-aggregate tensor contractions). While early results are promising, they largely focus on a narrow set of highly structured circuits and offer limited support for systematic database research, such as query optimization, physical design, and engine-level evaluation across a broad range of circuits.
We present \sys, a database-oriented benchmark for quantum circuit simulation. \sys generates \emph{general, compositional} circuits by assembling subcircuits from a set of circuit templates, emits each simulation task as an RDBMS-ready SQL workload, and extracts circuit and query features (static, graph, SQL, and dynamic) for workload characterization.
\sys also releases a large dataset of  202,975 circuits online, with a web-based viewer to support searching, filtering, and downloading circuits and feature records.
In experiments across RDBMS engines (PostgreSQL, SQLite, DuckDB, and \revisionB{Umbra}) and the widely used Qiskit Aer simulator, we find that  \revisionB{RDBMSs} achieve better peak memory usage than Qiskit Aer on more than \revisionB{50\%} of the circuits generated by \sys.
Moreover, using \sys features, lightweight machine learning models (linear and tree-based models) can accurately predict when SQL execution is preferable (with accuracy up to 95.6\% for runtime and 97.4\% for memory), enabling data-centric simulator selection and opening the door to principled optimization of SQL-based quantum circuit simulation.
\end{abstract}

\maketitle
\input{introduction}

\input{Preliminary}

\input{system}

\input{tool}

\input{evaluation}
\input{conclusion}

\section*{Acknowledgment}
This publication was supported in part by the Dutch Research Council (VI.Veni.222.439). We thank Tim Littau and Wenbo Sun for programming assistance and sharing code from their prior work, and Arun Pati and Giancarlo Gatti for discussion and feedback on quantum primitive composition.

\bibliographystyle{ACM-Reference-Format}
\bibliography{inferq_ref}

\input{appendix}
\end{document}

%% file: introduction.tex
\section{Introduction}

In the current noisy intermediate-scale quantum (NISQ) era, most quantum applications are developed and validated through classical simulation. Quantum programs are typically expressed in the circuit model as a sequence of gates, and the common practice is to execute the circuit on a simulator before running it on quantum hardware. 
Simulators produce the circuit’s (predicted) output, e.g., measurement outcomes or their probabilities. 
Efficient circuit simulation, therefore, remains a cornerstone for near-term quantum algorithm development. 
\revisionB{For database researchers, this raises a data management question: can circuit simulation be expressed as a relational workload whose execution can benefit from database management systems (DBMS) memory management, query optimization, and out-of-core processing? 
 }

\smallskip
\para{Data management challenge}
    Recent work \cite{einsteinSQL2023, hai2025quantum} argues that relational databases can serve as a competitive simulation \emph{engine} by compiling circuit simulation into relational workloads. A particularly promising result is that DuckDB can scale to \emph{millions of qubits} for sparse circuits. Concretely, under a 2GB memory limit, DuckDB simulates up to $4{,}000{,}862$ qubits for GHZ-state preparation and up to $53{,}017$ qubits for W-state preparation \cite{techreport}, which are two standard state-preparation routines that initialize many qubits into widely used multi-qubit entangled states.  
These results raise an obvious question for the database community: \emph{when does ``an RDBMS as a simulator'' actually work?}

The same study makes clear this scalability is highly workload-dependent: for a dense circuit such as Quantum Fourier Transform (QFT), the maximum supported qubit count is only $14$ under the same setting \cite{techreport}. Moreover, the largest improvements in those experiments arise from circuits with a specific structure. For instance, GHZ is a stabilizer circuit, which uses only Clifford gates (Hadamard, CNOT, Phase), and is known to be simulated efficiently on classical computers via the Gottesman–Knill theorem \cite{Aaronson_2004}. Moreover, a key systems gap is revealed in \cite{einsteinSQL2023, hai2025quantum}: existing approaches have not fully exploited RDBMS capabilities to handle simulation workloads through mechanisms such as query optimization, indexing, and materialization, emphasizing that closing this gap is important for an end-to-end database-driven simulation pipeline.
Altogether, these results motivate a database-centric benchmark: to move beyond isolated case studies, the community needs a systematic way to generate \emph{diverse} simulation workloads and study how SQL representations behave under query optimization and execution. \revisionB{Such a benchmark enables database researchers to study query planning, materialization, indexing, spill behavior, and backend engine selection for quantum circuit simulation.}
That is, these results point to a benchmark gap. 

\medskip
\para{Why this differs from existing tensor workloads in RDBMSs}
 When a quantum circuit is compiled into SQL, the underlying computation \emph{can} be expressed as tensor algebra:
applying gates corresponds to multiplying and contracting tensors (e.g., updating a state vector by small operators). 
%
At a computational level, circuit simulation in RDBMS  is a tensor workload: vector-matrix multiplication, matrix-matrix multiplication, and tensor contractions. One of the most frequent operations is matrix multiplication. Using SQL, matrix multiplication can be implemented as a join-and-aggregate operation: represent the matrices in long form as $A(i,k,a_{ik})$ and $B(k,j,b_{kj})$, compute $W=AB$ by joining rows of $A$ and $B$ on the shared inner index $k$, multiplying matching values, and summing over $k$:
\[
w_{ij}=\sum_k a_{ik}b_{kj},
\]
which corresponds to:
\begin{quote}\small
\texttt{SELECT A.i, B.j, SUM(A.val * B.val) AS w\_ij\\
FROM A JOIN B ON A.k = B.k\\
GROUP BY A.i, B.j;}
\end{quote}
This style of expressing tensor computation in SQL has been studied in the database community \cite{Khamis2020, boehm2023optimizing, Makrynioti2019survey, zhou2020database, 10107490}, including recent work on efficient Einstein summation in SQL \cite{einsteinSQL2023} and relational abstractions for tensor and linear-algebra workloads in ML \cite{chen2017towards, wang2020spores, luo2021automatic, yuan2021tensor, sun2026transql}. 

However, the simulation workload induced by quantum circuits differs from mainstream ML workloads in several fundamental ways. In transformer models, the core computation is dominated by dense linear algebra over matrices with dimensions in the thousands (e.g., attention uses dense $QK^\top$ and $QV$ products) \cite{vaswani2017attention}. In contrast, a quantum circuit acts on a state vector of dimension $2^{N}$: for qubit count $N=50$, the state has $2^{50}\approx 1.13\times 10^{15}$ amplitudes, which already requires about $16$~PB just to store a dense complex state vector. 
At the same time, each instruction applies only a tiny operator (typically $2\times 2$ or $4\times 4$ matrices), but it must be
applied across this exponentially large index space. 
The   structure of a quantum circuit determines a 
sequence of tensor contractions whose intermediate sparsity and entanglement can change abruptly across the gate sequence. 

This mismatch creates database-specific difficulties:
(i) intermediate results may be \emph{enormous} unless the representation stays sparse or factored;
(ii) whether sparsity survives depends on the \emph{gate sequence}, and entangling operations can rapidly densify the state and
cause abrupt blow-ups;
and (iii) the computation is a long, dependency-constrained sequence of small updates rather than a few large tensor computations,
so ``optimize one big join'' is not the right mental model. Consequently, optimizing SQL for quantum circuit simulation is not a
direct reuse of existing in-database tensor/ML techniques: the bottlenecks, cardinality growth, and the importance of
circuit structure (sparsity patterns, interaction graphs, composition rules) are qualitatively different.
A benchmark that systematically varies these properties is therefore essential for a principled optimizer and execution-engine
research in this space.

\medskip
\para{Gaps of existing benchmarks}
Existing quantum benchmarks such as SupermarQ \cite{tomesh2022supermarqscalablequantumbenchmark}, MQT Bench \cite{Quetschlich_2023}, and QASM Bench \cite{li2022qasmbenchlowlevelqasmbenchmark} provide valuable circuit suites for the quantum community, but they do not directly target database execution.
In particular, they generate circuits as quantum-circuit descriptions (e.g., QASM/Qiskit circuits) meant to be run by quantum toolchains, not as relational workloads. As a result, they are not designed to evaluate how an RDBMS behaves as a simulation engine, e.g., the impact of query optimization, physical design (indexes/partitioning), and execution strategies on circuit-simulation workloads. 

\begin{figure}[t]
  \centering
   \includegraphics[width=0.53\columnwidth]{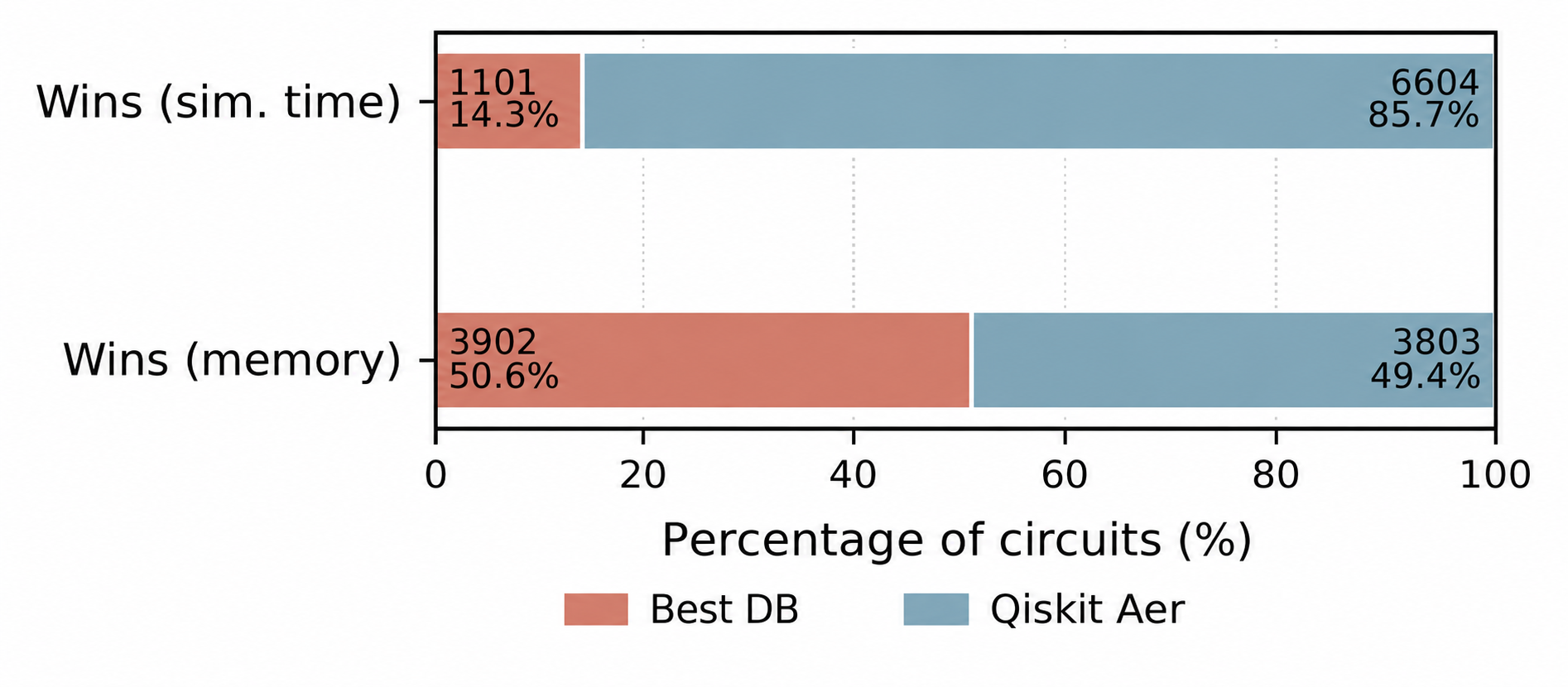}
  \caption{\revisionB{Across \num{7705} circuits, share of cases where RDBMSs (executing the emitted SQL workload)
  or Qiskit Aer 
  achieves the lowest runtime and the lowest
  peak memory. Detailed experimental settings are in Section~\ref{sec:eva}.}}
  \label{fig:aer-vs-sqlite-wins}
\end{figure}

\medskip

\para{Our approach}
We present \sys, a database-oriented benchmark for quantum circuit simulation.
Given a small user configuration (Table~\ref{tab:base_params}), \sys generates
\emph{compositional} circuits by assembling subcircuits from a set of quantum templates. Each circuit is accompanied by an
\emph{RDBMS-ready SQL workload} that expresses the simulation task as relational
operators. \revisionB{This makes simulation directly accessible to database research: each circuit becomes both a quantum artifact and a relational workload whose query shape, intermediate behavior, and backend performance can be analyzed.} Moreover, to support data-centric analysis, \sys also 
extracts four groups of features: \emph{static} and \emph{graph} features from the circuit
structure, \emph{SQL} features from the emitted query workload, and \emph{dynamic}
features from the simulation.
\revisionC{
Our experimental results further clarify why exposing simulation as database workloads is useful. 
Running the generated SQL inside DBMSs makes simulation subject to database execution and optimization choices, enabling better memory management, runtime improvements for certain circuits, and out-of-core execution for large circuits under memory limits. 
Thus, RDBMS-based simulation enables systematic optimization of quantum circuit simulation through DBMS memory management, spill-aware execution, and backend selection.
}

\para{Contributions}
We make the following contributions:
\begin{itemize}[leftmargin=*]
    \item 
    We propose \sys, a benchmark 
    that produces quantum circuit simulation workloads in \emph{relational form} (SQL),
    enabling direct evaluation inside RDBMS engines.

    \item 
    We design a template-based generator
    that \emph{scales} circuit size while preserving realistic structure via compositional circuit generation (Section~\ref{sec:benchmark}). 

     \item 
    From the generated circuits, we extract \emph{static, graph, SQL,} and \emph{dynamic} features per circuit, enabling
    systematic workload characterization and correlation with DB performance 
    (Section~\ref{ssec:features}).

    \item Using \sys workloads, we quantify when SQL-driven simulation is competitive with Qiskit Aer, \revisionB{including memory-efficient and out-of-core cases.}
    \sys features enable accurate, lightweight machine learning based selectors for deciding when to use an
    RDBMS engine (Section~\ref{sec:eva}).
\end{itemize}

%% file: Preliminary.tex
\section{Background }
\label{sec:pre}
\begin{figure}[t]
  \centering
  \includegraphics[width=0.65\linewidth]{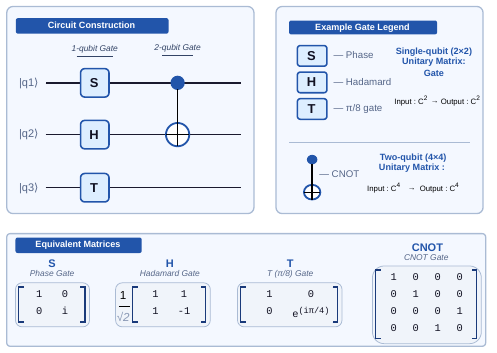}
  \caption{\revisionA{Quantum gates represented as matrices.  } }
  \label{fig:gates}
\end{figure}

\subsection{Quantum Bits, Gates, and Circuits}

A classical bit stores either 0 or 1. In contrast, a \emph{quantum bit} (qubit) can be represented as a unit vector in a two-dimensional complex vector space, typically described using the computational basis of $\ket{0}$ and $\ket{1}$:
$$
\ket{0}=\begin{bmatrix}1\\0\end{bmatrix},
\qquad
\ket{1}=\begin{bmatrix}0\\1\end{bmatrix}.
$$
An arbitrary single-qubit state can be written as a superposition of these basis states:
$$
\ket{\psi}=\alpha\ket{0}+\beta\ket{1},
\qquad
|\alpha|^2+|\beta|^2=1,
$$
where $\alpha$ and $\beta$ are complex amplitudes. 
If we perform a \emph{measurement} on the quantum state $\ket{\psi}$, it collapses to a classical bit of 0 or 1. That is, a measurement of $\ket{\psi}$ against the computational basis yields outcome $0$ with probability $|\alpha|^2$ and outcome $1$ with probability $|\beta|^2$. 
An $N$-qubit state can be expressed as a superposition over computational basis states indexed by bitstrings:
$$
\ket{\Psi}=\sum_{x\in\{0,1\}^N} c_x \ket{x},
\qquad
\sum_{x\in\{0,1\}^N} |c_x|^2=1.
$$
Fix an ordering of the computational basis states (e.g., lexicographic order). The \emph{state vector}
representation of $\ket{\Psi}$ is the column vector of its amplitudes in that order:
$
\left[\begin{smallmatrix}
c_{0\cdots 0}\\
c_{0\cdots 1}\\
\vdots\\
c_{1\cdots 1}
\end{smallmatrix}\right].
$
For example, when $N=2$, any two-qubit state $\ket{\Psi}$ can be represented as a linear combination of
$\ket{00}$, $\ket{01}$, $\ket{10}$, and $\ket{11}$, and
$|c_{00}|^2+|c_{01}|^2+|c_{10}|^2+|c_{11}|^2=1$.
The state vector representation of $\ket{\Psi}$ is
$
\left[\begin{smallmatrix}
c_{00}\\
c_{01}\\
c_{10}\\
c_{11}
\end{smallmatrix}\right].
$

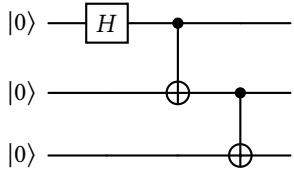
\begin{figure}[t]
  \centering
  \begin{quantikz}
        \lstick{$\ket{0}$} & \gate{H} & \ctrl{1} &  &\\
        \lstick{$\ket{0}$} &  &\targ{} & \ctrl{1} & \\
        \lstick{$\ket{0}$} &  & &\targ{} & \\
    \end{quantikz}
     \vspace{-0.4cm}
  \caption{GHZ state preparation circuit: it takes the all-zero state  $\ket{000}$ as input and prepares a maximally entangled state.}
 \vspace{-0.4cm}
  \label{fig:subcircuits}
  
\end{figure}

Quantum computation follows the circuit model, which is analogous to classical logic circuits. Computation is performed by applying quantum gates (unitary operators) acting on one or more qubits to transform the quantum state. A matrix $U$ is \emph{unitary} if
$
U^\dagger U = U U^\dagger = I
$. As shown in Figure~\ref{fig:gates}, a quantum gate acting on \(N\) qubits
is represented by a \(2^N \times 2^N\) unitary matrix \(U\). It maps an
input state \(\lvert\psi\rangle \in (\mathbb{C}^{2})^{\otimes N} \) to an output state
\(U\lvert\psi\rangle \in (\mathbb{C}^{2})^{\otimes N} \), where \(\mathbb{C}\) denotes
the complex numbers, e.g., one qubit corresponds to $\mathbb{C}^2$ and two qubits to $\mathbb{C}^4$.
A gate set is called universal if it can approximate any unitary operator to arbitrary precision \cite{nielsen2010quantum}. A widely used universal set is the Clifford gates (including the Pauli gates ($X,Y,Z$), the Hadamard gate $H$, the phase gate $S$, the controlled-NOT (CNOT) gate), and the $T$ gate. Figure~\ref{fig:gates} illustrates several of these gates and their matrix representations. A \emph{quantum circuit} is a sequence of quantum gates, and the overall computation corresponds to the composition of their unitary matrices. 

\emph{Circuit depth} is the number of sequential layers of quantum gates in a circuit, representing how many time steps are required to execute it when gates that act on different qubits can be applied in parallel. For example, in Figure~\ref{fig:subcircuits}, the circuit depth is 3. 
We can consider quantum circuits as higher-level operators composed of subcircuits. By abstracting away from hardware and physical implementation details, we can treat such structures modularly. Take the GHZ state preparation circuit in Figure~\ref{fig:subcircuits} for instance. It can serve as a state preparation subcircuit for larger algorithms, i.e., it prepares an input state for other algorithms.

\input{gap}

%% file: gap.tex
\subsection{Limitations of Existing Benchmarks}
\label{ssec:limit}

Several quantum circuit benchmarks are widely used in the quantum computing community, including, for example, 
SupermarQ~\cite{tomesh2022supermarqscalablequantumbenchmark}, MQT Bench~\cite{Quetschlich_2023}, and
QASM Bench~\cite{li2022qasmbenchlowlevelqasmbenchmark}. While these suites are valuable for evaluating
quantum circuits, they expose limited support for database-driven and data-centric studies of circuit simulation.

\medskip
\revisionC{\emph{Gap 1: No broad DBMS-executable benchmark workloads.}}
Existing benchmarks primarily distribute circuits as OpenQASM\footnote{\url{https://openqasm.com/}} and
Qiskit QPY\footnote{\url{https://quantum.cloud.ibm.com/docs/en/api/qiskit/qpy}} files. 
These formats are well-suited for quantum toolchains, but \revisionC{they do not expose the induced simulation computation as optimizer-visible relational workloads that can be directly executed, inspected, or optimized by an RDBMS.}
As a result, studying circuit simulation inside an RDBMS typically requires substantial manual translation and
non-trivial quantum expertise.
\revisionC{
The two recent DB-oriented simulation works discussed above \cite{einsteinSQL2023, hai2025quantum} are important first steps within this gap: they show that SQL/RDBMS execution can support quantum circuit simulation. 
However, their workload coverage is limited. \cite{einsteinSQL2023} targets general Einstein summation and evaluates SQL simulation on Google's Sycamore quantum-supremacy circuit, while \cite{hai2025quantum} studies W and GHZ state preparation and QFT circuits. 
Thus, these works motivate DBMS-backed simulation, but do not provide a benchmark-scale collection of diverse circuits with reusable SQL workloads and workload features.
}

We address this gap by having \sys generate \revisionC{general, diverse} circuits, each coming with an RDBMS-ready SQL workload (Section~\ref{sec:benchmark}). 
\revisionC{
This effort is important because it makes the simulation workload visible to database systems: tensor contractions become join-and-aggregate queries whose plans, indexes, materialization choices, and spill behavior can be studied and optimized. 
Consequently, \sys enables evaluating whether DBMS execution can improve memory usage, runtime, and scalability under memory limits, rather than merely translating circuits into another format.
}

\emph{Gap 2: Lack of structured compositional circuits.}
Real quantum programs 
are typically
assembled as pipelines of subroutines such as state preparation, algorithmic cores, estimation procedures, and
measurement. 
However, existing benchmarks largely treat circuits as independent, fixed templates for standalone algorithms.
They do not systematically compose building blocks into larger circuits, nor do they model dependencies between
subcircuits. In practice, the choice of one subroutine often constrains what can follow (e.g., Hamiltonian simulation
followed by phase estimation).
We address this gap in Section~\ref{sssec:seq_select} by modeling conditional, history-dependent template
composition as a Markov transition process.

\emph{Gap 3: Limited scalability for data-centric analysis.}
Because many existing benchmarks focus on fixed algorithm templates, their scalability is inherently limited:
the number of instances grows slowly, and structural diversity is constrained. This limits their usefulness for
data-centric workflows that require large corpora, such as applying machine learning models, cost modeling, and feature-driven analysis. 
We address this gap with scalable compositional generation in Section~\ref{sec:benchmark}, and release a large dataset of \num{202975} circuits to support large-scale quantum circuit simulation studies. We also validate scalability experimentally in Appendix~M of our online technical report \cite{inferqtech}.

\emph{Gap 4: Limited circuit-as-graph and SQL-level characterization.}
Quantum circuits are inherently graph-structured: qubits correspond to vertices and multi-qubit gates induce
interactions as edges~\cite{Markov+2008}. This structure is central to simulation cost and to the shape of the induced SQL 
workload. Nevertheless, existing benchmarks expose few graph-derived properties and provide limited support for
analyzing structural diversity across circuits.
We address this gap by explicitly extracting graph features (Section~\ref{ssec:graphf}) and
SQL features for the query representation (Section~\ref{ssec:sql}).

\para{Summary}
Overall, existing benchmarks fall short of supporting database-driven and data-centric research on quantum circuit
simulation. Addressing these gaps requires a benchmark that produces SQL workloads, supports conditional
composition of subcircuits, scales to large and diverse circuits, and exposes graph- and query-level structure for
systematic analysis.

%% file: system.tex
\section{\sys Benchmark}
\label{sec:benchmark}

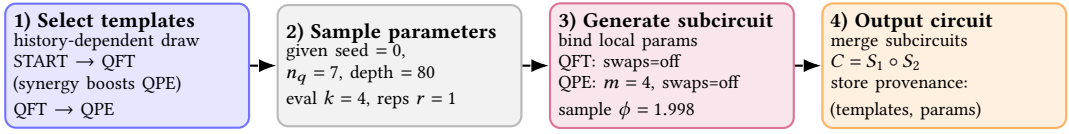
\begin{figure*}[t]
\centering

\tikzset{
  stagebox/.style={
    draw,
    rounded corners,
    thick,
    align=left,
    inner xsep=3.2pt,
    inner ysep=3.2pt,
    text width=30mm,
    minimum height=16mm
  },
  metricbox/.style={
    draw,
    rounded corners,
    thick,
    align=center,
    inner xsep=4pt,
    inner ysep=3pt,
    minimum width=16mm,
    minimum height=7mm
  },
  flowarrow/.style={
    ->,
    thick,
    dashed
  },
  outarrow/.style={
    ->,
    thick
  }
}

\begin{tikzpicture}[
  font=\footnotesize,
  >=Latex,
  node distance=6mm and 3.5mm
]


\node[stagebox, fill=blue!10, draw=blue!55] (sel) {%
  \textbf{1) Select templates}\\[-1pt]
  \scriptsize
  history-dependent draw\\
  START $\rightarrow$ QFT\\
  (synergy boosts QPE)\\
  QFT $\rightarrow$ QPE
};

\node[stagebox, fill=gray!10, draw=gray!55, right=of sel] (cfg) {%
  \textbf{2) Sample parameters}\\[-1pt]
  \scriptsize
  given seed $=0$,\\
  $n_q=7$, depth $=80$\\
  eval $k=4$, reps $r=1$
};

\node[stagebox, fill=purple!10, draw=purple!55, right=of cfg] (inst) {%
  \textbf{3) Generate subcircuit}\\[-1pt]
  \scriptsize
  bind local params\\
  QFT: swaps=off\\
  QPE: $m=4$, swaps=off\\
  sample $\phi=1.998$
};

\node[stagebox, fill=orange!12, draw=orange!60, right=of inst] (merge) {%
  \textbf{4) Output circuit}\\[-1pt]
  \scriptsize
  merge subcircuits\\
  $C=S_1 \circ S_2$\\
  store provenance:\\
  (templates, params)
};


\draw[flowarrow] (sel) -- (cfg);
\draw[flowarrow] (cfg) -- (inst);
\draw[flowarrow] (inst) -- (merge);


      
      



\end{tikzpicture}
\caption{\sys benchmark's quantum circuit generation pipeline, and  
running example in Section~\ref{sec:worked_example_qft_qpe}.
}
\label{fig:sec41_running_example}
\end{figure*}

\sys has two components. First, it generates quantum circuits via templates. Second, it extracts a structured set of circuit features. This section presents the \sys benchmark pipeline (Figure~\ref{fig:sec41_running_example}).
We describe the features in Section~\ref{ssec:features}.

\input{images/pipeline}
\medskip
\para{Design goal} \revisionC{
Quantum algorithms are typically built by composing reusable algorithmic blocks, often called \emph{primitives}. Examples include state preparation, Hamiltonian simulation, quantum Fourier transform (QFT), quantum phase estimation (QPE), amplitude amplification and estimation, and variational layers~\cite{abrams1999quantum, brassard2002quantum, farhi2014quantum, gilyen2019quantum, martyn2021grand}. For example, the Harrow--Hassidim--Lloyd (HHL) algorithm~\cite{harrow2009quantum} can be viewed, at a high level, as a composition of Hamiltonian simulation, QPE, and amplitude amplification. More broadly, many modern quantum algorithms are developed by composing primitive transformations in structured ways~\cite{gilyen2019quantum, martyn2021grand}. Thus, a simulation benchmark should not be limited to a fixed set of textbook circuits, but should support diverse and extensible workloads that can cover recent and future quantum algorithms.}

Database benchmarks such as TPC-H\footnote{\url{https://www.tpc.org/tpch/}}, TPC-DS\footnote{\url{https://www.tpc.org/tpcds/}}, and the Join Order Benchmark (JOB)~\cite{leis2015good} \revisionC{use parameterized templates to obtain controlled scale, coverage, and reproducibility. 
Inspired by this design, \sys generates circuits from a library of templates representing quantum primitives (Table~\ref{tbl:comb}) and composes them using feasibility constraints and history-dependent transition rules. In \sys, randomness is used to vary parameters, scale, and primitive combinations, while preserving meaningful composition patterns. Compared with existing quantum benchmarks centered on a fixed set of textbook circuits~\cite{tomesh2022supermarqscalablequantumbenchmark, Quetschlich_2023,li2022qasmbenchlowlevelqasmbenchmark}, \sys is more general and extensible: new primitives and transition rules can be added as quantum algorithms and simulation workloads evolve. This gives \sys broad benchmark coverage while preserving the structure needed to study realistic quantum circuit simulation workloads.}

\input{tables/Table_composition}

\subsection{Overview} 
\sys targets the compositional structure found in existing quantum circuits, where a
single circuit is assembled from multiple building blocks (Table~\ref{tbl:comb}).\revisionB{\footnote{Templates are compositional rather than disjoint: high-level primitives may include lower-level steps, but exposing common operations as reusable blocks supports flexible benchmarking. We use \textsc{Start}/\textsc{End} and keep \textsc{Measure} explicit, so every generated sequence has the uniform form \(\textsc{Start}\!\rightarrow\!\text{actual circuit templates}\!\rightarrow\!\textsc{Measure}\!\rightarrow\!\textsc{End}\).}}
This differs from template-only benchmarks that generate isolated, standalone circuits.
  \sys selects an ordered
sequence of \emph{circuit templates} and instantiates each template into a \emph{subcircuit}.
The resulting subcircuits are then composed into one benchmark circuit.

Concretely, \sys generates one circuit instance as follows:
\begin{enumerate}
    \item \textbf{Select templates.} 
    \sys sequentially samples a template sequence using the
    history-dependent distribution.
     \item \textbf{Sample parameters.}   \sys  samples user-configured  local parameters
    (e.g., qubit count and depth).
    \item \textbf{Generate  subcircuits.} For each selected template, \sys samples template
    parameters and generates the corresponding subcircuit.
    \item \textbf{Output circuit.} The circuit generation process stops when a stopping rule triggers, after which \sys composes
    all subcircuits into a single circuit.
\end{enumerate}

This design produces circuits that satisfy user constraints while supporting diverse
compositions and realistic template transitions.

\subsection{Select Templates}
\label{sssec:seq_select}
\revisionA{InferQ builds a circuit incrementally: at each step, it chooses one circuit template and later instantiates it as a subcircuit. The purpose of template selection is to keep this sequence valid while still producing diverse, realistic benchmark workloads.}
In \sys, we maintain a set of circuit templates \(F\).
Table~\ref{tbl:comb} lists representative templates, such as \famStatePrep; the full
template set is documented online.\footnote{\url{https://github.com/InfiniData-Lab/InferQ/blob/main/README.md}}
Given a selected template \(f\in F\), \sys instantiates a \emph{subcircuit} by sampling
concrete, circuit-specific parameter values. For example, from the  \famStatePrep $\ $ template we can
generate a GHZ state preparation subcircuit with 10 qubits.

\medskip
\para{Addressing Gap 2}
Database benchmarks typically instantiate each query template independently, yielding
a workload that is a set (or multiset) of queries without sequential dependencies among
template choices. In contrast, as discussed in Gap~2, the circuits targeted by \sys are
\emph{ordered} compositions of subcircuits, and the choice of the next subcircuit is often
constrained by (and semantically coupled to) the current one. 
\revisionA{InferQ therefore samples templates sequentially: the already-chosen template affects which templates are allowed next and how probabilities over possible choices are updated.}

\medskip
\para{Design element 1: Explicit synergy rules}
\sys provides a set of user-configurable \emph{synergy rules} that encode statistical
dependencies between templates and bias generation toward semantically coherent pipelines. For example,
\texttt{StatePrep}\(\rightarrow\)\texttt{HamSim}\(\rightarrow\)\texttt{QPE}, 
\texttt{ModularExp}\(\rightarrow\)\texttt{QPE}, and \texttt{A}\(\rightarrow\)\texttt{GroverIter}\(\rightarrow\)\texttt{AmpEst} (Table~\ref{tbl:comb}).
Intuitively, these rules specify how earlier template choices should increase or decrease
the probability of selecting specific templates later, making the sampling distribution
explicitly history-dependent.


\medskip
\para{Definitions}
Let \(\mathcal S\) denote the space of subcircuits, and let \(T\) be the number of
templates (equivalently, subcircuits) generated for one circuit instance. Let \(B\) denote the
\emph{context space}, i.e., the current circuit generation status (e.g., remaining allowed templates, and  synergy rules
in Table~\ref{tbl:comb}). 
A \emph{circuit template} \(f\in F\) is a parameterized generator
\begin{equation}
\label{eq:f-definition}
f:\Theta_f \times B \to \mathcal S.
\end{equation}
Let \(\mathsf{feas}(b,f)\in\{0,1\}\) be a hard feasibility predicate
for choosing template \(f\in F\) under context \(b\in B\). 

\sys generates an ordered template sequence \(f_1,f_2,\dots,f_T\). At step \(t\), after
selecting \(f_t\) and sampling parameters \(\theta_t\in\Theta_{f_t}\), \sys instantiates the
corresponding subcircuit as
\begin{equation}
s_t = f_t(\theta_t, b_{t-1}),
\end{equation}
where \(b_{t-1}\in B\) is the current context given historical $t-1$ steps.

At each step \(t\in\{1,\ldots,T\}\), \sys maintains a categorical distribution \(p_t\) over the
template set \(F\). We write \(p_t(f)\) for the probability   assigned to template \(f\in F\), so
\begin{equation}
\sum_{f\in F} p_t(f)=1.
\end{equation}
A larger value of \(p_t(f)\) means that template \(f\) is more likely to be selected at step \(t\).


\medskip
\para{Design element 2: Conditional composition as a Markov transition model}
At step \(t\), \sys samples the next template from a categorical distribution over \(F\).
Given the current context \(b_{t-1}\), it proceeds in two steps:

\smallskip
\noindent\textbf{(1) Sample a template.}
First apply feasibility masking and renormalize:
\begin{equation}
\label{eq:feasibility-mask}
\bar p_t(f)
=
\frac{p_t(f)\,\mathsf{feas}(b_{t-1},f)}
{\sum_{g\in F} p_t(g)\,\mathsf{feas}(b_{t-1},g)},
\qquad f\in F,
\end{equation}
then sample \(f_t\sim \bar p_t\).
\revisionA{Next, InferQ samples local parameters for \(f_t\), instantiates the corresponding subcircuit, appends it to the current circuit prefix, and updates the generation context.}

\smallskip
\noindent\textbf{(2) Reweight to obtain the next distribution.}
Let \(\mathcal R\) be the set of synergy rules, where each rule is a triple \((T,U,\beta)\) with
\(T\subseteq F\) (trigger), \(U\subseteq F\) (target), and \(\beta>0\).
After selecting \(f_t\), define the reweighted (unnormalized) probabilities
\begin{equation}
\label{eq:synergy-reweight}
\widehat p_{t+1}(f)
=
\bar p_t(f)\cdot
\prod_{(T,U,\beta)\in\mathcal R:\; f_t\in T,\; f\in U}\beta,
\qquad f\in F,
\end{equation}
and then apply feasibility under the updated context \(b_t\) and renormalize:
\begin{equation}
\label{eq:next-template-dist}
p_{t+1}(f)
=
\frac{\widehat p_{t+1}(f)\,\mathsf{feas}(b_t,f)}
{\sum_{g\in F}\widehat p_{t+1}(g)\,\mathsf{feas}(b_t,g)},
\qquad f\in F.
\end{equation}
\revisionA{Equations~\ref{eq:feasibility-mask}--\ref{eq:next-template-dist}
keep the generation process simple: feasibility masking removes disallowed templates that do not fit the current circuit, while synergy rules encourage quantum subcircuits (primitives) that commonly appear together. This lets InferQ generate circuits that follow quantum algorithm design principles without enumerating every full circuit by hand.}
\revisionC{
Allowed and disallowed compositions are captured by feasibility masking: a disallowed successor template receives probability zero in the current generation context and therefore cannot be sampled. 
The synergy rules are soft preferences over feasible successor circuit templates, guiding the generator toward meaningful algorithmic patterns, such as HamSim$\rightarrow$QPE, without hard-coding a fixed enumeration of complete circuits. 
This separation is important because quantum algorithms, quantum hardware, and their simulation needs are evolving rapidly. 
The current design keeps \sys configurable as new quantum primitives and workloads emerge.
}

\begin{example}[Single-step reweighting under a synergy rule]
\label{ex:pro_update}
\small
Assume the previously selected template is \texttt{StatePrep}. By Table~1,
the feasible successors are \(\{\texttt{HamSim},\texttt{ModularExp},\texttt{GroverIter},\texttt{Variational}\}\),
with a uniform distribution \(p_t(\cdot)=\tfrac{1}{4}\).
Let \(\mathcal{R}\) include \((T,U,\beta)=(\{\texttt{StatePrep}\},\\  \{\texttt{HamSim}\},2)\).
Then
\[
\begin{aligned}
\widehat p_{t+1}(\texttt{HamSim})&=\tfrac{1}{2}, &
\widehat p_{t+1}(\texttt{ModularExp})&=\tfrac{1}{4},\\
\widehat p_{t+1}(\texttt{GroverIter})&=\tfrac{1}{4}, &
\widehat p_{t+1}(\texttt{Variational})&=\tfrac{1}{4},
\end{aligned}
\]
and after normalization,
\[
\begin{aligned}
 p_{t+1}(\texttt{HamSim})&=\tfrac{2}{5}, &
 p_{t+1}(\texttt{ModularExp})&=\tfrac{1}{5},\\
p_{t+1}(\texttt{GroverIter})&=\tfrac{1}{5}, &
 p_{t+1}(\texttt{Variational})&=\tfrac{1}{5},
\end{aligned}
\]
\end{example}

\subsection{Sample Parameters}

Circuit generation is constrained by the global parameters in
Table~\ref{tab:base_params}, which are configured by users.
\sys requires users to provide ranges for key parameters, such as qubit count,
depth, 
repetitions count for repeatable subroutines, and number of qubits used during evaluation. A concrete value is drawn uniformly at random
from each range. If a user prefers a fixed value, they set the range endpoints equal
(e.g., \texttt{min\_qubits}=\texttt{max\_qubits}=10). This range-based interface provides more flexibility for users to construct workloads. \revisionA{For database evaluation over the generated simulation workload, these ranges play the role of scale factors: they control the size and shape of the generated SQL workloads.}

\subsection{Generate Subcircuits }
\label{sssec:subc}
Each selected circuit template \(f\) is instantiated by sampling parameters from its
parameter domain \(\Theta_f\). In \sys, we write one sampled
parameter tuple as
\begin{equation}
\theta = (n_q, d, k, r, \xi) \in \Theta_f,
\end{equation}
where \(n_q\) is the qubit count, \(d\) is the depth  allocated to this subcircuit,
\(k\) is the evaluation-qubit count, \(r\) is the repetition count for repeatable template
sections, and \(\xi\) is a tuple of template-specific configuration variables.
\revisionA{Moreover, \(n_q\) and \(d\) control the circuit size, while \(k\) and \(r\) control the evaluation and repetition settings. That is, these parameters determine the generated circuit structure and, therefore, the number and shape of relational operators in the generated SQL workload.}

The domain \(\Theta_f\) is derived from the  parameters  in Table~\ref{tab:base_params}. For example, the per-subcircuit depth \(d\) is sampled uniformly from the defined range between  min\_depth and max\_depth. 
\revisionA{Quantum-specific choices about \(\xi\) can be found in Appendix~K of our technical report \cite{inferqtech}.}
\begin{table}[t]
  \caption{Parameters.}
  \label{tab:base_params}
  \centering

  {\small
  \setlength{\tabcolsep}{4pt}
  \renewcommand{\arraystretch}{1.0}

  \begin{tabularx}{\linewidth}{@{}l l Y@{}}
    \toprule
    \textbf{Parameter} & \textbf{Type} & \textbf{Description} \\
    \midrule
    \texttt{min\_qubits}
      & int
      & Minimum number of qubits per generated circuit \\

    \texttt{max\_qubits}
      & int
      & Maximum number of qubits per generated circuit \\

    \texttt{min\_depth}
      & int
      & Minimum target depth of a generated quantum circuit \\

    \texttt{max\_depth}
      & int
      & Maximum target depth of a generated quantum circuit \\

    \texttt{min\_reps}
      & int
      & Minimum number of repetitions for repeatable subroutines
        (e.g., Grover iterations or variational layers) \\

    \texttt{max\_reps}
      & int
      & Maximum number of repetitions for repeatable subroutines \\

    \texttt{min\_eval\_qubits}
      & int
      & Minimum number of qubits used during evaluation \\

    \texttt{max\_eval\_qubits}
      & int
      & Maximum number of qubits used during evaluation \\

    \texttt{measure}
      & bool
      & Whether measurements are appended to the circuit \\

    \texttt{seed}
      & int
      & Global random seed to ensure full reproducibility of
        circuit generation \\
    \bottomrule
  \end{tabularx}
  }
\end{table}

\subsection{Output Circuit}
\label{sssec:merge}
Generation stops when a maximum number of templates is reached or when a probabilistic stopping condition triggers.
This prevents excessively long circuits while keeping circuit length variable across instances. 
After termination, the instantiated subcircuits \(S_1,\ldots,S_T\) are composed in order to
produce the final benchmark circuit:
\begin{equation}
C = S_1 \circ S_2 \circ \cdots \circ S_T.
\end{equation}
Alongside \(C\), \sys outputs a provenance record that stores the template sequence
\((f_1,\ldots,f_T)\), sampled parameters (including repetitions and \(\xi\)), and global
resource statistics (e.g., depth and gate counts). This record makes our generated circuit instances
easy to inspect, query, and reproduce. \revisionA{This provenance is useful for database benchmarking because the same randomized workload can be replayed and compared across engines.}

\medskip
\para{SQL query generation} Once a circuit is generated, we construct an equivalent SQL query using the implementation in \cite{hai2025quantum}. We extended their package by making it a dependency to InferQ. The package uses the einstein summation (einsum) representation of a quantum circuit with tensor algebra. It then converts tensor contractions into relational  joins as proposed by Blacher et al. \cite{einsteinSQL2023}. 

\medskip
\para{Reproducibility guarantee} To make sure 
all sampling in \sys is deterministic, we use seeded pseudo-random number generators\footnote{See
\url{https://docs.python.org/3/library/random\#random.seed} and
\url{https://numpy.org/doc/2.2/reference/random/generated/numpy.random.seed.html}.} and maintain a global random seed as part of the parameters in Table~\ref{tab:base_params}. Consequently, with the seed and the generated template
sequence, both the subcircuit structure and the numerical gate parameters are 
reproducible across runs.

\input{example}

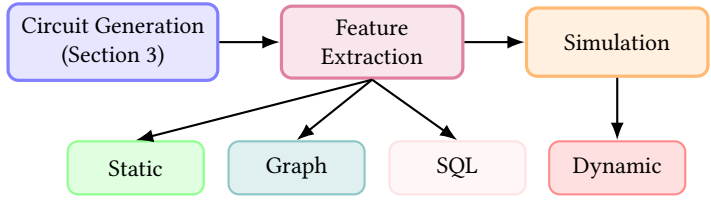
\begin{figure}[t]
\centering
\begin{tikzpicture}[
  font=\small,
  >=Latex,
  node distance=8mm and 8mm,
  main/.style={
    draw,
    rounded corners,
    very thick,
    align=center,
    inner sep=5pt,
    minimum width=24mm,
    minimum height=9mm
  },
  outputbox/.style={
    draw,
    rounded corners,
    thick,
    align=center,
    inner sep=4pt,
    minimum width=18mm,
    minimum height=7mm
  },
  arrow/.style={->, thick}
]

\node[
  main,
  fill=blue!10,
  draw=blue!50
] (gen) {Circuit Generation\\(\revisionB{Section 3})};

\node[
  main,
  fill=purple!10,
  draw=purple!50,
  right=of gen
] (feat) {Feature\\Extraction};

\node[
  main,
  fill=orange!12,
  draw=orange!55,
  right=of feat
] (sim) {Simulation};

\node[
  outputbox,
  fill=teal!12,
  draw=teal!45,
  below=of feat,
  xshift=-10mm
] (graph) {Graph};

\node[
  outputbox,
  fill=green!12,
  draw=green!45,
  left=3mm of graph
] (stat) {Static};

\node[
  outputbox,
  fill=pink!12,
  draw=pink!45,
  right=3mm of graph
] (sql) {SQL};

\node[
  outputbox,
  fill=red!12,
  draw=red!45
] (dyn) at (sql -| sim) {Dynamic};

\draw[arrow] (gen.east) -- (feat.west);
\draw[arrow] (feat.east) -- (sim.west);

\draw[arrow] (feat.south) -- (stat.north);
\draw[arrow] (feat.south) -- (graph.north);
\draw[arrow] (feat.south) -- (sql.north);
\draw[arrow] (sim.south) -- (dyn.north);

\end{tikzpicture}

\caption{\sys feature extraction workflow. Static, graph-based, and
SQL-derived features are extracted prior to simulation, whereas dynamic
features can only be obtained after the simulation completes.}
\label{fig:feat_overview}
\end{figure}

\section{Circuit Feature Extraction}
\label{ssec:features}

When we push quantum circuit simulation into an RDBMS, it induces a sequence of
relational operators (e.g., joins and aggregations implementing tensor
contractions) over large intermediate relations. In this setting, performance
depends on cardinality growth, join selectivity, skew, and the size of
intermediate states. However, such properties are not explicit in the quantum 
circuit representation. 
To make systematic progress on query optimization,
physical design (e.g., indexing/partitioning), and \emph{routing} (deciding when
an RDBMS backend is beneficial), we need a set of circuit features that
indicate what the induced relational query workload looks like. Ideally, a
feature vector that approximates workload characteristics \emph{before}
executing the full simulation.
Therefore, we include a feature extraction component in \sys and aim to cover
different sources of cost, including problem size (number of qubits, circuit
depth), interaction structure,  state evolution, and the generated SQL queries. 
As shown in Figure~\ref{fig:feat_overview}, \sys models each circuit with four types
of features: \emph{static}, \emph{graph}, \emph{dynamic}, and \emph{SQL} features, as listed in Table~\ref{tbl:static}--\ref{tbl:dyn}, respectively. Such a
separation makes \sys \textbf{extensible}: new features can be added easily.  

Notably, as shown in Figure~\ref{fig:feat_overview},
feature extraction is an optional post-processing step: \sys can generate circuits
without extracting features, and computes them only when needed for analysis. 



\subsection{Static Features}
Static features describe basic properties of a quantum circuit. These features
are crucial for constructing heuristics for efficient classical simulation.
For example, dense state-vector representations are only feasible for circuits
with small width and a small number of qubits, as memory requirements grow
exponentially with these parameters. For an RDBMS-backed simulator, gate counts
and depth largely determine the number and types of relational operators
generated.

\begin{table}[t]
  \caption{Notation used in circuit feature extraction.}
  \label{tab:notation}
  \centering
  \small
  \setlength{\tabcolsep}{3pt}
  \renewcommand{\arraystretch}{1.05}
  \begin{tabularx}{0.88\columnwidth}{@{}l>{\footnotesize}X@{}}
    \toprule
    \textbf{Notation} & \textbf{Description} \\
    \midrule
    $C$ & Quantum circuit (ordered list of gate instructions). \\
    $G$ & Number of circuit instructions (gate count). \\
    $N$ & Number of qubits in the circuit. \\
    $W$ & Circuit width (maximum simultaneously active qubits). \\
    $G_I=(V,E)$ & Qubit interaction graph induced by multi-qubit gates. \\
    $V$ & Vertex set of $G_I$ (qubits; $|V|=N$). \\
    $E$ & Edge set of $G_I$ (qubit interactions; $|E|$ is the edge count). \\
    $2^N$ & State-space dimension (length of a full statevector). \\
    $\mathbf{p}=(p_1,\ldots,p_{2^N})$ & Probability vector derived from the saved statevector. \\
    $H(\mathbf{p})$ & Shannon entropy of $\mathbf{p}$ (superposition). \\
    $\rho(\mathbf{p})$ & Sparsity / support density: fraction of entries above threshold (superposition). \\
    $S_{\mathrm{vN}}^{(k)}$ & von Neumann entropy across qubit line $k$ (entanglement). \\
    $\tau$ & Threshold for treating near-zero entries as zero in sparsity computation. \\
    $\mathbb{I}(\cdot)$ & Indicator function ($1$ if condition holds, else $0$). \\
    $N_{\mathrm{AST}}$ & Total number of nodes in the SQL abstract syntax tree (AST). \\
    $S$ & Number of \texttt{SELECT} statements in the query. \\
    $P$ & Number of predicates appearing in \texttt{WHERE} clauses. \\
    $J$ & Number of distinct relations participating in join conditions. \\
    \bottomrule
  \end{tabularx}
    \vspace{-0.4cm}
\end{table}

As shown in Table~\ref{tbl:static}, static features can be computed efficiently.
We treat a circuit as an ordered list of instructions, where each instruction
applies a gate to one or more qubits. This view allows us to analyze the
complexity of computing static features. Let $G$ be the number of gates
(instructions), $N$ the number of qubits, and $W$ the circuit width (maximum
simultaneously active qubits). We assume $W \le N$ and typically $G \gg N$. We
summarize the notation in Table~\ref{tab:notation}. All static features in
Table~\ref{tbl:static} are computable in at most linear time in $G$ with modest
memory overhead, making them inexpensive yet informative metadata for the \sys
benchmark.

\begin{table}[t]
\centering
\caption{Time and space complexities for static features.}
\label{tbl:static}

{\small
\setlength{\tabcolsep}{4pt}
\renewcommand{\arraystretch}{1.0}
\begin{tabularx}{0.8\columnwidth}{@{}lccY@{}}
\hline
\textbf{Feature} & \textbf{Time} & \textbf{Space} & \textbf{Description} \\
\hline
num\_qubits & $O(1)$ & $O(1)$ & Total number of qubits \\
width & $O(G)$ & $O(W)$ & Maximum simultaneous active qubits \\
depth & $O(G)$ & $O(W)$ & Longest dependency chain \\
circuit\_size & $O(1)$ & $O(1)$ & Total gate count \\
pauli\_gate\_count & $O(G)$ & $O(1)$ & Number of Pauli gates (X, Y, Z) \\
two\_qubit\_gate\_count & $O(G)$ & $O(1)$ & Number of two-qubit gates \\
two\_qubit\_gate\_percentage & $O(G)$ & $O(1)$ & Ratio of two-qubit gates \\
locality\_ratio & $O(G)$ & $O(1)$ & Local vs.\ non-local gate ratio \\
idling\_score & $O(G)$ & $O(W)$ & Idle qubit utilization \\
density\_score & $O(G)$ & $O(G)$ & Gate density feature \cite{bandic2024profilingquantumcircuitsefficient} \\
gate\_counts & $O(G)$ & $O(G)$ & Per-gate-type histogram \\
\hline
\end{tabularx}
}
\end{table}

\subsection{Graph Features}
\label{ssec:graphf}
Graph features are computed from the \emph{qubit interaction graph}, where
vertices are qubits and edges represent interactions induced by multi-qubit
gates. This abstraction enables the computation of classical graph properties,
similar to existing works on quantum circuit compilation \cite{Bandic_2023,bandic2024profilingquantumcircuitsefficient}
and graph-based circuit characterization \cite{Hernndez2015ClassificationOG}.

Graph features make circuit structure directly accessible to database analysis:
they indicate what the join structure and intermediate-result growth may look
like in relational tensor computation. Intuitively, circuits whose interaction
graphs are highly connected (e.g., with diameter-reducing shortcuts, high node
degree, and strong centralization) tend to induce more global correlations and
fewer separable subproblems, which often translates into harder contractions and
less opportunity for decomposition.
\begin{table}[t]
\centering
\caption{Time and space complexities for graph features.}
\label{tbl:graph}
{\small
\setlength{\tabcolsep}{4pt}
\renewcommand{\arraystretch}{1.0}
\begin{tabularx}{\columnwidth}{@{}lccY@{}}
\hline
\textbf{Feature} & \textbf{Time} & \textbf{Space} & \textbf{Description} \\
\hline
edge\_count & $O(1)$ & $O(1)$ & Number of interaction edges \\
max\_degree & $O(N)$ & $O(1)$ & Maximum node degree \\
min\_cut & $O(NE)$ & $O(V+E)$ & Minimum cut \\
diameter & $O(N^3)$ & $O(N^2)$ & Graph diameter \\
radius & $O(N^3)$ & $O(N^2)$ & Minimum eccentricity \\
average\_degree & $O(N)$ & $O(1)$ & Mean node degree \\
average\_clustering\_coefficient & $O(N^3)$ & $O(1)$ & Local clustering tendency \\
average\_shortest\_path\_length & $O(N^3)$ & $O(N^2)$ & Mean shortest path length \\
central\_point\_of\_dominance & $O(NE)$ & $O(N)$ & Graph centralization measure \\
std\_dev\_adjacency\_matrix & $O(N^2)$ & $O(1)$ & Adjacency matrix variance \\
\hline
\end{tabularx}
}
\end{table}
As summarized in Table~\ref{tbl:graph}, graph features can be computed in
polynomial time and space with respect to qubit count $N$, with worst-case
complexity bounded by $O(N^3)$. Let $E$ be the number of edges in the
interaction graph. Under a gate set dominated by one- and two-qubit gates, each
gate introduces at most one interaction edge; hence $E=O(G)$, with worst-case
space $O(N^2)$ for a weighted, undirected interaction graph. Building an
adjacency-list representation takes $O(N+E)$ time and space. While some
features require $O(N^3)$ time in the worst case, they remain practical for the
qubit ranges where graph characterization is useful, and can be computed
offline.

\subsection{SQL Features}
\label{ssec:sql}
SQL features are extracted from the generated SQL queries, which are relevant to the performance of RDBMS engines for quantum circuit simulation. 
 Table \ref{tab:sql_complexity}  summarizes the complexities to calculate these features from the circuit SQL query representation. To calculate the various joins and other clauses, we view the query tree as an abstract syntax tree (AST) and then do a traversal. Getting to this SQL representation of a quantum circuit is not expensive. Using existing methods \cite{Littau_2025, hai2025quantum}, we can generate these queries by translating the quantum circuit into Einstein summation (einsum) notation representing the tensor contraction sequence.







\begin{table}[t]
\centering
\caption{Time and space complexities for extracting SQL features (excluding parsing).}
\label{tab:sql_complexity}
\resizebox{\columnwidth}{!}{%
\begin{tabular}{lccc}
\hline
\textbf{Feature} & \textbf{Time} & \textbf{Space} & \textbf{Description} \\
\hline
num\_and\_clauses 
& $O(N_\text{AST})$ 
& $O(1)$ 
& Counts logical AND operators during AST traversal \\

num\_joins 
& $O(S \cdot P + J^2)$ 
& $O(J^2)$ 
& Counts explicit and implicit joins and stores join edges \\

num\_select\_columns 
& $O(N_\text{AST})$ 
& $O(1)$ 
& Counts column expressions in SELECT clauses \\

num\_agg\_funcs 
& $O(N_\text{AST})$ 
& $O(1)$ 
& Counts aggregate functions (e.g., COUNT, SUM, AVG) \\

num\_where\_clauses 
& $O(N_\text{AST})$ 
& $O(1)$ 
& Counts WHERE clause nodes in the AST \\

num\_eq\_predicates 
& $O(N_\text{AST})$ 
& $O(1)$ 
& Counts equality predicates used in filters and joins \\

\hline
\end{tabular}%
}
\end{table}


\subsection{Dynamic Features}
As shown in Table~\ref{tbl:dyn}, dynamic features depend on the circuit's
\emph{output state} and cannot be derived from circuit structure alone. They
are crucial for database-backed simulation because relational approaches
typically rely on \emph{sparse} or \emph{factorized} state representations: if
the state remains sparse, relational operators can operate over small relations
and avoid exponential blow-up; if entanglement grows quickly, intermediate
results densify and the advantage diminishes.  We therefore extract three dynamic
features that capture these effects: (i) Shannon entropy, (ii) von Neumann entropy, and (iii) a sparsity (non-zero support) measure of the
saved statevector.

\begin{table}[t]
\caption{Time and space complexities for dynamic features.}
\label{tbl:dyn}
\centering

{\small
\setlength{\tabcolsep}{4pt}
\renewcommand{\arraystretch}{1.0}
\begin{tabularx}{\columnwidth}{@{}lccY@{}}
\hline
\textbf{Feature} & \textbf{Time} & \textbf{Space} & \textbf{Description} \\
\hline
Shannon\_entropy& $O(2^N)$ & $O(2^N)$ & Shannon Entropy of stored statevector\\
von\_neumann\_entropy& $O(N \cdot 2^N)$& $O(N \cdot 2^N)$&Von Neumann Entropy of stored statevector\\
sparsity & $O(2^N)$ & $O(2^N)$ & Sparsity of stored statevector \\
\hline
\end{tabularx}
}
\end{table}

Using the vector representation for the quantum state defined in
Section~\ref{sec:pre}, we define sparsity and entropies as follows:

Let $\mathbf{p} = (p_1,\dots,p_{2^N})$ denote the probability distribution obtained from an $N$-qubit statevector in the computational basis. We compute the Shannon entropy, analogous to its use in classical and quantum information, as
\begin{equation}
H(\mathbf{p}) = - \sum_{i=1}^{2^N} p_i \log_2 p_i.
\end{equation}

To quantify entanglement, we compute the von Neumann entropy across a partition defined by a cut along a qubit line $k$,
\begin{equation}
S_{\mathrm{vN}}^{(k)} = -\mathrm{Tr}\!\left(\rho_{A_k}\log_2\rho_{A_k}\right),
\qquad
\rho_{A_k} = \mathrm{Tr}_{B_k}(\rho),
\end{equation}
where the bipartition corresponds to separating qubit line $k$ from the remaining qubit lines. We define the sparsity
\begin{equation}
\rho(\mathbf{p}) = \frac{1}{2^N} \sum_{i=1}^{2^N} \mathbb{I}(p_i > \tau),
\end{equation}
which we use synonymously with the support density, i.e., the fraction of basis states carrying non-negligible probability mass. 

Here, the Shannon entropy $H(\mathbf{p})$ and sparsity $\rho(\mathbf{p})$ characterize the degree of superposition in the computational basis, with higher values indicating broader support over basis states, while the von Neumann entropy $S_{\mathrm{vN}}^{(k)}$ measures quantum entanglement across each qubit line. We store the von Neumann entropy per qubit line to capture individual entanglement, rather than treating the full state as a combined vector, which being a pure state, has total entanglement entropy as zero. Characterizing both superposition and entanglement is essential for understanding the mechanisms underlying quantum advantage.

All dynamic features are expensive to compute and require exponential time and memory in $N$ for exact computation due to the full pass over the length-$2^{N}$ vector. In particular, computing the von Neumann entropy across each of the $N$ qubit lines incurs an additional factor of $N$ on top of the full vector pass against the rest of the system, resulting in $O(N \cdot 2^N)$ time and memory. This motivates an additional use case for InferQ in Section~\ref{ssec:uc2}: enabling accurate dynamic-feature estimators without requiring full simulation of the circuit. 

In addition, we highlight InferQ’s extensibility in supporting the addition of new features, which facilitates a wide range of use cases and experimental analyses.


%% file: tables/Table_composition.tex
\begin{table*}[t]
\caption{\revisionA{Representative circuit templates and example compositions. }}
\label{tbl:comb}
\centering
\scriptsize
\setlength{\tabcolsep}{3pt}
\renewcommand{\arraystretch}{1.35}

\begin{tabularx}{\textwidth}{@{} P{1.6cm} P{2.4cm} P{2.2cm} X P{3.2cm} @{}}
\toprule
\textbf{Current template} & \textbf{Full circuit template name} & \textbf{Next possible templates} & \textbf{Description} & \textbf{Example compositions} \\
\midrule

\famSTART
& (start symbol)
& \famStatePrep\candsep \famOracleA\candsep \famQFT\candsep \famVariational
& Common entry points (prepare input/eigenstate/ansatz), enabling realistic multi-block pipelines.
& \vis{\famSTART\qarrow\famStatePrep\\[-1pt]
       \famSTART\qarrow\famOracleA\\[-1pt]
       \famSTART\qarrow\famQFT} \\
\midrule 

\famStatePrep
& State Preparation
& \famHamSim\candsep \famModularExp\candsep \famGroverIter\candsep \famVariational
& Prepares \(|\psi\rangle\)/\(|b\rangle\) or initial superpositions before an algorithmic primitive.
& \vis{\famStatePrep\qarrow\famHamSim\\[-1pt]
       \famStatePrep\qarrow\famVariational} \\
\famHamSim
& Hamiltonian Simulation
& \famQPE\candsep \famVariational\candsep \famRandom
& Controlled time-evolution unitaries are typically read out via QPE (incl.\ IQFT decoding) in eigenvalue/spectral workflows.
& \vis{\famHamSim\qarrow\famQPE} \\       
\famQFT
& Quantum Fourier Transform
& \famQPE\candsep \famMeasure\candsep \famRandom
& Fourier-basis transform and dense benchmark block. 
& \vis{\famQFT\qarrow\famQPE\\[-1pt]
       \famQFT\qarrow\famMeasure} \\ 
\famOracleA
& Oracle/Algorithm-\(A\) Wrapper 
& \famGroverIter\candsep \famVariational\candsep \famMeasure
& Encodes a problem-dependent subroutine \(A\) (e.g., oracle/feature map/encoding) that is naturally followed by Grover-style amplification.
& \vis{\Ablock\qarrow\famGroverIter\qarrow\famAmpEst} \\
\famModularExp
& Modular Exponentiation
& \famQPE
& Order-finding and Shor-style structure: modular arithmetic unitaries followed by phase estimation (incl.\ IQFT) to extract phases/periods.
& \vis{\famModularExp\qarrow\famQPE} \\
\famGroverIter
& Grover Iteration 
& \famGroverIter\ (repeat)\candsep \famAmpEst\candsep \famMeasure
& Amplification layers typically repeat and then transition to an estimation wrapper or terminate with measurement.
& \vis{\famGroverIter\qarrow\famGroverIter\\[-1pt]
       \famGroverIter\qarrow\famAmpEst\\[-1pt]
       \famGroverIter\qarrow\famMeasure} \\
\famAmpAmpl
& Amplitude Amplification
& \famMeasure
& Boosts the probability of a desired or postselected outcome. 
& \vis{\famAmpAmpl\qarrow\famMeasure} \\
\famQPE
& Quantum Phase Estimation
&\famAmpAmpl\candsep \famMeasure\candsep \famVariational\candsep \famRandom
& Phase readout/decoding stage (typically includes an IQFT); often followed by measurement or optional continuations.
& \vis{\famQPE\qarrow\famAmpAmpl\\[-1pt]
       \famQPE\qarrow\famMeasure\\[-1pt]
       \famQPE\qarrow\famVariational\\[-1pt]
       \famQPE\qarrow\famRandom} \\

\famVariational
& Variational Circuit Layer/ Ansatz
& \famVariational\ (repeat)\candsep \famMeasure\candsep \famRandom
& Layered ansatz patterns; measurement used for objective estimation; can be interleaved for hybrid workloads.
& \vis{\famVariational\qarrow\famVariational\\[-1pt]
       \famVariational\qarrow\famMeasure\\[-1pt]
       \famVariational\qarrow\famRandom} \\

\famRandom
& Random Circuit Block
& \famRandom\candsep \famMeasure
& Diversity/stress-test tail; may extend circuit length or terminate by measurement.
& \vis{\famRandom\qarrow\famRandom\\[-1pt]
       \famRandom\qarrow\famMeasure} \\

\famAmpEst
& Amplitude Estimation
& \famMeasure
& Estimates a success amplitude or probability. It normally terminates the quantum subroutine by measurement and classical post-processing.
& \vis{\famAmpEst\qarrow\famMeasure} \\
\midrule 


\famMeasure
& Measurement  
& \famEND
& Final step that yields classical outcomes and ends the whole circuit generation process.
& \vis{\famMeasure\qarrow\famEND} \\
\famEND
& (end symbol)
& ---
& 
It marks the end of the template sequence.
& \vis{\famEND} \\
\bottomrule
\end{tabularx}

\end{table*}

%% file: example.tex
\subsection{Running Example}
\label{sec:worked_example_qft_qpe}

Figure~\ref{fig:sec41_running_example} illustrates one end-to-end circuit instance generated by \sys.
The workflow is intentionally analogous to a database benchmark that (i)  selects a template sequence, (ii) samples a scale/configuration,
(iii) instantiates a small sequence of templates with parameters, and  (iv) materializes both an artifact and its metadata.

\textbf{Step 1 (Template selection).} Starting from START, \sys samples a short
template sequence using the history-dependent distribution. Here the sequence is
\emph{QFT} $\rightarrow$ \emph{QPE}.

\textbf{Step 2--3 (Parameters \& subcircuits generation).}  Given a user seed, \sys samples an instance configuration
$(n_q, d, k, r, \xi)$. In this example, $n_q{=}7$ qubits, depth  $d{=}80$, evaluation  $k{=}4$,
and repetition count $r{=}1$. Each template is then instantiated by binding
template-local parameters (e.g., boolean flags and numeric values such as $\phi{=}1.998$). All choices are deterministic given the seed.

\textbf{Step 4 (Merge \& provenance).} The instantiated subcircuits are composed in order to produce the
final circuit artifact $C = S_1 \circ S_2$. Alongside $C$, \sys stores a provenance record containing the
template sequence and all sampled parameters, enabling exact replay and database-style inspection.


%% file: tool.tex
\begin{figure}[t]
    \centering
    \includegraphics[width=0.7\linewidth]{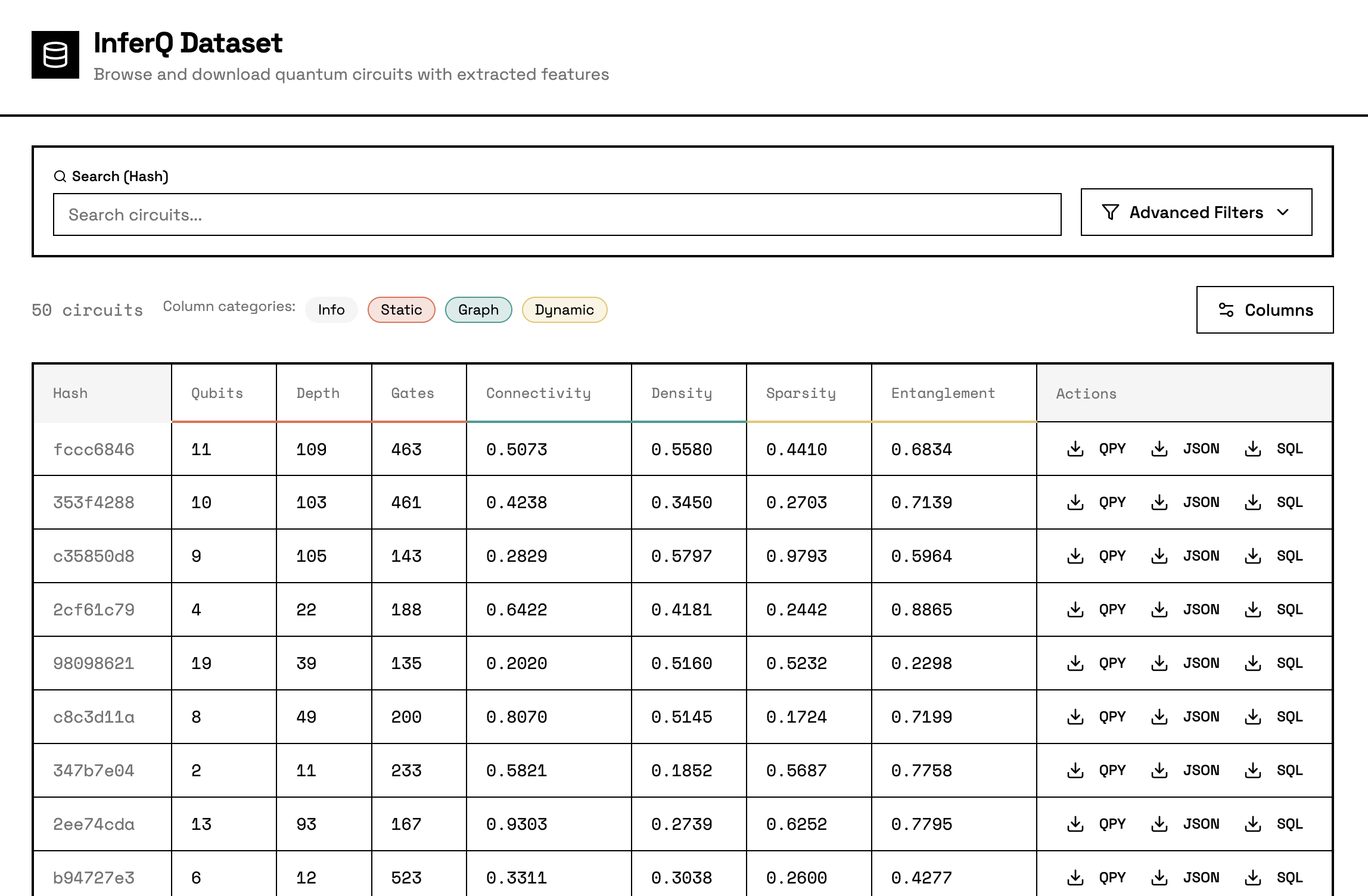}
    \caption{InferQ's web-based UI for exploring dataset and circuit features. }
    \label{fig:inferq-web}
\end{figure}

\section{Implementation and Artifacts}
\label{sec:impl}

\sys ships as (i) a Python framework that turns a user configuration
(Table~\ref{tab:base_params}) into \emph{circuits, features, and an RDBMS-ready SQL
workload} for simulation, and (ii) a web-based UI for interactive dataset exploration
(Fig.~\ref{fig:inferq-web}). Concretely, a database researcher can provide a small
configuration file (e.g., qubit count) and run the toolkit to obtain  SQL
queries (plus circuit artifacts in JSON and QPY formats) that are ready to execute in an
RDBMS engine. \revisionM{Detailed descriptions of all circuit templates and the corresponding SQL code are provided in the online documentation.\footnote{\url{https://github.com/InfiniData-Lab/InferQ/blob/main/generators/sql-templates.md}}}
In addition to on-demand generation, \sys releases a large dataset of \num{202975} circuits online.\footnote{\url{https://github.com/InfiniData-Lab/InferQ/blob/main/analysis/training_data/estimator_training_data.parquet}}
Both database researchers and quantum scientists can search, filter, and download
circuits and feature records through the web UI.

The framework implements the pipeline described in
Section~\ref{sec:benchmark} and Section~\ref{ssec:features}.
Each circuit is assigned a \emph{content-based hash} computed from its structure and
sampled parameters. This hash serves as the primary key and links the circuit artifact,
the SQL workload, and all extracted feature records. For each circuit instance, \sys stores:
(i) the generated SQL queries (as \texttt{.sql} scripts) that represent
the simulation workload,
(ii) the circuit in a portable Qiskit serialization format (e.g., \texttt{.qpy}), and
(iii) a JSON record containing the template provenance and extracted features. 
\revisionB{InferQ includes a metadata analysis function that aggregates circuit statistics from JSON files, such as gate mix, depth, width, and interaction-graph properties. Details are provided in Appendix~J of \cite{inferqtech} and in the online repository.\footnote{\url{https://github.com/InfiniData-Lab/InferQ/blob/main/analysis/distributions/distributions.py}}}

 \medskip
\para{Reproducibility} The code of our benchmark can be found online.\footnote{\url{https://github.com/InfiniData-Lab/InferQ/blob/main/README.md}}
\sys is designed to be reproducible. Given the same code revision, dependency
lockfile, and global seed (Table~\ref{tab:base_params}), the framework deterministically
reproduces the same template sequence, sampled parameters, circuit artifact, emitted SQL,
and extracted features. The content-based hash and provenance record make each benchmark
instance easy to reference, regenerate, and audit. We include more implementation and reproducibility
details in our technical report \cite{inferqtech}.

%% file: evaluation.tex
\section{Evaluation}
\label{sec:eva}

 
In this section, we use \sys to quantify and explain when an RDBMS engine is a competitive simulation engine for \emph{general, compositional} quantum circuits. First, in Section~\ref{ssec:simu} we compare the \sys-emitted SQL workloads on \revisionB{four} RDBMS engines (PostgreSQL, SQLite, DuckDB, \revisionB{and Umbra}) against Qiskit Aer for wall-clock runtime and peak memory. We have also trained lightweight selectors (linear and tree models) that predict when RDBMS should be chosen over Qiskit Aer. We validate the feature design via feature-importance analysis in Section~\ref{ssec:feat_val} and demonstrate additional data-centric tasks enabled by \sys in Section~\ref{ssec:uc2}. 

\subsection{Experimental Settings}

All experiments were run on a  workstation equipped with an Intel Core
i9--12900KF (24 threads), 64\,GB RAM, and an NVIDIA RTX~3090 GPU, running
Ubuntu~20.04.6~LTS with Linux kernel~5.15.0. To ensure a fair comparison between
different RDBMS engines and other quantum circuit simulators, we disabled GPU and other
hardware acceleration; all simulations were executed on CPU.

We implemented our prototype in Python~3.12, with all dependencies pinned in
\texttt{pyproject.toml}. Core numerical routines use NumPy~2.2.5, SciPy~1.15.3, and
scikit--learn~(\(\ge\)1.8). For in-memory quantum-state simulation, we use
Qiskit~2.0.1 together with Qiskit Aer~0.17.0. For database-backed simulation, we
generate SQL using the \texttt{sql-einsum} code generator\footnote{\url{https://github.com/ti2-group/sql-einsum/tree/main/generate_sql_code}}
and build on the RDBMS simulator introduced by \cite{hai2025quantum}. 
PostgreSQL is version~12.22 
accessed via
\texttt{psycopg2-binary}~2.9.10. SQLite is version~3.50.4, and DuckDB is version~1.1.3. \revisionB{
Umbra (\texttt{v0.2-1665-gfeeb4bb5d}) is accessed through its PostgreSQL-compatible (18.1) server interface using
\code{psycopg2-binary}~2.9.10.}
\revisionA{For the out-of-core experiments, memory limits are enforced with Docker Engine~($\ge$24.0) using per-container cgroups.}
Unless otherwise stated, we report the mean values across
five repetitions.


\input{temp2}

\begin{table}[t]
\caption{Sparsity and Entropy (Dynamic metrics) estimators performance for Random Forest and Linear Regression models. Reported are coefficient of determination ($R^2$) and root mean squared error (RMSE) comparing predictions to actual values.}
\label{tab:estimators}
\begin{tabular}{lcc}
\toprule
Model & Sparsity Model & Entropy Model \\
\midrule
Random Forest $R^2$ & 0.7737 & 0.8936 \\
Random Forest RMSE  & 0.1928 & 1.9931 \\
\midrule
Linear Regression $R^2$ & 0.3934 & 0.6956 \\
Linear Regression RMSE  & 0.3156 & 3.3714 \\
\bottomrule
\end{tabular}
\end{table}

\begin{figure}[t]
  \centering
  \begin{minipage}{\linewidth}
   \centering
    \includegraphics[width=0.7\linewidth]{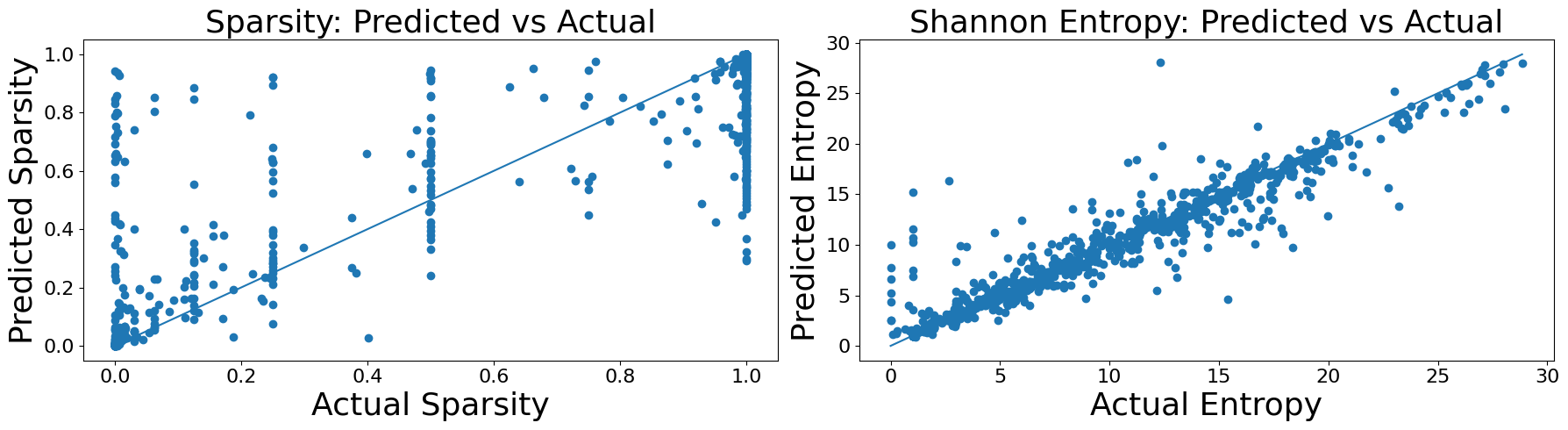}
  \end{minipage}
  \caption{Predicted vs Actual value for the dynamic feature RF regression estimators for subset (1000 entries) of test set.}
  \label{fig:estimators}
  \vspace{0.2cm}
\end{figure}


\begin{figure}[t]
  \centering
  \includegraphics[width=0.75\linewidth]{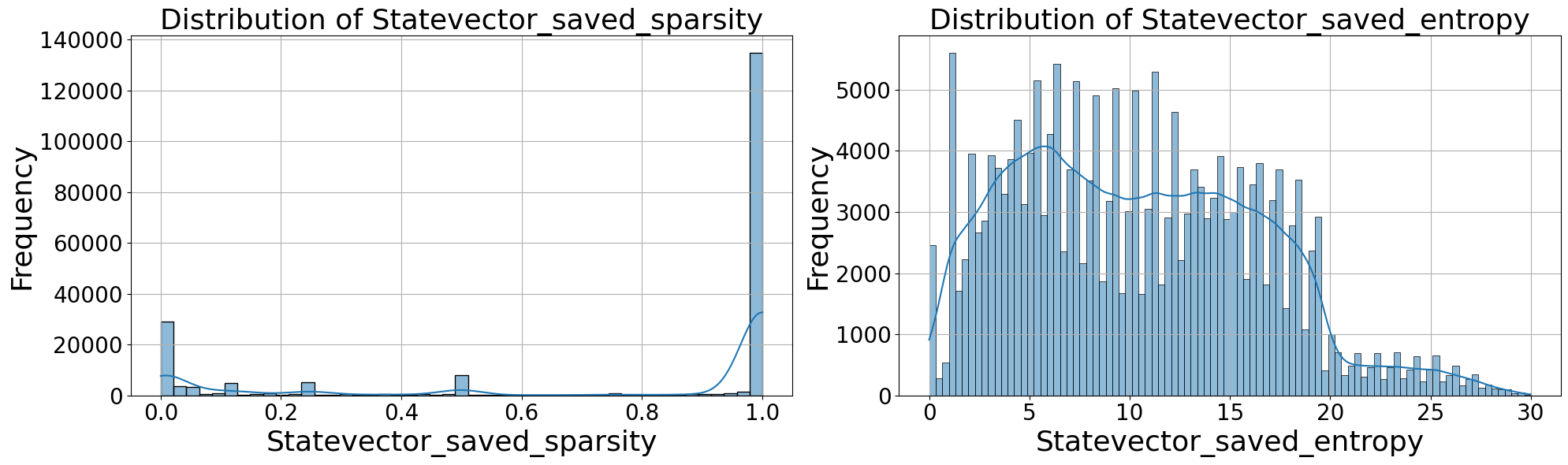}
  \caption{Distribution of dynamic features over 202k entries in InferQ with Dynamic features.}
    \vspace{0.2cm}
  \label{fig:distribution}
\end{figure}

\subsection{Use case 2: Predicting Entropies and Sparsity}
\label{ssec:uc2}
Dynamic features from Section~\ref{ssec:features} (Shannon entropy, von Neumann entropy, and sparsity) are only available \emph{after} simulation, since they are computed from the output state. This limits their use for larger circuits and motivates a second, data-centric use case: learning inexpensive estimators that predict these dynamic properties from static/graph/SQL features. Besides being useful for database researchers (Use case 1), these metrics are also fundamental to quantum analysis.

\begin{figure}[t]
  \centering
  \begin{minipage}{0.37\linewidth}
    \centering
    \includegraphics[width=\linewidth]
      {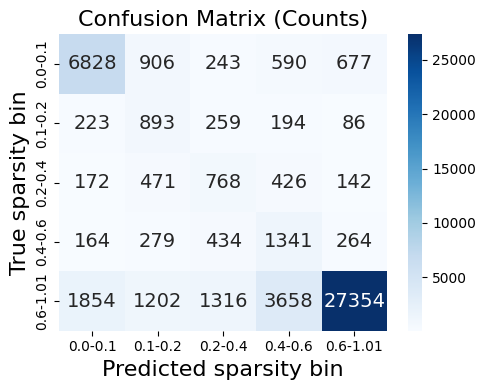}
  \end{minipage}%
  \hspace{0.02\linewidth}%
  \begin{minipage}{0.37\linewidth}
    \centering
    \includegraphics[width=\linewidth]
      {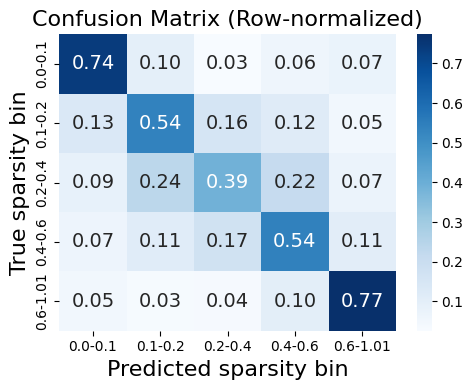}
  \end{minipage}
  \caption{Confusion matrices with counts and normalization for the random forest classifier used for sparsity-band prediction.}
  \label{fig:confusion}
\end{figure}

\vspace{0.2cm}
\noindent\emph{Q5: Can we estimate complex quantum properties using \sys features?}

\vspace{0.2cm}
We construct estimators for \feat{sparsity} and
\feat{shannon_entropy} using \num{202975} circuits for which
dynamic features are available.\footnote{Estimator training data (Parquet) available at \url{https://github.com/InfiniData-Lab/InferQ/blob/main/analysis/training_data/estimator_training_data.parquet}.}
We train (i) a Random Forest regressor and
(ii) a linear regression baseline, 
and evaluate
prediction quality using $R^2$ and RMSE as shown in Table~\ref{tab:estimators}.



As Table~\ref{tab:estimators} shows, Random Forest substantially outperforms the
linear model for both targets. Shannon entropy is modeled well by the feature
set, while sparsity is harder to regress accurately. This is also visible in the
prediction-versus-actual plots in Figure~\ref{fig:estimators}.

To understand why sparsity is challenging, we inspect its empirical distribution
(Figure~\ref{fig:distribution}). Sparsity is not smoothly distributed. Instead,
it concentrates in distinct bands. This suggests treating sparsity as a
coarse-grained classification problem. Concretely, we bin sparsity into
$[0.0,0.1),[0.1,0.2),[0.2,0.4),[0.4,0.6),[0.6,1.01]$ and train a Random Forest
classifier. This yields approximately $77\%$ accuracy, and the normalized confusion matrix
is concentrated near the diagonal (Figure~\ref{fig:confusion}), indicating that
\sys features capture sparsity  even when exact regression
is difficult.

Finally, feature-importance analysis for the sparsity-band classifier highlights
\texttt{pauli\_gate\_count} as a stronger predictor than \texttt{num\_qubits}.
This suggests that (in our generated workload) sparsity is driven more by the
gate mix than by state dimension alone, offering a concrete, data-derived signal
that can guide both simulator design and feature engineering.

\medskip
  \vspace{0.2cm}
\noindent
\setlength{\fboxsep}{6pt}
\fcolorbox{black!15}{black!4}{%
  \parbox{\dimexpr\linewidth-2\fboxsep-2\fboxrule\relax}{%
\para{Take-away}
\textbf{(1)} \sys enables learning \emph{cheap estimators} for expensive dynamic properties (e.g., entropy and sparsity) using only features that do not require simulation.
\emph{This enables follow-up work on fast, simulation-free predictors of state complexity for larger circuits.}\\[2pt]  
\textbf{(2)} The learned models reveal structure in the data (e.g., sparsity bands) and highlight which circuit characteristics are most predictive.
\emph{This lays a practical foundation for improving estimators (e.g., hybrid classification/regression) and for connecting circuit structure to quantum-state behavior in a data-centric manner.}
}}%
  \vspace{0.2cm}

%% file: temp2.tex
\begin{figure}[]
    \centering
    \includegraphics[width=0.7\linewidth]{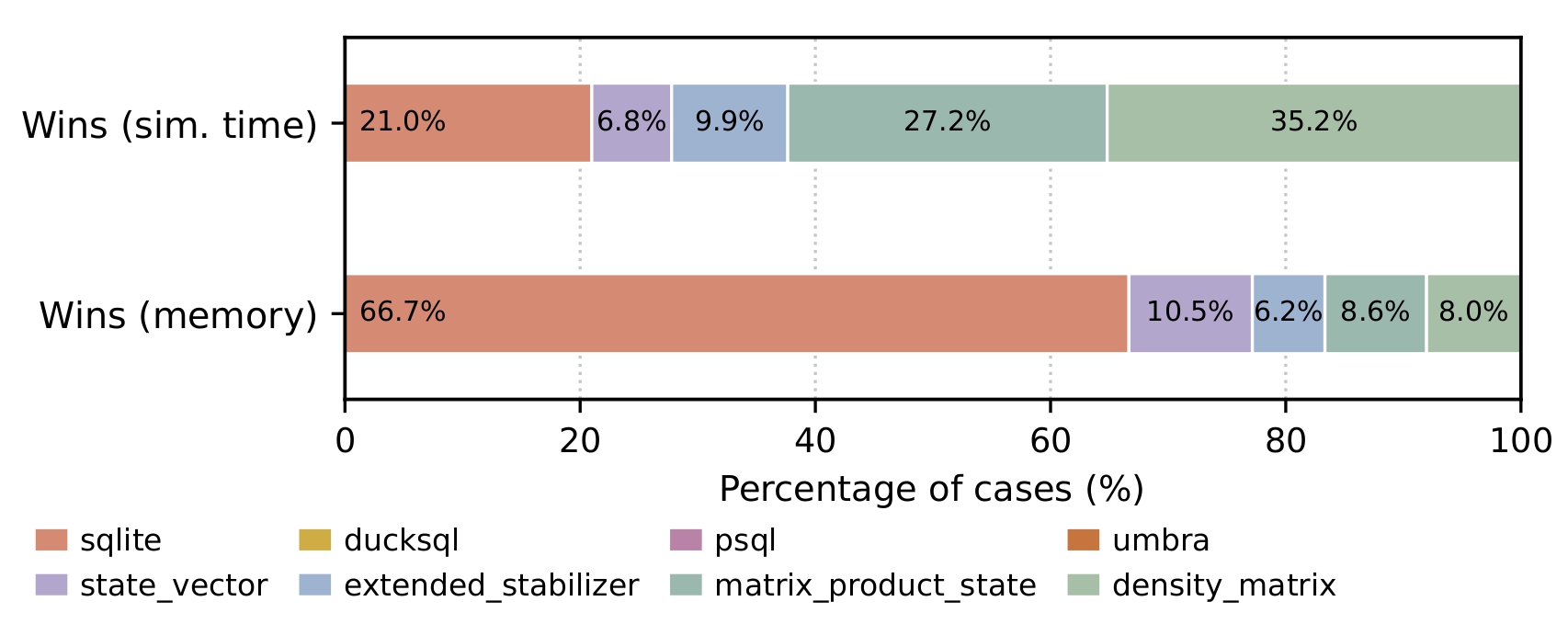}
    \caption{\revisionB{Execution time and memory usage comparison between RDBMSs (SQLite, DuckDB, PostgreSQL, and 
Umbra) and Qiskit Aer (state vector, stabilizer, matrix product state, density matrix) over 162 circuits generated by \sys}.
    }
    \label{fig:rdbmsqiskitcomparison}
\end{figure}

\begin{table*}[t]
  \centering
  \caption{Routing performance on the 20\% held-out test set. Models are evaluated using accuracy and normalized confusion-matrix entries. The columns represent true label $\rightarrow$ prediction label. 
  }
  \label{tab:rdbms_router_combined}
 \begin{subtable}[t]{0.48\linewidth}
    \centering
    \caption{Time-optimal routing}
    \label{tab:rdbms_time_router}
    \resizebox{\linewidth}{!}{
    \begin{tabular}{lccccc}
      \toprule
      Model & Accuracy & rdbms$\rightarrow$rdbms & qiskit$\rightarrow$qiskit & rdbms$\rightarrow$qiskit & qiskit$\rightarrow$rdbms \\
      \midrule
      Logistic Regression & 0.887 & 0.94 & 0.88 & 0.06 & 0.12  \\
      Linear SVM & 0.908 & 0.93 & 0.90 & 0.07 & 0.10\\
      \midrule
      Decision Tree & 0.940 & 0.78 & 0.97 & 0.22 & 0.03 \\
      Random Forest & 0.956 & 0.81 & 0.98 & 0.19 & 0.02 \\
      XGBoost & 0.953 & 0.88 & 0.96 & 0.12 & 0.04 \\
      \bottomrule
    \end{tabular}}
  \end{subtable}
  \hfill
  \begin{subtable}[t]{0.48\linewidth}
    \centering
    \caption{Memory-optimal routing}
    \label{tab:rdbms_memory_router}
    \resizebox{\linewidth}{!}{
    \begin{tabular}{lccccc}
      \toprule
      Model & Accuracy & rdbms$\rightarrow$rdbms & qiskit$\rightarrow$qiskit & rdbms$\rightarrow$qiskit & qiskit$\rightarrow$rdbms \\
      \midrule
      Logistic Regression & 0.935 & 0.93 & 0.94 & 0.07 & 0.06 \\
      Linear SVM & 0.940 & 0.93 & 0.95 & 0.07 & 0.05 \\
      \midrule
      Decision Tree & 0.962 & 0.94 & 0.98 & 0.06 & 0.02 \\
      Random Forest & 0.973 & 0.95 & 0.99 & 0.05 & 0.01 \\
      XGBoost & 0.974 & 0.95 & 0.99 & 0.05 & 0.01 \\
      \bottomrule
    \end{tabular}}
  \end{subtable}

\end{table*}

\subsection{Use case 1: RDBMS as Simulation Backend}
\label{ssec:simu}
\emph{Q1: Is an RDBMS an efficient simulation engine?}

To answer this question, we generated 162 circuits\footnote{\url{https://github.com/InfiniData-Lab/InferQ/blob/main/analysis/training_data/rdbms_all_methods_training_data.parquet}} using \sys. For each circuit, we
measured wall-clock runtime and peak memory when simulating it with (i) 
RDBMS engines (PostgreSQL, SQLite, DuckDB, and \revisionB{Umbra}) executing the SQL workload emitted by
\sys, and (ii) Qiskit Aer using four built-in simulation methods\footnote{Qiskit Aer Simulation Methods: \url{https://qiskit.github.io/qiskit-aer/tutorials/1_aersimulator.html##Simulation-Method-Option}}
(\feat{statevector}, \feat{extended_stabilizer}, \feat{matrix_product_state}, and
\feat{density_matrix}).

Figure~\ref{fig:rdbmsqiskitcomparison} reports, for each metric, the fraction of circuits
for which a backend achieves the best result (minimum execution time or minimum peak
memory) among all evaluated options. As expected, Qiskit Aer attains the best execution
time on most circuits. 
SQLite
achieves the lowest peak memory on 66.7\% of the circuits (108/162). This suggests that
RDBMS engines can be competitive simulation backends, particularly for \emph{memory}. This result 
motivates  further work on SQL workload optimization. 
Moreover, it is interesting to explore learning-based routing in simulation, i.e., predicting when an RDBMS engine is likely to outperform conventional simulators. This leads to our next question.

\begin{figure}[t]
  \centering
  \includegraphics[width=0.7\columnwidth]{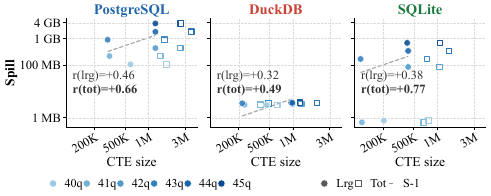}
  \caption{\revisionA{Out-of-core spill versus intermediate relation sizes.}}
  \label{fig:ooc-expander-spill-cte}
\end{figure}

\begin{figure}[t]
  \centering
  \includegraphics[width=0.65\columnwidth]{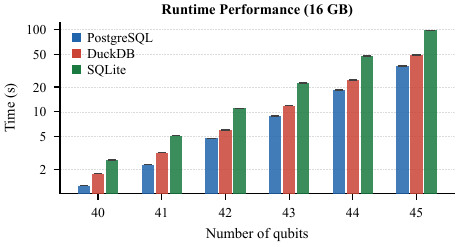}
  \caption{\revisionA{Runtime versus \#qubits.}}
  \label{fig:ooc-expander-runtime-qubits}
\end{figure}

\begin{figure}[t]
  \centering
  \includegraphics[width=0.65\columnwidth]{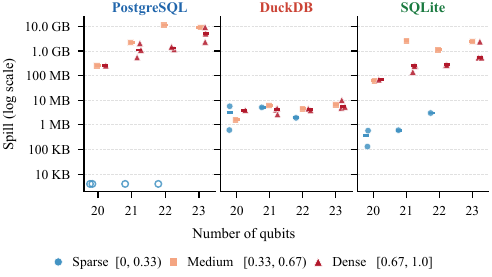}
  \caption{\revisionA{Spill versus \#qubits on sampled InferQ circuits.}}
  \label{fig:ooc-inferq-spill-qubits}
   \vspace{-0.4cm}
\end{figure}

\begin{figure}[t]
  \centering
  \includegraphics[width=0.7\columnwidth]{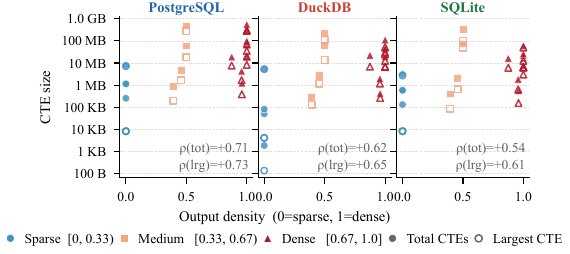}
    \vspace{-0.4cm}
  \caption{\revisionA{Intermediate CTE sizes versus output density on sampled InferQ circuits.}}
  \label{fig:ooc-inferq-cte-density}
    \vspace{-0.4cm}
\end{figure}

  \vspace{0.2cm}
\noindent\emph{Q2: How to decide whether to choose an RDBMS as a simulation engine?} 
  \vspace{0.2cm}

We model this choice as a \emph{binary selection} problem: given a circuit, predict
whether an RDBMS engine achieves lower runtime and/or lower peak memory than a
conventional simulator. Prior database-oriented studies of SQL-based simulation report
strong results on a small set of highly structured circuits (e.g., sparse state-preparation
workloads)~\cite{hai2025quantum,einsteinSQL2023}. Using \sys, we move beyond these
constrained cases and quantify when an RDBMS is beneficial over a substantially broader
set of 
general circuits.
 We have created a training dataset of \num{7705} circuits\footnote{Training data (Parquet) available at \url{https://github.com/InfiniData-Lab/InferQ/blob/main/analysis/training_data/rdbms_training_data.parquet}. The results reported in Figure~\ref{fig:aer-vs-sqlite-wins} also used this dataset.} and train five standard 
models as selectors: logistic regression,
linear SVM,  Decision Tree, random forest (RF), and XGBoost. 
For each circuit, we have logged runtime and peak
memory for Qiskit Aer and the RDBMS backends. This yields ground-truth labels
(\emph{RDBMS wins} vs.\ \emph{Qiskit wins}) for both objectives.

Table~\ref{tab:rdbms_time_router} shows that
these models achieve high accuracy using the features extracted in
Section~\ref{ssec:features}, despite class imbalance (since SQLite is faster only for a
subset of circuits). 
Random Forest achieves the highest runtime-routing accuracy (95.6\%), while XGBoost follows closely (95.3\%) and better identifies RDBMS-win cases.
The best linear model, linear SVM, too achieves 90.8\% accuracy with very low off diagonal confusion entries on the test set, suggesting that much of the
time-selection boundary is close to linear in our feature space. Given the strong performance of these lightweight models, we did not evaluate more
complex models that require substantially higher training and tuning cost.

For memory efficiency
(Table~\ref{tab:rdbms_memory_router}), performance is similarly strong: XGBoost achieves the highest accuracy (97.4\%), while random forest follows closely (97.3\%). Linear SVM also remains  highly competitive (94.0\%
accuracy). Overall, \sys features in Section~\ref{ssec:features} make the RDBMS selection decision
predictable with simple, practical models.

\revisionA{
\para{Routing overhead}
The learned selector is trained offline on labeled workloads generated by \sys
and deployed only for inference. 
For XGBoost, which provides the best accuracy--efficiency trade-off among the models in Tables~\ref{tab:rdbms_time_router} and~\ref{tab:rdbms_memory_router}, inference takes only 13.9\,ms for runtime routing and 13.5\,ms for memory routing per batch.  Appendix~C 
 of our technical report \cite{inferqtech} reports the full training and inference times for all models. Our goal is therefore not to add heavyweight ML components to simulation, but to study a lightweight learned routing policy for backend selection, aligned with recent database work on learned decision-making and cost models~\cite{DBLP:journals/pacmmod/HeinrichLWKB25, DBLP:journals/pacmmod/YangWZDLC23}.}

\revisionA{The previous experiments compare runtime and peak memory only when all backends finish successfully. For larger circuits, conventional in-memory simulators such as Qiskit Aer can fail once the state representation exceeds available memory. RDBMS engines can instead execute queries out of core by spilling intermediate results to disk. This motivates the following question:}

  \vspace{0.2cm}
\noindent\revisionA{\emph{Q3: Can RDBMS-based simulation complete workloads that no longer fit in memory?}}
  \vspace{0.2cm}
  


\revisionA{We evaluate Q3 using a family of large, sparse circuits with up to 45 qubits under a 16~GB memory limit and one CPU core. Full experimental details are in Appendix~A of   our technical report. Under this memory limit, Qiskit Aer’s exact \texttt{statevector} method fails from 30 qubits onward, while PostgreSQL, DuckDB, and SQLite complete all runs by spilling temporary data to disk. Figure~\ref{fig:ooc-expander-spill-cte} shows that intermediate relations remain MB-scale, but spill reaches GB-scale and correlates more with the total intermediate volume than with the single largest intermediate, indicating that cumulative joins and group-bys lead to externalization. Figure~\ref{fig:ooc-expander-runtime-qubits} reports runtime: PostgreSQL is fastest, DuckDB is second, and SQLite is slower but robust. We have also observed that runtime is strongly correlated with spill volume. 
}

\revisionA{To understand out-of-core performance and the characteristics of the simulation workload, we have also sampled 17 InferQ circuits (20--23 qubits) with different output state sparsity. Figures~\ref{fig:ooc-inferq-spill-qubits} and~\ref{fig:ooc-inferq-cte-density} show that sparse outputs usually incur less spill, but output state sparsity alone does not determine spill: several medium-density circuits spill more than dense ones due to intermediate-relation growth and optimizer choices (e.g., join ordering and materialization). Full settings and additional results are reported in \cite{inferqtech}, Appendix~A.}

\medskip
\noindent
\setlength{\fboxsep}{6pt}
\fcolorbox{black!15}{black!4}{%
  \parbox{\dimexpr\linewidth-2\fboxsep-2\fboxrule\relax}{%
\para{Take-away}
\textbf{(1)} Across a wide range of \emph{general, compositional} circuits, RDBMS engines can be
competitive simulation engines, especially in peak memory, and in some cases also in runtime.
\emph{This enables follow-up work on optimizing the emitted SQL to further improve RDBMS performance.}\\[2pt]
\textbf{(2)} \sys enables data-centric routing at scale: it generates diverse circuits, labels each circuit
with observed runtime and peak memory, and exposes features that allow simple models to decide when an
RDBMS engine should be used.
\emph{This provides a foundation for learning-based simulator selection and optimization, trained on \sys-generated circuits.}


\revisionA{(3) For large circuits that exceed memory, RDBMS backends can still complete simulation via out-of-core execution. \emph{This highlights a distinct advantage of database-backed simulation and enables follow-up work on spill-aware planning and optimization.}
}}}
\vspace{0.2cm}
\begin{table*}[t]
  \centering
  \caption{Routing performance using only SQL-derived features on the 20\% held-out test set. Models are evaluated using accuracy and normalized confusion-matrix entries. The columns represent true label $\rightarrow$ prediction label. 
  }
\label{tab:sql_only_router_combined}
  \vspace{-0.2cm}
  \begin{subtable}[t]{0.48\linewidth}
    \centering
    \caption{Time-optimal routing}
    \label{tab:sql_only_routing}
    \resizebox{\linewidth}{!}{
    \begin{tabular}{lccccc}
      \toprule
      Model & Accuracy & rdbms$\rightarrow$rdbms & qiskit$\rightarrow$qiskit & rdbms$\rightarrow$qiskit & qiskit$\rightarrow$rdbms \\
      \midrule
      Logistic Regression & 0.787 & 0.88 & 0.77 & 0.12 & 0.23 \\
      Linear SVM & 0.803 & 0.88 & 0.79 & 0.12 & 0.21 \\
      \midrule
      Decision Tree & 0.915 & 0.76 & 0.94 & 0.24 & 0.06 \\
      Random Forest & 0.932 & 0.75 & 0.96 & 0.25 & 0.04 \\
      XGBoost & 0.924 & 0.81 & 0.94 & 0.19 & 0.06 \\
      \bottomrule
    \end{tabular}}
  \end{subtable}
  \hfill
  \begin{subtable}[t]{0.48\linewidth}
    \centering
    \caption{Memory-optimal routing}
    \label{tab:sql_only_memory}
    \resizebox{\linewidth}{!}{
    \begin{tabular}{lccccc}
      \toprule
      Model & Accuracy & rdbms$\rightarrow$rdbms & qiskit$\rightarrow$qiskit & rdbms$\rightarrow$qiskit & qiskit$\rightarrow$rdbms \\
      \midrule
      Logistic Regression & 0.735 & 0.77 & 0.71 & 0.23 & 0.29 \\
      Linear SVM & 0.737 & 0.77 & 0.72 & 0.23 & 0.28 \\
      \midrule
      Decision Tree & 0.836 & 0.81 & 0.85 & 0.19 & 0.15 \\
      Random Forest & 0.866 & 0.82 & 0.90 & 0.18& 0.10 \\
      XGBoost & 0.842 & 0.84 & 0.85 & 0.16 & 0.15 \\
      \bottomrule
    \end{tabular}}
  \end{subtable}

\end{table*}

\subsection{Feature Validation}
\label{ssec:feat_val}


Next, we want to go deeper and understand:

\noindent\emph{Q4: Which properties determine whether an RDBMS engine is an efficient simulation engine?}

Our goal is to validate that the four feature groups introduced in Section~\ref{ssec:features}
(\emph{static, graph, dynamic, SQL}) capture the key signals behind the routing outcome in
Section~\ref{ssec:simu}. 
We therefore analyze feature importance for two high-performing tree-based models: XGBoost for \emph{runtime} and Random Forest for \emph{peak memory}.
We additionally report a
Linear SVM to expose the \emph{direction} of influence via signed coefficients. \revisionB{We also include an ablation study.}

\subsubsection{SQL-only features.}
\label{ssec:sqlonly}
We first restrict to SQL-derived features extracted from output SQL queries representing the generated circuits
(Section~\ref{ssec:sql}, Table~\ref{tab:sql_complexity}).
Table~\ref{tab:sql_only_routing} shows that these SQL-only features already yield strong routing
accuracy, especially for time-optimal routing (XGBoost exceeds 92\% accuracy on the held-out
test set). This indicates that the engine choice is strongly correlated with \emph{query shape},
without requiring circuit-specific features. \revisionM{We have included more results on dominant SQL signals in Appendix~L 
of \cite{inferqtech}}.

\begin{figure}[t]
    \centering
    \includegraphics[width=0.8\linewidth]{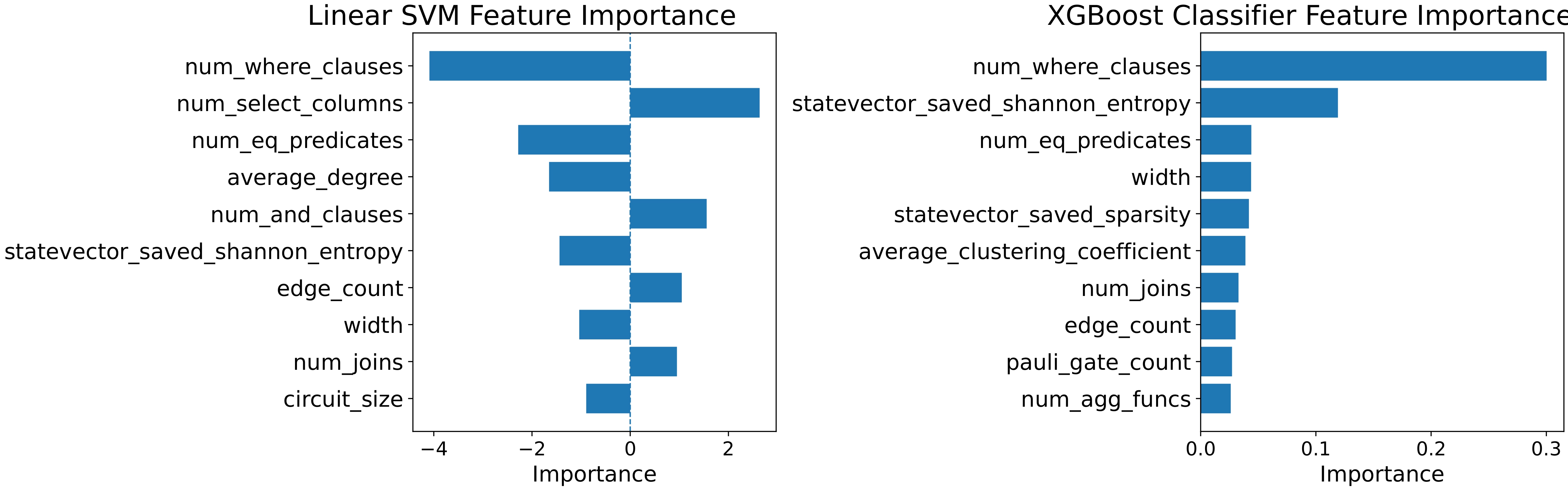}
    \caption{Importance of top 10 InferQ numeric features for SVM and XGBoost trained for optimizing execution time.}
    \label{fig:timeoptallrdbmsimp}
\end{figure}
 \begin{figure}[t]
    \centering
    \includegraphics[width=0.8\linewidth]{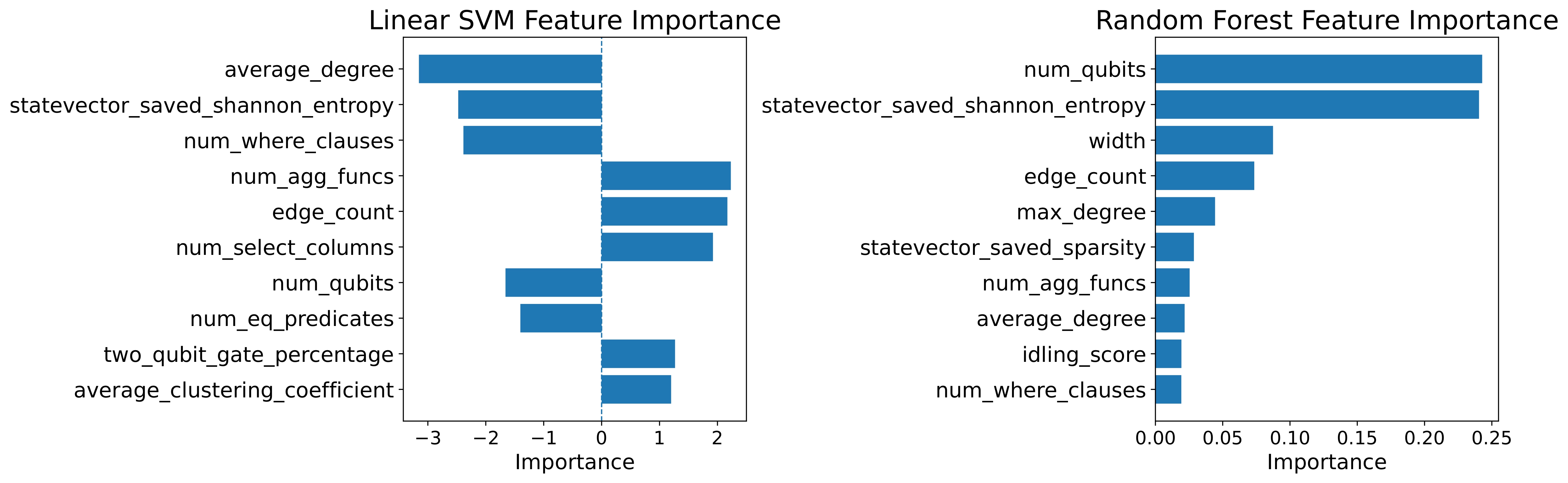}
    \caption{Importance of top 10 InferQ numeric features for SVM and Random Forest trained for optimizing memory.}
    \label{fig:memoptallrdbmsimp}
\end{figure}

\subsubsection{All \sys features.}
We then include the full feature set from Section~\ref{ssec:features}.
For runtime (Figure~\ref{fig:timeoptallrdbmsimp}), for both the Linear SVM and XGBoost, 
features associated with state growth and interaction intensity are more important, including
\feat{shannon_entropy}, \feat{width}, and \feat{edge_count}, together with the
SQL-derived features. With XGBoost, in addition,  graph-related features are important, 
such as \feat{average_clustering_coefficient}, 
\feat{average_shortest_path_length}, and \feat{average_degree}, indicating that irregular interaction structure matters for
runtime differences.

For peak memory (Figure~\ref{fig:memoptallrdbmsimp}), the Linear SVM and Random Forest agree on
the main features: \feat{num_qubits} and \feat{shannon_entropy} dominate,
with additional contributions from \feat{width}, \feat{edge_count}, degree statistics, and
sparsity measures. The Random Forest captures non-linear thresholds in these features,
separating regimes where sparse relational execution substantially reduces peak memory. 


\revisionB{\subsubsection{Ablation study}
\label{sssec:abl}
To quantify the contribution of each feature group, we further run a feature-group
ablation on the routing task. Table~\ref{tab:feature_group_ablation_main} reports  the results for XGBoost on the 7705 dataset. The full results with all models are reported in Appendix~B 
of \cite{inferqtech}. In Section~\ref{ssec:sqlonly}, we have observed that SQL-only features already provide a useful
indication, especially for runtime routing, showing that the shape of the generated SQL query
workload is informative. Here, we see that
the full feature set gives the best overall accuracy and F1 score.}

\begin{table}[t]
\centering
\small
\caption{\revisionB{Feature-group ablation for routing (XGBoost)}}
\label{tab:feature_group_ablation_main}
\begin{tabular}{lcccc}
\toprule
\multirow{2}{*}{Feature set} &
\multicolumn{2}{c}{Runtime} &
\multicolumn{2}{c}{Memory} \\
\cmidrule(lr){2-3}\cmidrule(lr){4-5}
 & Acc. & F1 & Acc. & F1 \\
\midrule
\textbf{Static + Graph + SQL + Dynamic} & \textbf{0.95} & \textbf{0.84} & \textbf{0.98} & \textbf{0.98} \\
Static + Graph + SQL & 0.93 & 0.78 & 0.97 & 0.96 \\
Static + Graph & 0.92 & 0.74 & 0.96 & 0.95 \\
Static & 0.92 & 0.74 & 0.95 & 0.94 \\
\bottomrule
\end{tabular}
\end{table}

\medskip
\noindent
\setlength{\fboxsep}{6pt}
\fcolorbox{black!15}{black!4}{%
  \parbox{\dimexpr\linewidth-2\fboxsep-2\fboxrule\relax}{%
\para{Take-away}
\textbf{(1)} SQL-only features are already strong predictors: the SQL query workload shape (joins, predicates, aggregates) explains much of when an RDBMS should be chosen.
\emph{This enables follow-up work on SQL-level cost modeling and query optimization tailored to quantum circuit simulation workloads.}\\[2pt]
\textbf{(2)} When using the full feature set, the top features are consistent with the underlying bottlenecks: state growth indicators such as entropies, circuit scale like qubit count, and interaction structure (graph features).
    \emph{This supports future work on data-centric models that learn when to use an RDBMS engine and how to optimize execution from \sys-generated training data.}
}}%
 \vspace{0.2cm}

%% file: conclusion.tex
\section{Conclusion}
\sys makes SQL-based quantum circuit simulation accessible as a database benchmarking problem: it generates compositional circuits as RDBMS-ready SQL workloads, attaches features that explain relational workload difficulty, and provides both an on-demand toolkit and a large dataset for reproducible analysis. Our evaluation shows that RDBMS engines can be competitive, especially in peak memory across a broad set of general circuits, and that lightweight models trained on \sys features can reliably predict when an RDBMS engine will be preferable. We expect \sys to enable follow-up work in (i) query optimization and physical design for emitted simulation SQL, and (ii) data-centric routing and cost modeling for simulator selection, trained at scale on \sys-generated circuits and features. 

\revisionB{\para{Outlook} As future work, we plan to further optimize RDBMS-based simulation and develop array-native execution for tensor-aware DBMSs.} 
\revisionC{\sys can also be extended from circuit generation to end-to-end validation for
quantum algorithm design, covering correctness checks, iterative test--debug cycles, and
hardware-aware constraints such as native gate sets and device connectivity.}


%% file: appendix.tex
\clearpage
 
\begin{appendices}

\appendix

\noindent Table~\ref{tab:appendix-roadmap} provides a compact roadmap of the appendix material.

\begin{table}[H]
\centering
\caption{Appendix roadmap.}
\label{tab:appendix-roadmap}
\renewcommand{\arraystretch}{1.08}
\begin{tabularx}{\columnwidth}{@{}p{0.17\columnwidth}X@{}}
\toprule
Appendix & Main content \\
\midrule
A & Out-of-core evaluation and spill behavior. \\
B--C & Routing ablations and training/inference overhead. \\
D--E & Tensor-aware DBMS discussion and support for external benchmark suites. \\
F--H & SQLite--Aer profiling, DBMS-specific tuning, and physical-design discussion. \\
I--K & Reproducibility details, generated-workload characterization, and subcircuit generation. \\
L--O & SQL-only results, preprocessing scalability, implementation details, Qiskit Aer selector, and feature schema. \\
\bottomrule
\end{tabularx}
\end{table}

\input{chapters/outofcore}
\input{chapters/Ablation}
\input{chapters/scidb}
\input{chapters/benchmark-suites}

\section{Subcircuit generation}
\label{sssec:circuittemplate}
In Section~\ref{sec:benchmark} we mentioned how to generate a subcircuit. Here, we provide more details.

Each selected circuit template \(f\) is instantiated by sampling parameters from its
parameter domain \(\Theta_f\). In \sys, we write one sampled
parameter tuple as
\begin{equation}
\theta = (n_q, d, k, r, \xi) \in \Theta_f,
\end{equation}
where \(n_q\) is the qubit count, \(d\) is the depth  allocated to this subcircuit,
\(k\) is the evaluation-qubit, \(r\) is the repetition count for repeatable template
sections, and \(\xi\) is a tuple of template-specific configuration variables.

The domain \(\Theta_f\) is derived from the  parameters  in Table~\ref{tab:base_params}. For example, the per-subcircuit depth \(d\) is sampled uniformly from the defined range between  min\_depth and max\_depth.

The template-specific
configuration variables \(\xi\) captures local design choices supported by \(f\).
Typical examples include whether \texttt{QFT} includes final swaps, the choice of feature map
and ansatz in variational templates, the number of layers in QAOA-style templates, or the
number of steps and coin operators in quantum-walk templates.

\paragraph{Structural vs. numerical parameters.}
In \sys, template instantiation is staged. \sys first fixes the \emph{structure} of the subcircuit
(e.g., qubit assignment, gate topology, repetition pattern, and depth contribution) from
\((n_q,d,k,r,\xi)\). If the template contains parametrized gates,  \sys then samples the
\emph{numerical} values (e.g., rotation angles) from template-specific distributions. This
separation lets \sys explore structural diversity independently from numerical variability.

\input{chapters/feature}

\input{scalability}

\section{Implementation}
\label{sec:imp}

InferQ is supported by two complementary tools:
(1) a \textit{generation and feature-extraction framework} used to construct the dataset, and
(2) a \textit{web-based dataset browser} for interactive exploration and retrieval.

\subsection{InferQ Generation and Feature Extraction Framework}

The parameters in Table~\ref{tab:base_params} are stored in Python dataclass. 

The InferQ framework implements the circuit generation and metric extraction pipeline described in Section~3. It produces combinational quantum circuits and computes their static, graph, SQL, and dynamic features in a reproducible and structured manner.

Each generated circuit is assigned a \emph{content-based hash} that uniquely identifies its structure and parameters. This hash serves as the primary key across the dataset and links the circuit to all extracted feature records.

For every circuit, InferQ stores:
\begin{itemize}
    \item the quantum circuit in a portable serialization format compatible with Qiskit, and
    \item a feature record in JSON format containing all static, graph, and dynamic metrics.
\end{itemize}

Feature extraction is modular and category-based. Static features are derived directly from the circuit representation, graph features are computed from the qubit interaction graph, and dynamic features are obtained from full-state simulation. This separation allows InferQ to be extended with new feature families without recomputing or modifying existing entries.

\subsection{InferQ Web Dataset Viewer}

InferQ is accompanied by a web-based dataset browser inspired by MQTBench\cite{Quetschlich_2023} that enables interactive inspection, filtering, and download of circuits and their features.

The interface presents the dataset as a table where each row corresponds to a circuit identified by its hash. Columns are grouped into the four InferQ feature categories: \textit{static}, \textit{graph}, \textit{SQL}, and \textit{dynamic}.

Users can enable or hide individual feature columns or entire feature groups, apply range-based filters on any visible numeric feature, and explore circuits based on structural, topological, or physical properties.

Each circuit entry includes download links for both the quantum circuit file and its corresponding feature JSON record, allowing users to retrieve exactly the data selected through the interface and use it directly in simulators, learning pipelines, or benchmarking workflows.

\para{Reproducibility and Code} 
InferQ is fully reproducible: given the same code revision, dependency lockfile, and random seed, the framework deterministically regenerates identical circuit structures, parameters, and extracted features. Each circuit is assigned a content-based hash that uniquely identifies its structure and serves as the primary key linking the circuit artifact to all feature records.

\paragraph{Code and environment.}
InferQ is released as open-source software with separate modules for circuit generation, feature extraction, and dataset management. The repository includes a locked Python environment via \texttt{uv.lock} and a declared interpreter version. For exact reproduction, users should report the git commit hash, Python version, and lockfile version.

\paragraph{Cloud-agnostic storage.}
Circuit binaries and feature files are stored in a cloud object store. While Azure Blob Storage is used in the reference deployment, all storage operations are routed through a backend-agnostic interface, allowing any S3-compatible or local filesystem to be used. Each circuit is stored in its own directory
\texttt{Circuits/\{circuit\_hash\}}, containing a serialized \texttt{.qpy} circuit file and a \texttt{.json} metadata file that records generator history and extracted features.

\paragraph{Deterministic circuit generation.}
Circuit construction follows an incremental process in which subcircuits (“generators”) are sampled and appended until a stochastic stopping condition is met. All randomness is controlled by a single global seed, which governs instance-level parameters (e.g., qubit count and depth), generator selection, generator-internal choices, and gate parameters.

\paragraph{Synergy-based generator sampling.}
Generator selection is history-dependent rather than i.i.d. After each generator is chosen, the categorical distribution over the remaining generators is reweighted using predefined \emph{synergies} between generator families. These synergies favor complementary compositions (e.g., state preparation followed by algorithmic cores and optional variational layers), discourage long runs of the same family, and enforce feasibility constraints such as depth and qubit limits. The updated distribution is then renormalized before sampling the next generator. Because these updates are deterministic functions of the previous generator sequence and the global seed, the entire generator sequence is exactly reproducible.

\paragraph{Feature reproducibility.}
For each circuit, InferQ extracts static, graph-based, and dynamic (simulation-based) features. Feature extractors are modular and deterministic given the same simulator backend, numeric precision, and hardware configuration. For fair comparison, users should report simulator version and floating-point settings when dynamic features are used.

\subsection{Qiskit Aer Data Representation Selector}
\label{ssec:selector_RESULTS}
Most quantum applications will be run using a simulator framework like IBM Qiskit \cite{IBMQiskit} with Qiskit Aer\cite{QiskitAerDocs} as a simulation backend. Qiskit Aer offers various simulation backends corresponding to different quantum data representations. Each performs better or worse in terms of time and memory depending on the circuit to be simulated. This leads to the research question: \textit{what is the optimal Qiskit Aer data representation for classical simulation?} 

InferQ can be used to train a selector for the Qiskit data representations for optimal simulation. In our experiment we trained various ML models  to select the best representation within the classes of ["statevector", "density\_matrix", "extended\_stabilizer", "matrix\_product\_state"] provided by Qiskit Aer. We recorded the execution time for 93{,}567 entries in InferQ on our machine from Section~\ref{sec:impl}.\footnote{Qiskit Selector training data (Parquet) available at \url{https://github.com/InfiniData-Lab/InferQ/blob/main/analysis/training_data/selector_training_data.parquet}} For this experiment, we used the framework from Qymera \cite{Littau_2025}.\footnote{We thank the authors for sharing their code and helping us set up the experiments.}

By defining the best representation to be the one taking lowest execution time, we could train various multiclass classifier models to select the best method. 

Training models of complexities and  evaluating them on a test set of size 20\% outperformed Qiskit Aer's own ``automatic'' data representation selector by a margin of up to 28\%.

The accuracy in table  \ref{tab:selectoraccuracy} is the percentage of cases in the test set that the predicted best method matched the actual best method. Upon inspection of the features, it became clear why Naive Bayes performs weakly. There is multimodality within feature distributions rather than Gaussian structures.  In addition, the classes are highly imbalanced, as shown in Figure \ref{fig:selectortestset}.

\begin{table}[tb]
\centering
\begin{minipage}{\columnwidth}

\caption{Test-set accuracy of InferQ-trained selectors compared with Qiskit Aer's built-in ``automatic'' method selector.}
\label{tab:selectoraccuracy}

\centering
\resizebox{\columnwidth}{!}{%
\begin{tabular}{cccccccc}
\toprule
 & Qiskit ``Automatic'' & \multicolumn{6}{c}{Trained on InferQ} \\
Metric & Accuracy & \textbf{XGBoost} & Random Forest & Decision Tree & KNN & Logistic Regression & Naive Bayes \\
\midrule
Accuracy & 36\% & \textbf{64\%} & 64\% & 63\% & 61\% & 61\% & 55\% \\
\bottomrule
\end{tabular}
}

\vspace{0.6em}

\centering
\includegraphics[width=0.7\linewidth]{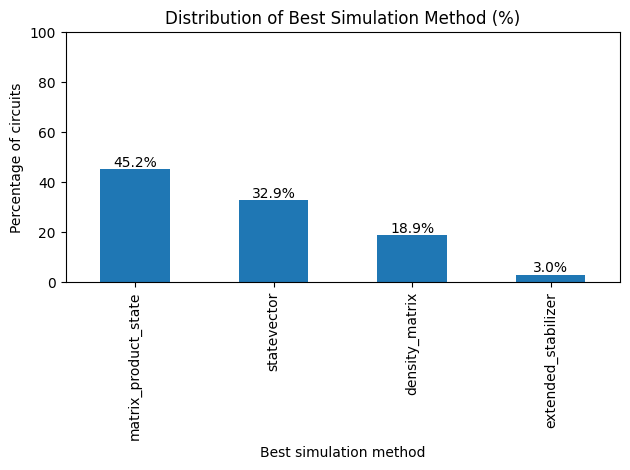}
\captionof{figure}{Bar chart displaying the imbalance between the classes being the best simulation method for minimum execution time.}
\label{fig:selectortestset}

\end{minipage}
\end{table}

\begin{figure*}[tbh]
  \centering
  \includegraphics[width=\linewidth]{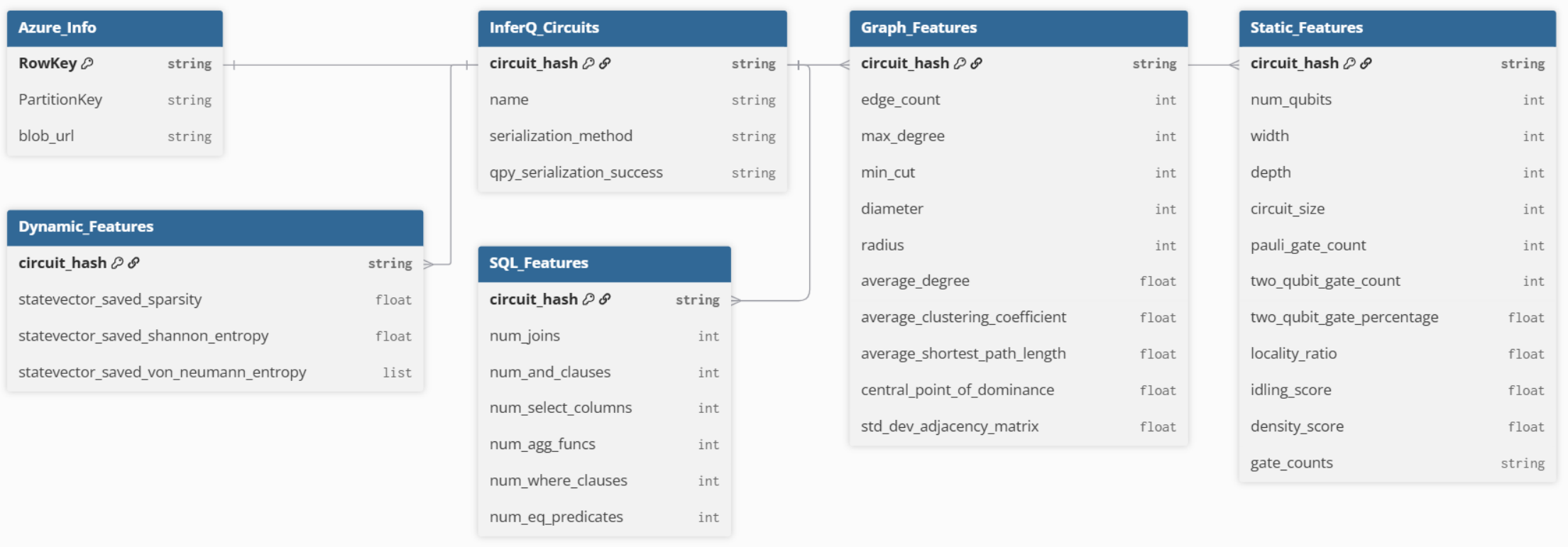}
  \caption{Schema for storing extracted circuit features. 
  \sys circuit data model schema. Each circuit is stored with a unique
  hash (primary key) and associated feature classes. \sys entries are centered
  around circuits in the \texttt{InferQ\_Circuits} table, which reference
  static, graph, SQL and dynamic features, as well as storage metadata in Azure DB. 
      }
  \label{fig:schema}
\end{figure*}

\begin{figure*}[tb]
    \centering
    \includegraphics[width=\linewidth]{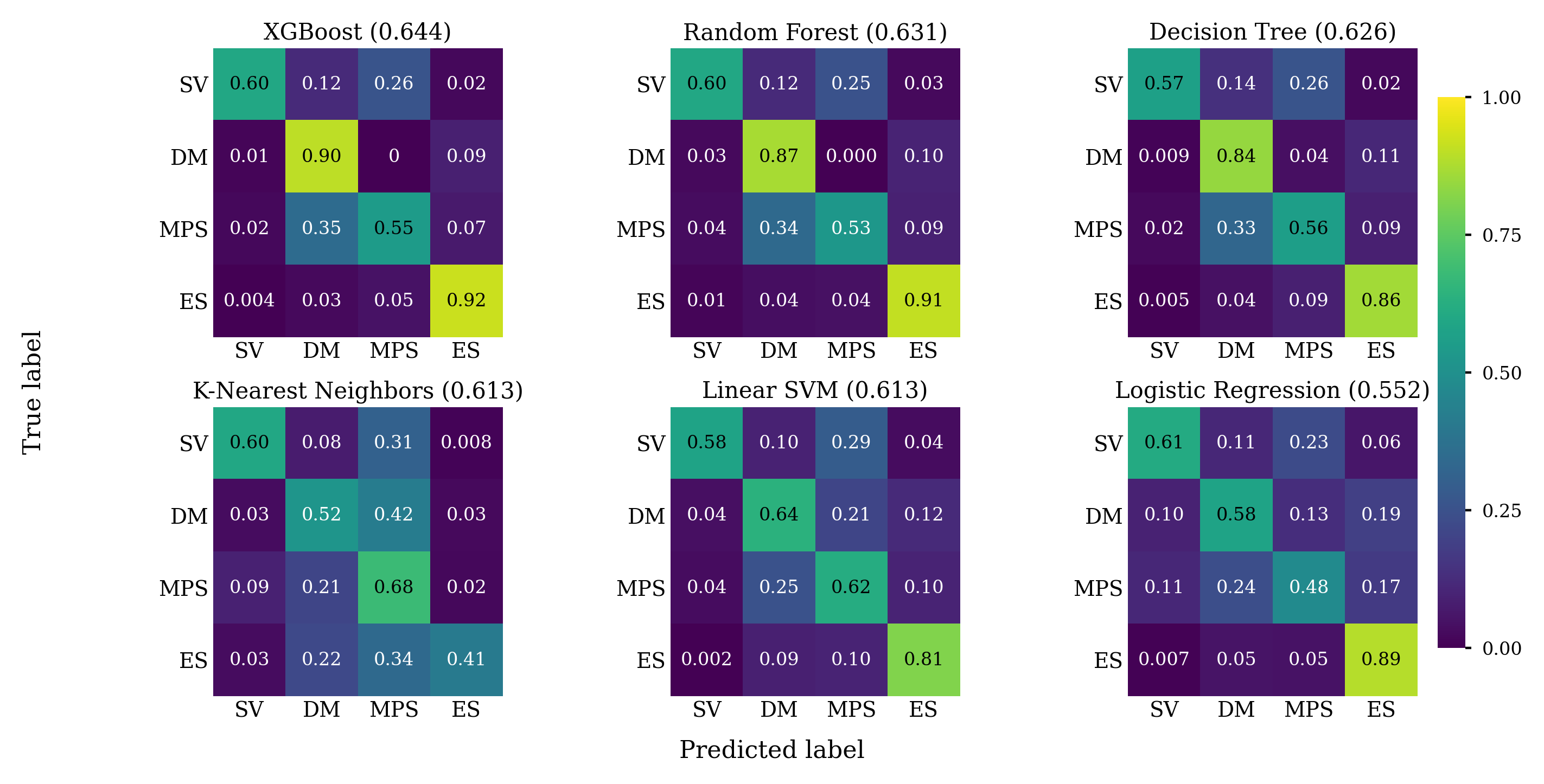}
    \caption{Confusion matrices, normalised over the true classes, for each model's performance on the test set.
    SV = \texttt{statevector}, DM = \texttt{density\_matrix}, MPS = \texttt{matrix\_product\_state}, ES = \texttt{extended\_stabilizer}.
    Panels are ordered by decreasing accuracy and share a common colour scale.}
    \label{fig:selectorconfusion}
\end{figure*}

Despite the K-Nearest Neighbour (KNN)'s relatively high accuracy, it is clear from 
Figure \ref{fig:selectorconfusion} 
that it is unable to predict the smaller classes well, as 42\% of the ``density\_matrix'' and 34\% ``extended\_stabilizer'' methods are being classified as ``matrix\_product\_state''. The Decision Tree is much better overall for all the classes, except it is overcompensating for the minority classes and hence has low accuracy for the dominant ``matrix\_product\_state''. Logistic regression has lowest accuracy so we are left with tree models and linear SVM.

In order to break this tie, we can look at the distribution in which the simulation methods come 2nd, 3rd and 4th place. From Table \ref{tab:method_ranks} it becomes clear that ``matrix\_product\_state'', ``density\_matrix'' and "statevector" dominate in 2nd position too. Hence, we can choose XGBoost which performs the best for these 3 dominant classes as can be seen from confusion plots of Figure 33. Although the accuracy of XGBoost is 64.4\%, it still outperforms Qiskit Aer's own "automatic" method which is only able to select correctly in 35.9\% of the cases. Even simpler models like logistic regression and linear SVMs are able to leverage InferQ features to perform relatively well against the current Qiskit Aer "automatic" selector. 

\begin{table}[h]
\caption{Number of entries where each simulation method is ranked 1st, 2nd, 3rd, and 4th by the XGBoost-based selector, ordered by predicted execution time.}
\label{tab:method_ranks}
\centering
\resizebox{\columnwidth}{!}{%
\begin{tabular}{lcccc}
\toprule
 & \multicolumn{4}{c}{Rank Position} \\
\cmidrule(lr){2-5}
Method & 1 & 2 & 3 & 4 \\
\midrule
\texttt{statevector}            & 30802 & 34743 & 25131 & 2891 \\
\texttt{density\_matrix}        & 17704 & 20340 & 7625  & 948  \\
\texttt{matrix\_product\_state} & 42288 & 35959 & 14537 & 783  \\
\texttt{extended\_stabilizer}   & 2773  & 2525  & 46274 & 41995 \\
\bottomrule
\end{tabular}%
}
\end{table}

We thus show that InferQ can be used to optimize Qiskit Aer Workloads by building a selector, demonstrating that a complex model like XGBoost surpasses its simpler competitors in predictions amongst the highly imbalanced classes of representations. By outperforming Qiskit Aer's own "automatic" selector with all models experimented with, it validates our data-centric approach to this problem.

Moreover, the experiment  showcases that data management techniques can be used to make classical simulation of quantum computation more efficient without incorporating the domain specific techniques of quantum science. It also completes a pipeline to solve the mutli-class problem of which classical simulation method is optimal given a circuit incorporating the novel approach of RDBMS.

\section{InferQ Feature Schema}
\label{sec:inferqschemafull}
Figure~\ref{fig:schema} illustrates the schema of how \sys stores the extracted circuit features.

\end{appendices}

%% file: chapters/outofcore.tex
\section{Out-of-Core Evaluation}
\label{app:qft-cap-sensitivity}

In this section, we study \emph{out-of-core} execution for DBMS-backed SQL simulation and
address the following question:

\smallskip
\noindent\textbf{Q3:} Can an RDBMS backend complete simulation queries once the workload no
longer fits in main memory?

\smallskip
To answer Q3, we conduct two experiments. First, in
Section~\ref{app:ooc-expander}, we generate a family of large circuits with an increasing number of qubits up to 45. Under a fixed memory limit, Qiskit Aer cannot complete the simulation, whereas RDBMS engines can spill
temporary data to secondary storage and complete execution, demonstrating a clear case
\emph{when} an RDBMS backend is beneficial. Second, in
Section~\ref{app:ooc-inferq-sparsity}, we use \sys to generate circuits with controlled
variation in sparsity and study how sparsity, intermediate results, and spill behavior
interact.

\subsection{Large Circuits Simulation}
\label{app:ooc-expander}

\para{Experiment setup}
We generate a family of circuits with a fixed structure and similar sparsity, while
increasing the qubit count $N$ (one circuit for each $N$ from 1 to 45). These circuits
are designed so that the simulation state can be represented compactly as a relation of
non-zero entries. These circuits are representative in practice and are a common scenario for structured and sparse quantum circuit simulation workloads (e.g., stabilizer, amplitude amplification, and sparse linear systems). 
Quantum-specific construction details are provided in Section~\ref{app:expander-details}.

We execute the generated SQL queries on PostgreSQL, DuckDB, and SQLite under a 16~GB memory cap. In SQL-based simulation, the tensor-contraction sequence is written as a chain of Common Table Expressions (CTEs), each encoding one contraction step (join + aggregate). 
For each query, we record (i) median spill volume over five runs and (ii) the
maximum and total intermediate relation sizes across the CTE chain. The engine-specific
spill configuration is described in
Section~\ref{app:spill-instrumentation}.

\para{Results}
Under the 16~GB limit, Qiskit Aer fails to complete exact simulation starting at 30 qubits.
This is expected. Recall in Sec. 1 and 2.1 we have explained that 
an $N$-qubit state requires $2^N$
complex entries. If we store each complex entry in double precision, this is approximately $16 \cdot 2^N$ bytes,
which reaches roughly 16~GB at $N{=}30$ (ignoring runtime overhead) and grows
exponentially thereafter. For the 40--45 qubit circuits evaluated here, Qiskit would require approximately 16--512~TB (this is known as
statevector in Qiskit for exact simulation). 

In contrast, all three RDBMS engines
complete all runs because they store the state as tuples and can exploit sparsity: storage
scales with the number of non-zero entries rather than with $2^N$.

\begin{figure}[t]
    \centering
    \includegraphics[width=1.0\linewidth]{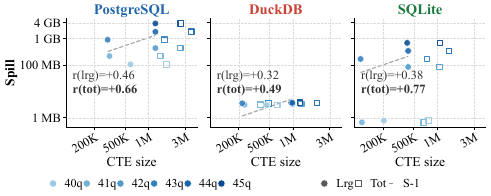}
    \caption{Spill volume of RDBMS engines (PostgreSQL, DuckDB, SQLite). Filled points (\texttt{tot}) denote total intermediate relation size across the CTE chain; open points (\texttt{lrg}) denote the maximum single-step intermediate size.}
    \label{fig:expander_ooc}
\end{figure}

Figure~\ref{fig:expander_ooc} summarizes out-of-core behavior for the 40--45 qubit circuit simulation under the 16~GB limit.
Across engines, intermediate relation sizes remain in the megabyte range, consistent with RDBMSs' support for storing sparse tensors.\footnote{We observe different spill behavior across DBMS
engines. DuckDB spills the least in this experiment, while SQLite and PostgreSQL are
comparable on several circuits. Note that PostgreSQL exposes query-level temporary I/O
counters, so we can compute spill volume exactly. DuckDB and SQLite do not expose a
stable, query-level temp-bytes-written counter in the same way, so we use a conservative
operating-system level proxy for them. Details are provided in
Section~\ref{app:spill-instrumentation}.}
At the same time, total spill volume is often orders of magnitude larger than the sizes of the
intermediate relations. This is likely due to the overhead during joins and
group-bys (e.g., hash tables, sorting or grouping state, and temporary structures), as
well as repeated materialization of intermediates along the CTE chain.

We further inspect the Spearman correlation values between spill volume and the maximum
single-step intermediate size ($r(\texttt{lrg})$) and the total intermediate size across
all CTEs ($r(\texttt{tot})$). In all three engines, $r(\texttt{tot}) > r(\texttt{lrg})$,
indicating that spill is more strongly associated with cumulative intermediate volume than
with the peak single-step intermediate. This suggests that optimization opportunities
include reducing repeated intermediate materialization and improving reuse, for example by
caching selected intermediates when they are used multiple times.

\begin{figure}[t]
    \centering
    \includegraphics[width=1.0\linewidth]{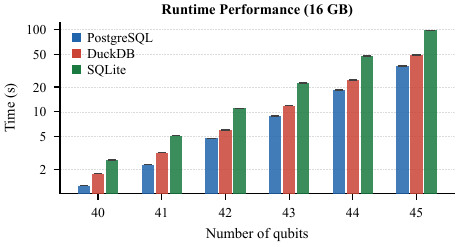}
    \caption{Runtime of RDBMS engines (PostgreSQL, DuckDB, SQLite).}
    \label{fig:expander_walltime_qubits}
\end{figure}

\begin{figure}[t]
    \centering
    \includegraphics[width=1.0\linewidth]{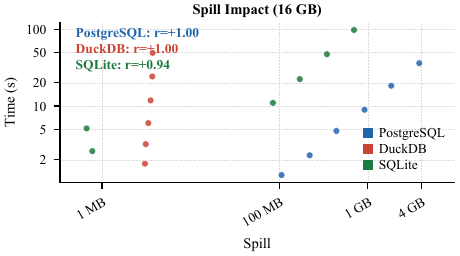}
    \caption{Runtime versus spill volume for RDBMS engines (PostgreSQL, DuckDB, SQLite).}
    \label{fig:expander_walltime_spill}
\end{figure}

\begin{figure*}[t]
    \centering
    \includegraphics[width=0.55\linewidth]{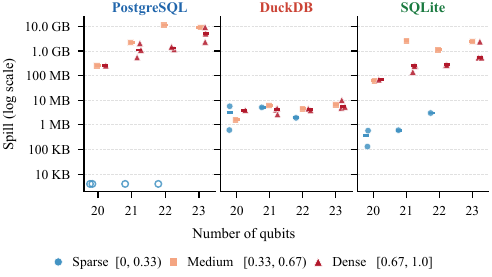}
    \caption{Spill volume with varying numbers of qubits (log scale).}
    \label{fig:dbspillqubits_repeat}
\end{figure*}
\begin{figure*}[t]
    \centering
    \includegraphics[width=0.6\linewidth]{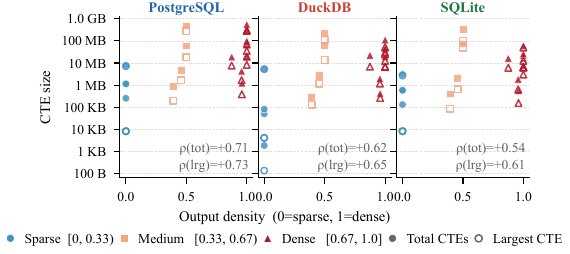}
    \caption{Maximum and total intermediate CTE sizes against output state sparsity (log scale).}
    \label{fig:dbspillctesparse_repeat}
\end{figure*}

Figure~\ref{fig:expander_walltime_qubits} reports query runtime. 
Runtime increases rapidly with the number of qubits for all engines. PostgreSQL is consistently the fastest in this experiment, DuckDB is second,
and SQLite is the slowest. Figure~\ref{fig:expander_walltime_spill} shows runtime and
spill volume are related, indicating that externalization overhead is a major
reason for end-to-end query time once operator workspaces spill to disk.

\medskip
\noindent
\setlength{\fboxsep}{6pt}
\fcolorbox{black!15}{black!4}{%
  \parbox{\dimexpr\linewidth-2\fboxsep-2\fboxrule\relax}{
\para{Takeaways}
This experiment demonstrates \emph{when} DBMS-backed simulation is particularly useful:
it enables exact simulation for large circuits that exceed the in-memory limits of
Qiskit Aer by combining a sparse tuple-level representation with out-of-core
execution. At the same time, spill volume and runtime are closely tied to relational
operator workspace and intermediate relation growth, which motivates RDBMS
optimization opportunities such as join ordering, operator selection, and materialization
control to reduce externalization overhead.
}}

\begin{figure*}[t]
    \centering
    \includegraphics[width=0.5\linewidth]{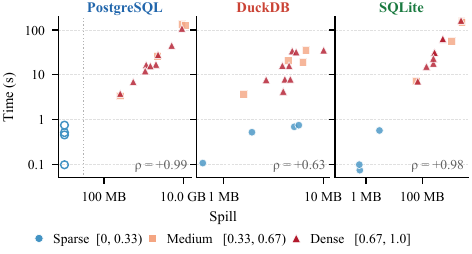}
     \vspace{-0.3cm}
    \caption{Execution time versus spill volume for sampled InferQ circuits (log-log).}
    \label{fig:dbwalltimespill_repeat}
    \vspace{-0.4cm}
\end{figure*}
\subsection{Experiment on Varying Sparsity}
\label{app:ooc-inferq-sparsity}

\begin{table*}[!ht]
  \centering
  \caption{Ablation study for \emph{time-optimal method} prediction
    (80\% - 20\% train-test split).
    \textbf{Bold}: best overall result (XGB, full feature set);
    \underline{underline}: second best (RF, full feature set).
    Acc: accuracy; F1: macro F1-score.}
  \label{tab:ablation_time}
  \scriptsize
  \setlength{\tabcolsep}{3pt}
  \renewcommand{\arraystretch}{1.05}
  \begin{tabular}{l cc cc cc cc cc}
    \toprule
    & \multicolumn{2}{c}{\textbf{LR}}
    & \multicolumn{2}{c}{\textbf{SVM}}
    & \multicolumn{2}{c}{\textbf{DT}}
    & \multicolumn{2}{c}{\textbf{RF}}
    & \multicolumn{2}{c}{\textbf{XGB}} \\
    \cmidrule(lr){2-3}\cmidrule(lr){4-5}\cmidrule(lr){6-7}
    \cmidrule(lr){8-9}\cmidrule(lr){10-11}
    \textbf{Feature Set}
      & Acc & F1
      & Acc & F1
      & Acc & F1
      & Acc & F1
      & Acc & F1 \\
    \midrule
    SQL
      & 0.79 & 0.54
      & 0.80 & 0.57
      & 0.91 & 0.70
      & 0.93 & 0.75
      & 0.92 & 0.75 \\
    Static
      & 0.80 & 0.55
      & 0.80 & 0.55
      & 0.90 & 0.68
      & 0.92 & 0.72
      & 0.92 & 0.74 \\
    Static + Graph
      & 0.84 & 0.60
      & 0.81 & 0.56
      & 0.90 & 0.66
      & 0.92 & 0.73
      & 0.92 & 0.74 \\
    Static + SQL
      & 0.86 & 0.64
      & 0.86 & 0.65
      & 0.93 & 0.76
      & 0.94 & 0.78
      & 0.94 & 0.80 \\
    Static + Dynamic
      & 0.86 & 0.66
      & 0.86 & 0.65
      & 0.92 & 0.73
      & 0.94 & 0.77
      & 0.94 & 0.79 \\
    Static + Graph + SQL
      & 0.87 & 0.67
      & 0.84 & 0.62
      & 0.92 & 0.74
      & 0.94 & 0.78
      & 0.93 & 0.78 \\
    Static + Graph + SQL + Dyn.
      & 0.89 & 0.70
      & 0.88 & 0.69
      & 0.94 & 0.80
      & \underline{0.95} & \underline{0.83}
      & \textbf{0.95} & \textbf{0.84} \\
    \bottomrule
  \end{tabular}
\end{table*}

The previous experiment indicates that sparse representations can make DBMS-backed
simulation more useful under a memory limit.   
We therefore conduct a second experiment on
InferQ-generated workloads to study how spill volume, intermediate CTE sizes, and query
runtime relate to the sparsity of a circuit.
This is possible because InferQ can generate circuits
with controlled variation in sparsity.

We first explain the sparsity of the final output state of a circuit simulation.
For an $n$-qubit circuit, the output is a length-$2^n$ amplitude vector
$\mathbf{p}$. We define \emph{output density} as the fraction of non-zero entries in the vector $\mathbf{p}$ against the vector size $|\mathbf{p}|$:
\[
\mathrm{density}(\mathbf{p}) \;=\; \frac{\left|\{\, i \mid p_i \neq 0 \,\}\right|}{|\mathbf{p}|}.
\]
A smaller density indicates fewer non-zero tuples in the DBMS representation, while a
density close to 1 indicates a near-dense output.

\medskip
\para{Experiment setup}
We construct a sample of 17 InferQ circuits\footnote{\url{https://github.com/InfiniData-Lab/InferQ/blob/main/analysis/sample_csvs/final_sparse_sample.csv}}
spanning three density bands: Sparse $[0,0.33)$, Medium $[0.33,0.67)$, and Dense
$[0.67,1.0]$. The circuits span a number of qubits from 20 to 23. 
We execute the
SQL simulation on PostgreSQL, DuckDB, and SQLite under a 16~GB memory cap. For each run,
we record spill volume, wall-clock runtime, and the maximum and total intermediate CTE
sizes observed over the CTE chain.

\medskip
\para{Results}
Figure~\ref{fig:dbspillqubits_repeat} plots spill volume against the number of qubits. Across
all engines, circuits in the Sparse band tend to incur the least spill, and PostgreSQL
often shows negligible spill for the sparsest cases. This matches the intuition that a
sparse output can be represented as a smaller relation, reducing memory pressure and
temporary-workspace needs.

From Figure~\ref{fig:dbspillqubits_repeat}, we observe that output density alone does not determine spill volume. For PostgreSQL and SQLite, several
Medium-density circuits spill more than some Dense circuits at similar qubit counts. This indicates that spill is strongly influenced by
\emph{intermediate} results produced during query execution, which depend on plan choices
such as join ordering and materialization behavior, rather than being determined solely
by the final output density.

Figure~\ref{fig:dbspillctesparse_repeat} reports the maximum and total intermediate CTE sizes as
a function of output density. The correlations are consistently
positive across engines, and the maximum-intermediate correlation is slightly stronger
than the total-intermediate correlation. At the same time, intermediate sizes vary across
DBMS engines for the same circuit, which is expected because optimizers may choose
different join orders and physical operators, leading to different intermediate growth.

Finally, Figure~\ref{fig:dbwalltimespill_repeat} shows runtime versus spill volume. PostgreSQL and
SQLite show a very strong positive association between execution time and spill (rank
correlations shown in the figure), and DuckDB exhibits a weaker but
still positive association. This result indicates that externalization cost is a dominant
reason for end-to-end runtime once operators spill to disk. 

\medskip
  \vspace{0.2cm}
\noindent
\setlength{\fboxsep}{6pt}
\fcolorbox{black!15}{black!4}{%
  \parbox{\dimexpr\linewidth-2\fboxsep-2\fboxrule\relax}{
\para{Takeaways}
It shows that DBMS out-of-core behavior is governed primarily by
intermediate relation growth and operator workspace, which depend on workload structure
and DBMS plan choices, rather than only on the density of the final output. It motivates
optimizing SQL-based simulation using standard DBMS techniques such as improving join
ordering, controlling materialization, and reducing intermediate blow-up. It also
suggests that spill-aware cost models are useful for predicting when a DBMS   will
benefit from out-of-core execution and when externalization overhead will dominate in future work.
}}

%% file: chapters/Ablation.tex
\section{Ablation Study of InferQ Feature groups}
\label{appendix:abl}

\begin{table*}[!ht]
  \centering
  \caption{Ablation study for \emph{memory-optimal method} prediction
    (80\% - 20\% train-test split).
    \textbf{Bold}: best overall result (XGB, full feature set);
    \underline{underline}: second best (RF, full feature set).
    Acc: accuracy; F1: macro F1-score.}
  \label{tab:ablation_mem}
  \scriptsize
  \setlength{\tabcolsep}{3pt}
  \renewcommand{\arraystretch}{1.05}
  \begin{tabular}{l cc cc cc cc cc}
    \toprule
    & \multicolumn{2}{c}{\textbf{LR}}
    & \multicolumn{2}{c}{\textbf{SVM}}
    & \multicolumn{2}{c}{\textbf{DT}}
    & \multicolumn{2}{c}{\textbf{RF}}
    & \multicolumn{2}{c}{\textbf{XGB}} \\
    \cmidrule(lr){2-3}\cmidrule(lr){4-5}\cmidrule(lr){6-7}
    \cmidrule(lr){8-9}\cmidrule(lr){10-11}
    \textbf{Feature Set}
      & Acc & F1
      & Acc & F1
      & Acc & F1
      & Acc & F1
      & Acc & F1 \\
    \midrule
    SQL
      & 0.75 & 0.72
      & 0.75 & 0.72
      & 0.83 & 0.79
      & 0.86 & 0.82
      & 0.85 & 0.81 \\
    Static
      & 0.87 & 0.85
      & 0.87 & 0.84
      & 0.93 & 0.92
      & 0.95 & 0.93
      & 0.95 & 0.94 \\
    Static + Graph
      & 0.90 & 0.87
      & 0.89 & 0.87
      & 0.93 & 0.91
      & 0.96 & 0.94
      & 0.96 & 0.95 \\
    Static + SQL
      & 0.89 & 0.87
      & 0.89 & 0.87
      & 0.94 & 0.93
      & 0.96 & 0.95
      & 0.97 & 0.96 \\
    Static + Dynamic
      & 0.94 & 0.92
      & 0.94 & 0.93
      & 0.95 & 0.94
      & 0.97 & 0.96
      & 0.97 & 0.97 \\
    Static + Graph + SQL
      & 0.90 & 0.88
      & 0.90 & 0.88
      & 0.94 & 0.92
      & 0.97 & 0.96
      & 0.97 & 0.96 \\
    Static + Graph + SQL + Dyn.
      & 0.94 & 0.93
      & 0.94 & 0.93
      & 0.96 & 0.95
      & \underline{0.98} & \underline{0.97}
      & \textbf{0.98} & \textbf{0.98} \\
    \bottomrule
  \end{tabular}
\end{table*}

We conduct a feature-group ablation study on the 7705 dataset. 
Tables~\ref{tab:ablation_time} and~\ref{tab:ablation_mem} report the performance of five models (Logistic Regression, Linear SVM, Decision Tree, Random Forest, and XGBoost) for runtime- and memory-optimal routing, respectively. 
We observe that combining all feature groups yields the best overall performance. 
In some cases, such as Logistic Regression in Table~\ref{tab:ablation_mem}, adding dynamic features also leads to a noticeable improvement.

\section{Training and Inference Time}
\label{sec:traintime}
We report the computational cost of each classifier under the all-features setting.
Table~\ref{tab:ablation_cost_all_features} reports training and inference time separately for runtime- and memory-optimal routing, with all values in seconds.
Training time is measured over 6,164 training samples, whereas inference time is measured over 1,541 test samples; therefore, the two measurements should be interpreted independently rather than compared directly.

Combining these results with the accuracy and F1 results in Tables~\ref{tab:ablation_time} and~\ref{tab:ablation_mem}, we observe that XGBoost provides the best overall trade-off between predictive performance and inference efficiency.
Although Decision Trees have the lowest inference time for both routing tasks, requiring 0.0102 s for runtime-optimal routing and 0.0098 s for memory-optimal routing, XGBoost achieves substantially higher accuracy and F1 score while remaining highly efficient at inference time.
Compared with Random Forest, the second strongest classifier in predictive performance, XGBoost is considerably faster during inference: 0.0139 s versus 0.1319 s for runtime-optimal routing and 0.0135 s versus 0.1076 s for memory-optimal routing, measured over the same 1,541 inference samples.
XGBoost also maintains moderate training cost, requiring 0.36 s and 0.34 s for runtime- and memory-optimal routing, respectively.
These results further support our use of XGBoost as the routing model.

\begin{table}[thb]
  \centering
  \caption{Training and inference time under the all-features setting
    using an 80\%--20\% train-test split.
    All values are total elapsed time in seconds.
    Training was measured over 6,164 samples, while inference was measured
    over 1,541 samples.
    \textbf{Bold}: lowest value;
    \underline{underline}: second-lowest value within each section and routing setting.}
  \label{tab:ablation_cost_all_features}
  \scriptsize
  \setlength{\tabcolsep}{5pt}
  \renewcommand{\arraystretch}{1.08}
  \begin{tabular}{lcc}
    \toprule
    \multicolumn{3}{c}{\textbf{Training time over 6,164 samples (s)}} \\
    \midrule
    \textbf{Model}
      & \textbf{Runtime-Optimal Routing}
      & \textbf{Memory-Optimal Routing} \\
    \midrule
    LR
      & \underline{0.12}
      & \textbf{0.07} \\
    SVM
      & 1.80
      & 0.90 \\
    DT
      & \textbf{0.10}
      & \underline{0.10} \\
    RF
      & 0.79
      & 0.72 \\
    XGB
      & 0.36
      & 0.34 \\
    \midrule
    \multicolumn{3}{c}{\textbf{Inference time over 1,541 samples (s)}} \\
    \midrule
    \textbf{Model}
      & \textbf{Runtime-Optimal Routing}
      & \textbf{Memory-Optimal Routing} \\
    \midrule
    LR
      & 0.0154
      & 0.0139 \\
    SVM
      & 0.1951
      & 0.0943 \\
    DT
      & \textbf{0.0102}
      & \textbf{0.0098} \\
    RF
      & 0.1319
      & 0.1076 \\
    XGB
      & \underline{0.0139}
      & \underline{0.0135} \\
    \bottomrule
  \end{tabular}
\end{table}

%% file: chapters/scidb.tex
\section{Tensor-Aware DBMSs}
\label{app:tensor-aware-dbms}

\begin{figure*}[t]
    \centering
    \includegraphics[width=\linewidth]{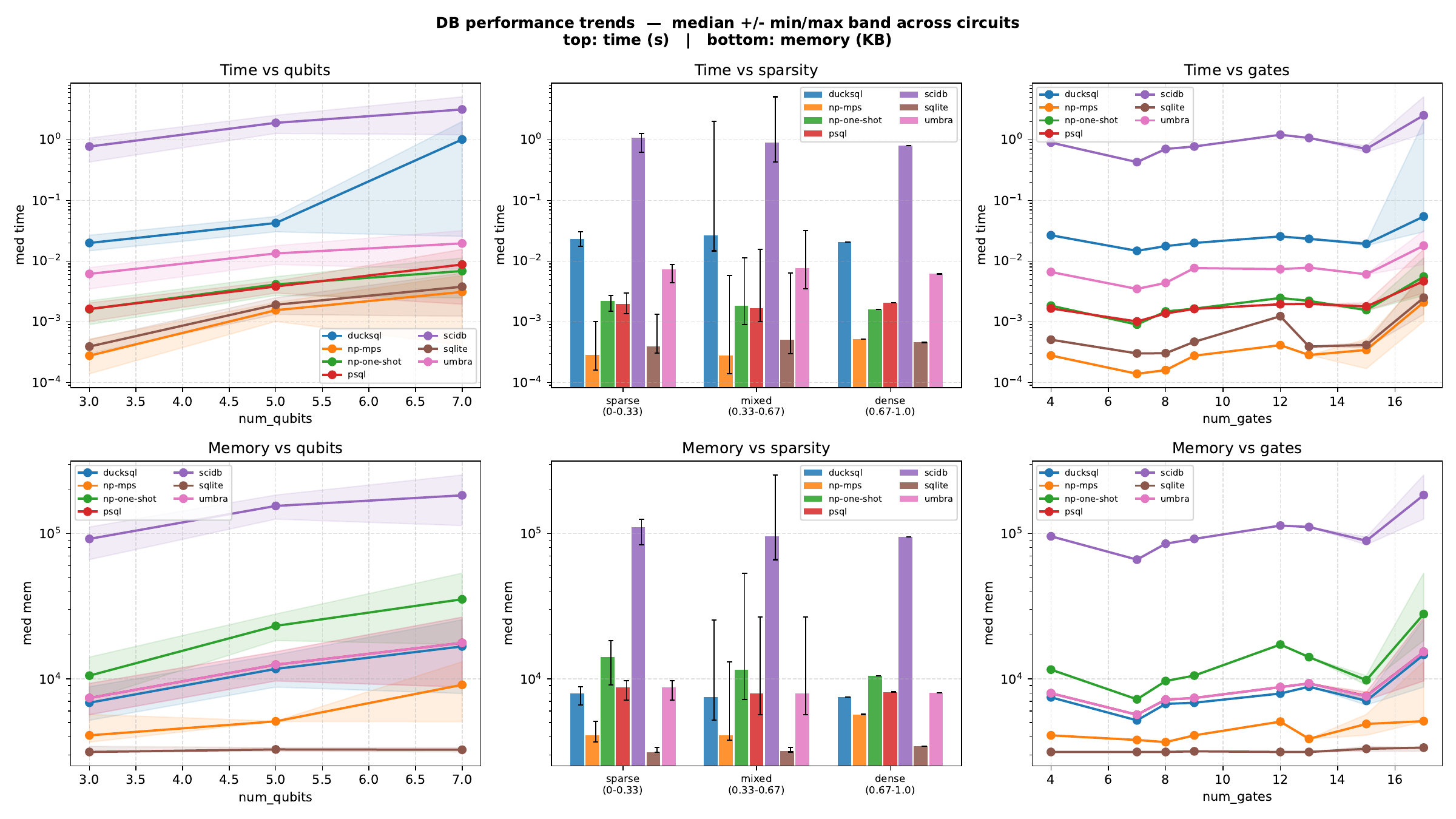}
    \caption{Additional DB comparison on 11 small \sys circuits with varying output density. 
    Top row: runtime; bottom row: peak memory. The plots group circuits by qubit count,
    output-density bin, and gate count.}
    \label{fig:addscidbcomp}
\end{figure*}

Beyond the RDBMS engines used in the main evaluation, tensor-aware DBMSs
such as TileDB and SciDB are natural candidates for quantum circuit simulation.
They expose array-oriented storage and execution abstractions that appear well matched
to tensor contractions. However, using them fairly requires significant effort because \sys currently
emits SQL workloads whose core operations are joins and aggregations.

TileDB is an array storage engine for dense and sparse multidimensional arrays.\footnote{
TileDB documentation: \url{https://docs.tiledb.com/main/}. TileDB-Py API:
\url{https://tiledb-inc-tiledb.readthedocs-hosted.com/projects/tiledb-py/en/stable/python-api.html}.}
A direct TileDB evaluation would require an array-native implementation of the simulator:
one would need to choose array schemas, chunk sizes, qubit-index layouts, and contraction
kernels, rather than executing the same SQL workload used by PostgreSQL, SQLite, and
DuckDB. The closest SQL-oriented possibility, \texttt{TileDB-MariaDB/MyTile}, is deprecated and supports
only limited pushdown, which makes it unsuitable as a fair drop-in replacement for our
join-and-aggregate SQL workload.\footnote{
TileDB-MariaDB/MyTile documentation:
\url{https://github.com/TileDB-Inc/TileDB-MariaDB}.}
We therefore do not include TileDB in the timing comparison. We view a TileDB-native
implementation as important future work rather than a direct backend substitution.

SciDB is closer to our setting because it provides array operators and a query interface
for composing them.\footnote{
SciDB-Py documentation: \url{https://paradigm4.github.io/SciDB-Py/guide.html}.}
Since \sys emits SQL, we made a preliminary SciDB implementation by translating each tensor
contraction into SciDB's Array Functional Language (AFL)-style array operations. In principle, an entire circuit could be
encoded as one nested array expression. In practice, the generated expressions become
large and fragile for full circuits, so our prototype splits the simulation at tensor
contraction boundaries and materializes intermediate arrays. This makes the prototype
robust, but it removes some global optimization opportunities available to RDBMSs when
they optimize a CTE chain. Thus, please consider the results below as preliminary
results rather than as results from a fully optimized SciDB implementation.

\medskip
\para{Experimental setting}
Figure~\ref{fig:addscidbcomp} reports a comparison on 11 \sys circuits \footnote{\url{https://github.com/InfiniData-Lab/InferQ/blob/main/analysis/sample_csvs/small_sample_scidb.csv}} with
3--7 qubits and 4--17 gates, spanning sparse, mixed, and dense output-density bins.
We compare PostgreSQL, DuckDB, SQLite, Umbra, SciDB, and two NumPy baselines:
\texttt{np-one-shot}, an exact tensor-contraction baseline, and \texttt{np-mps}, an
MPS-based approximate baseline. 

\medskip
\para{Results}
We have two main observations from Figure~\ref{fig:addscidbcomp}. First, among DB-backed
methods, SQLite remains the strongest backend. It beats
\texttt{np-one-shot} in peak memory for all 11 circuits and in runtime for 4 of the
11 circuits. PostgreSQL and Umbra remain competitive in memory but are slower than
SQLite. Second, the direct SciDB prototype has substantial overhead: it is consistently
slower and uses more memory than the RDBMS backends on these small circuits. The
\texttt{np-mps} baseline is often fast, but it is approximate, whereas the DB-backed
methods and \texttt{np-one-shot} are exact strong simulators.

\medskip
  \vspace{0.2cm}
\noindent
\setlength{\fboxsep}{6pt}
\fcolorbox{black!15}{black!4}{%
  \parbox{\dimexpr\linewidth-2\fboxsep-2\fboxrule\relax}{
\para{Take-away}
Tensor-aware DBMSs are promising, but they are not plug-in replacements for the SQL
workload studied in this paper. To make TileDB or SciDB competitive for quantum circuit
simulation, future work should co-design the simulator with the array engine, including:
(i) qubit-index layout and chunking, (ii) sparse-versus-dense array selection,
(iii) contraction fusion to avoid per-step materialization, and (iv) cost models for
array-native execution. It is an interesting future direction to explore which types of simulation workloads 
array DBMSs would be most promising for.
  }}

%% file: chapters/benchmark-suites.tex
\section{Support for Existing Benchmarks}
\label{app:benchmark-suites}

In addition to circuits generated by \sys, the evaluation pipeline can ingest circuits
from established benchmark suites: SupermarQ~\cite{tomesh2022supermarqscalablequantumbenchmark},
MQT Bench~\cite{Quetschlich_2023}, and QASM
Bench~\cite{li2022qasmbenchlowlevelqasmbenchmark}. These suites provide workloads that complement our generated corpus. The compositional circuit simulation workload introduced in Sec.~\ref{sec:benchmark}
remains necessary for controlled simulation scale, feature-space coverage, and reproducible
randomized workload generation, while external suites provide an additional validation
source based on named benchmark circuits.

\begin{table}[t]
  \centering
  \caption{External benchmark-suite circuit entries supported by the \sys ingestion pipeline. Per-suite counts are before cross-suite deduplication.}
  \label{tab:benchmark-suite-ingestion}
  \footnotesize
  \begin{tabular}{@{}lr@{}}
    \toprule
    \textbf{Benchmark suite} & \textbf{\# circuit entries} \\
    \midrule
    SupermarQ & 8 \\
    MQT Bench & 34 \\
    QASM Bench & 67 \\
    \midrule
    Total (after deduplication) & 85 \\
    \bottomrule
  \end{tabular}
\end{table}

All imported circuits are normalized to Qiskit \texttt{QuantumCircuit} objects and
tagged with their source suite and benchmark name. The ingestion pipeline then applies
the same SQL generation, feature extraction, and the rest of the workflow used for \sys-generated
circuits. As a result, external-suite circuits can be queried, filtered, and evaluated
together with generated circuits under the same RDBMS, simulator-selection, and
out-of-core workflows.

The per-benchmark counts in Table~\ref{tab:benchmark-suite-ingestion} are before deduplication. Since
the three benchmark suites contain overlapping algorithm families and problem-size variants, \sys
also records a deduplicated inventory; after deduplication, the external benchmark
inventory covers 85 unique algorithm families. The complete per-suite circuit-family
list and ingestion commands are provided in the \emph{Benchmark Algorithms} table in the online documentation.\footnote{\url{https://github.com/InfiniData-Lab/InferQ/blob/main/scripts/benchmark_suites/README.md}}

\input{chapters/profiling} 

\begin{table*}[t]
  \centering
  \caption{Database-specific tuning settings used for the 7{,}705-circuit evaluation. 
  }
  \label{tab:finetuned-rdbms-profiles}
  \footnotesize
  \setlength{\tabcolsep}{4pt}
  \begin{tabularx}{\textwidth}{@{}llX@{}}
    \toprule
    \textbf{Engine} & \textbf{Profile} & \textbf{Parameters} \\
    \midrule
    DuckDB
      & \texttt{small\_size}
      & \texttt{memory\_limit=10780MB}; \texttt{threads=2};
        \texttt{preserve\_insertion\_order=false};
        \texttt{max\_temp\_directory\_size=256GB} \\
    DuckDB
      & \texttt{large\_size}
      & \texttt{memory\_limit=7409MB}; \texttt{threads=1};
        \texttt{preserve\_insertion\_order=true};
        \texttt{max\_temp\_directory\_size=128GB} \\
    SQLite
      & \texttt{small\_size}
      & \texttt{cache\_mb=64}; \texttt{db\_path=:memory:};
        \texttt{mmap\_mb=0}; \texttt{temp\_store=MEMORY};
        \texttt{threads=1} \\
    SQLite
      & \texttt{large\_size}
      & \texttt{cache\_mb=128}; file-backed database;
        \texttt{mmap\_mb=128}; \texttt{temp\_store=MEMORY};
        \texttt{threads=1} \\
    PostgreSQL
      & \texttt{small\_size}
      & \texttt{work\_mem=256MB}; \texttt{temp\_buffers=128MB};
        \texttt{effective\_cache\_size=8GB};
        \texttt{hash\_mem\_multiplier=2.0};
        \texttt{max\_parallel\_workers\_per\_gather=4};
        \texttt{join\_collapse\_limit=1}; \texttt{from\_collapse\_limit=1};
        \texttt{jit=off} \\
    PostgreSQL
      & \texttt{large\_size}
      & \texttt{work\_mem=55MB}; \texttt{temp\_buffers=65MB};
        \texttt{effective\_cache\_size=2GB};
        \texttt{hash\_mem\_multiplier=1.94};
        \texttt{max\_parallel\_workers\_per\_gather=0};
        \texttt{join\_collapse\_limit=4}; \texttt{from\_collapse\_limit=1};
        \texttt{jit=off} \\
    \bottomrule
  \end{tabularx}
\end{table*}

\section{DBMS-Specific Tuning}
\label{app:finetuned-rdbms}

To understand how DBMS configuration affects simulation performance, we have added
DBMS-specific tuning for PostgreSQL, DuckDB, and SQLite.\footnote{We did not include Umbra in the DBMS-specific tuning study because its publicly available documentation for the released Docker image only describes basic access through the PostgreSQL-compatible server and command-line interface, and we could not identify a stable, documented set of engine-level tuning parameters needed to define a reproducible configuration-search space: \url{https://hub.docker.com/r/umbradb/umbra }} We keep the circuit set and
\sys’s SQL generation pipeline unchanged and only vary DBMS-level configuration.

\medskip
\para{Automated tuning via configuration search}
We initially explored manual tuning to understand the impact of major parameters, but the
configuration space is large, engine-specific, and not scalable across thousands of
circuits. We therefore adopt an automated configuration-search workflow based on MLOS
(Machine Learning for Operating Systems)\footnote{\url{https://github.com/microsoft/MLOS}}:
the workload is fixed, and the tuner searches over DBMS parameters to minimize runtime.
For reproducibility, we document the parameter spaces, seed profiles, and runner details in
Section~\ref{app:tuning-repro}.

\medskip
\para{Workload-specific tuning}
In preliminary experiments, we observed that a single global configuration is suboptimal:
circuits induce heterogeneous query shapes (e.g., short vs.\ long CTE chains) and
different bottlenecks (overhead-dominated vs.\ memory- and I/O-dominated). We therefore
introduce workload stratification by circuit size. For the 162-circuit all-engine MLOS
validation set,\footnote{\url{https://github.com/InfiniData-Lab/InferQ/blob/main/analysis/training_data/rdbms_all_methods_training_data.parquet}}
we split circuits by the median \texttt{circuit\_size} (33 gates): circuits with
\texttt{circuit\_size} $\le 33$ use a \texttt{small\_size} profile (82 circuits) and the
remaining circuits use a \texttt{large\_size} profile (80 circuits). This allows the tuner
to select configurations appropriate for structurally different workloads of small and large circuits.
Table~\ref{tab:finetuned-rdbms-profiles} summarizes the final
profiles. We observed that SQLite did not obtain much performance gain from automatic tuning, so we have tuned it manually\footnote{Tuning results for the 7705 dataset can be found here: \url{https://github.com/InfiniData-Lab/InferQ/blob/main/analysis/finetuned/rdbms7705_sqlite_static_contr/rdbms7705_sqlite_static_contr_results.csv}}.

\begin{table}[t]
  \centering
  \caption{Tuned DB performance on the 162 circuits. Time and peak-memory wins are counted against the best successful Qiskit Aer method per circuit.}
  \label{tab:finetuned-rdbms-results}
  \small
  \setlength{\tabcolsep}{6pt}
  \begin{tabular}{lrr}
    \toprule
    \textbf{Backend} & \textbf{Time wins} & \textbf{Memory wins} \\
    \midrule
    SQLite      & 34/162  & 108/162 \\
    DuckDB      & 0/162   & 0/162 \\
    PostgreSQL  & 0/162   & 0/162 \\
    Umbra       & 0/162   & 0/162 \\
    \midrule
    RDBMS total     & 34/162  & 108/162 \\
    Qiskit Aer total & 128/162 & 54/162 \\
    \bottomrule
  \end{tabular}
\end{table}

\para{Results}
Table~\ref{tab:finetuned-rdbms-results} reports results after tuning 
on the 162-circuit dataset, which are also the results for Figure~\ref{fig:rdbmsqiskitcomparison}. Overall, Qiskit Aer
remains the runtime winner on most circuits (128/162). Among the evaluated DBMS backends,
only SQLite achieves runtime wins (34/162); DuckDB, PostgreSQL, and Umbra do not win on
runtime on this subset. In contrast, SQLite is the peak-memory winner on 108/162 circuits ($66.7\%$),
while Qiskit Aer wins peak memory on 54/162 circuits.
We have also evaluated the tuned configurations on the full \num{7705}-circuit dataset
(Figure~\ref{fig:aer-vs-sqlite-wins}). The result is similar: RDBMS backends achieve the
lowest runtime on \num{1101} circuits (14.3\%), and all of these runtime wins come from the SQLite, for which tuning did not have much improvement. For peak memory, RDBMS backends win on \num{3902}
circuits (50.6\%), dominated by tuned SQLite (\num{3901} wins) with one additional win
by Umbra. Relative to the untuned setting, tuning increases SQLite's peak-memory win count
from \num{3196} to \num{3901} circuits.

\medskip
  \vspace{0.2cm}
\noindent
\setlength{\fboxsep}{6pt}
\fcolorbox{black!15}{black!4}{%
  \parbox{\dimexpr\linewidth-2\fboxsep-2\fboxrule\relax}{
\para{Take-away}
Taken together with Appendix~A (out-of-core) and Appendix~F (memory profiling), these
results suggest two concrete future optimization opportunities for DBMS-backed simulation. 
\\i) For
\emph{large} circuits where memory is the bottleneck, DBMS execution is uniquely useful
because it can externalize operator workspaces and intermediate relations. For such simulation,
end-to-end performance is largely driven by spill and intermediate growth, motivating
spill-aware planning (join ordering, operator selection, and materialization control) and  physical design studies over
intermediate tables (indexes, partitioning and layout, and caching reused intermediates) as discussed in Appendix~H.
\\ii) For \emph{small} circuits where Aer finishes in milliseconds, DBMS runtime is often
dominated by fixed overheads (SQL compilation, planning, and repeated query execution),
motivating workload-level optimizations such as batching,  reuse, and
handling repeated subcircuit computations across many runs. 
\\
\sys is a convenient tool
for future database research in these directions, because it generates controlled workloads
(small or large, sparse or dense) and emits database-native SQL along with query-shape and
circuit metadata needed for systematic evaluation. 
}}

\section{Indexes, partitioning, and CTE query structure}
\medskip
 
In \sys, each circuit is compiled into a \emph{single SQL statement} of the form
\texttt{WITH ... SELECT}, where each tensor contraction step is represented as a Common Table Expressions (CTE) and
implemented as a join followed by an aggregate, consistent with prior SQL-based
formulations \cite{hai2025quantum, einsteinSQL2023}. This CTE-chain representation is
useful because it makes the contraction order explicit and exposes the workload to the
DBMS optimizer as a sequence of standard relational operators (joins and group-bys).
However, it also limits classic physical design actions: intermediate CTE results are
 not persistent relations, so users cannot directly create secondary indexes, define
table partitioning, or maintain materialized views over those intermediates. Instead, the
engine may build transient internal structures (e.g., SQLite's automatic indexes and temporary B-tree structures), but these are optimizer-managed and not user-controlled. To enable explicit
physical design, we added a second query generator which lowers the same contraction pipeline
into a sequence of \texttt{CREATE TEMP TABLE AS} statements; each intermediate becomes a
first-class relation on which indexes and partitioning can be applied. 
Note that this is not part of \sys's standard pipeline; we added it to support the
experiments in this section. Systematically exploring index and partition selection for
these intermediate relations is a natural direction for future work.

\medskip
\para{Implications for physical design}
The tuned profiles in Table~\ref{tab:finetuned-rdbms-profiles} provide a reproducible
baseline that isolates DBMS-level execution effects (memory budgeting, temporary-workspace
management, parallel execution, and planner constraints) for this workload. Combined with
\texttt{split} execution, they also define a concrete path for future physical design
studies using standard database methodology. For example, one can treat intermediate
tables as design objects and evaluate: (i) index selection on join keys for contraction
steps (e.g., B-tree or hash indexes depending on the engine and operator), (ii)
partitioning and layout decisions to improve join locality and reduce spill (where the engine
supports partitioned tables and partitionwise planning), and (iii) selective
materialization and caching of frequently reused intermediates (e.g., repeated subcircuits)
as materialized relations. Because \sys exposes circuit features and query-shape metadata
(e.g., number of contraction steps (CTEs), runtime, and spill proxies), these choices can be
studied systematically as physical design problems across engines and workload regimes.

\medskip
\para{Preliminary Results}
We have conducted an experiment similar to Appendix~\ref{app:qft-cap-sensitivity} under a
4~GB memory limit. We choose a dense circuit from existing benchmarks, namely QFT, and vary
its qubit count from 3 to 26. Except for DuckDB at 26 qubits, all simulations completed.
We observe that the new query generator, which forces CTE materialization, reduces runtime
by roughly $2.1\times$ for DuckDB and $1.8\times$ for PostgreSQL; SQLite is essentially
unchanged. This shows that CTE materialization can be a potentially effective approach
for improving DBMS-backed simulation on dense workloads under tight memory limits.

\section{Additional Configuration Details on Experiments for Reproducibility}
\label{app:tuning-repro}

This section documents how we measure spilling performance and how we execute tuned runs.

\subsection{Circuit Construction Details}
\label{app:expander-details}
\begin{figure}[t]
    \centering
    \includegraphics[width=1.0\linewidth]{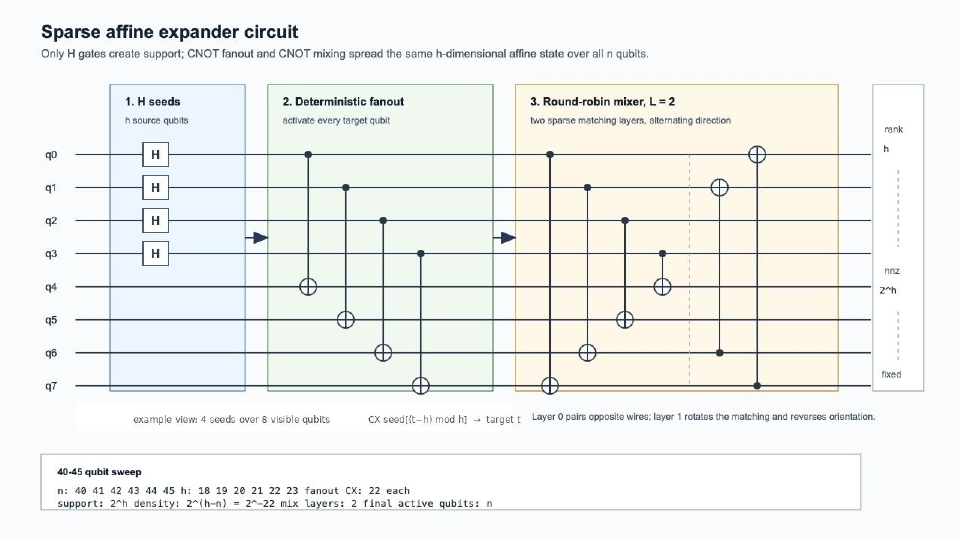}
    \caption{Expander circuit structure: Hadamards on $h$ seed qubits followed by CNOT-only
    layers that expand connectivity while preserving sparse support.}
    \label{fig:expander_circuit}
\end{figure}

We design a circuit family with large qubit
counts but fixed sparse values.
Each circuit is parameterized by $(n,h)$, where $n$
is the number of qubits and $h<n$ is a small seed size. The construction applies Hadamards
only to the $h$ seed qubits and then applies only CNOT layers, which preserve support
size; therefore the final state has exactly $2^h$ non-zero amplitudes in the computational
basis. We
evaluate a sample of Expander circuits with $n \in [40,45]$, where a dense statevector
would require approximately 16--512~TB.  
An Expander circuit on $n$ qubits is parameterized by a seed size $h<n$. We apply
Hadamards to the first $h$ qubits to create a superposition over $2^h$ basis states and
initialize the remaining qubits to $\ket{0}$. We then apply only CNOT layers, which are
reversible linear transformations and preserve the number of basis states in support.
Consequently, the final state remains a uniform superposition over an affine subspace of
dimension $h$ and has exactly $2^h$ non-zero amplitudes in the computational basis (each
with magnitude $2^{-h/2}$). Figure~\ref{fig:expander_circuit} illustrates the circuit
structure.

Such sparse or highly structured states occur in practice in affine and stabilizer-style subroutines and
in workloads where computation is restricted to a low-dimensional subspace. Moreover,
even when the final state is dense, intermediate tensors and relations during contraction
can be much smaller than $2^n$. Separately, amplitude-amplification procedures (e.g.,
Grover-style iterations) tend to \emph{concentrate probability mass} onto a small subset
of basis states as they iterate, and sparse Hamiltonians and sparse linear systems (as in
eigensolver and HHL-style routines) often induce structured tensors. These regimes
motivate studying DBMS backends with sparse representations and out-of-core execution.

\subsection{Spill Configuration for Out-of-core experiments}
\label{app:spill-instrumentation}

We run SQL queries on PostgreSQL, DuckDB, and SQLite, sweeping Docker memory limit, e.g.,  16~GB. 
All containers use a one-CPU quota, swap is disabled by setting Docker
\texttt{--memory-swap} equal to the memory cap, and host page cache is not
explicitly dropped between runs. DuckDB uses one thread and sets
\texttt{memory\_limit} to the cap minus a 1024~MB headroom pad (3072, 7168,
and 15360~MB for the 4, 8, and 16~GB caps, respectively). SQLite uses a
64~MB cache budget for both main and temporary schemas. PostgreSQL uses the
tuned PostgreSQL~12.22 container, with the server container capped at the
experiment memory limit and \texttt{temp\_file\_limit} set to 64~GB. Spill is
measured as \texttt{postgres\_explain\_temp\_written} for PostgreSQL and as
\texttt{proc\_io\_write\_bytes} for DuckDB and SQLite.

We report spill volume using engine-specific sources when available, and otherwise use a
conservative operating system level proxy.

\noindent\textbf{PostgreSQL.} We compute exact temporary spill bytes from \texttt{EXPLAIN} (ANALYZE, BUFFERS, FORMAT JSON) by converting the reported temp written/read blocks to bytes using the 8192-byte block size; we treat this value as the per-query externalization metric.

\noindent\textbf{DuckDB.} We use \texttt{proc\_\allowbreak io\_\allowbreak write\_\allowbreak bytes} as the primary spill/workspace proxy, and when available we additionally parse DuckDB profiling JSON fields such as \texttt{temporary\_\allowbreak storage\_\allowbreak bytes} and \texttt{spilled\_\allowbreak bytes} for diagnostics only, since the profile schema can vary across versions.

\noindent\textbf{SQLite.} Similar to DuckDB, here we use \texttt{proc\_\allowbreak io\_\allowbreak write\_\allowbreak bytes} as the spill/workspace proxy. SQLite does not expose a stable query-level temp-byte counter through Python’s \texttt{sqlite3} interface, so we intentionally do not report a DB-native temp metric.


\subsection{Configuration for DB tuning}
\label{app:mlos-runner}
We add a dedicated runner under \texttt{scripts/finetuned\_rdbms} that reuses the same
InferQ circuits and SQL generation path, but executes the emitted SQL under
engine-specific configurations rather than untuned defaults. The same directory contains
an automated tuning script (\texttt{mlos\_tune\_rdbms.py}) based on MLOS.

MLOS keeps the workload fixed (each circuit is compiled once into a SQL
query) and searches only over DBMS configuration parameters. For each engine, the tuner
evaluates seed profiles (e.g., \texttt{balanced}, \texttt{fast}, \texttt{spill\_safe})
and then uses a black-box optimizer backend (e.g., FLAML or SMAC) to propose additional
configurations. Its objective is the median runtime on a pilot set; timeouts and
execution failures receive a large penalty. All trials are recorded in
\texttt{mlos\_trials.csv}, and the best configuration per engine is stored in
\texttt{mlos\_best\_configs.json}. These winning configurations are then validated on the
full discovered circuit set via the fine-tuned runner using \texttt{--tuning-json}. Thus,
the reported improvements correspond to \emph{DBMS-level physical execution choices}
rather than changes to the circuit generator or SQL compilation.

Operationally, the runner generates SQL once per circuit, records query-shape metadata
(e.g., number of CTEs/contraction steps), and then executes the workload under the
selected engine profiles with warm-up and timed runs, configurable timeouts, and resume
support. To reduce total tuning cost, we optionally use SQLite as a lightweight gate:
if SQLite times out for a circuit under the tuned profile, we mark other engines as
\texttt{skipped\_sqlite\_timeout} for that circuit (i.e., we do not spend additional
compute on circuits already rejected by the gate). This produces a controlled comparison
of tuned physical execution while keeping the workload and SQL generation fixed.

Table~\ref{tab:finetuned-rdbms-profiles} reports the best configurations selected and validated by our tuning workflow. For DuckDB, \texttt{small\_size} allocates a larger memory budget and modest parallelism for short query plans, while disabling insertion-order preservation to enable spill-capable operator choices; \texttt{large\_size} reduces memory and parallelism to limit concurrent memory pressure and temp-space exposure for deeper plans where planner/executor overhead can dominate. For SQLite, \texttt{small\_size} keeps the database and temporaries fully in-memory to avoid filesystem overhead, whereas \texttt{large\_size} uses a file-backed database with a modest mmap window to stabilize larger intermediate results while still keeping temporaries in memory. For PostgreSQL, \texttt{small\_size} increases per-node memory (\texttt{work\_mem}, \texttt{temp\_buffers}) and enables parallel workers to accelerate small contractions, while constraining join enumeration (\texttt{join\_collapse\_limit}/\texttt{from\_collapse\_limit}) to preserve the emitted join order and disabling JIT; \texttt{large\_size} caps per-node memory and disables parallel gather to reduce worker amplification and make externalization behavior more predictable. Finally, we include Umbra via its PostgreSQL-compatible frontend to broaden the engine comparison to a modern compiled analytical DBMS without changing the SQL workload.

\section{Generated Circuit Characterization}
\label{app:circuit-characterization}

\begin{figure*}[t]
    \centering
    \begin{subfigure}[t]{0.48\textwidth}
        \centering
        \includegraphics[width=\linewidth]{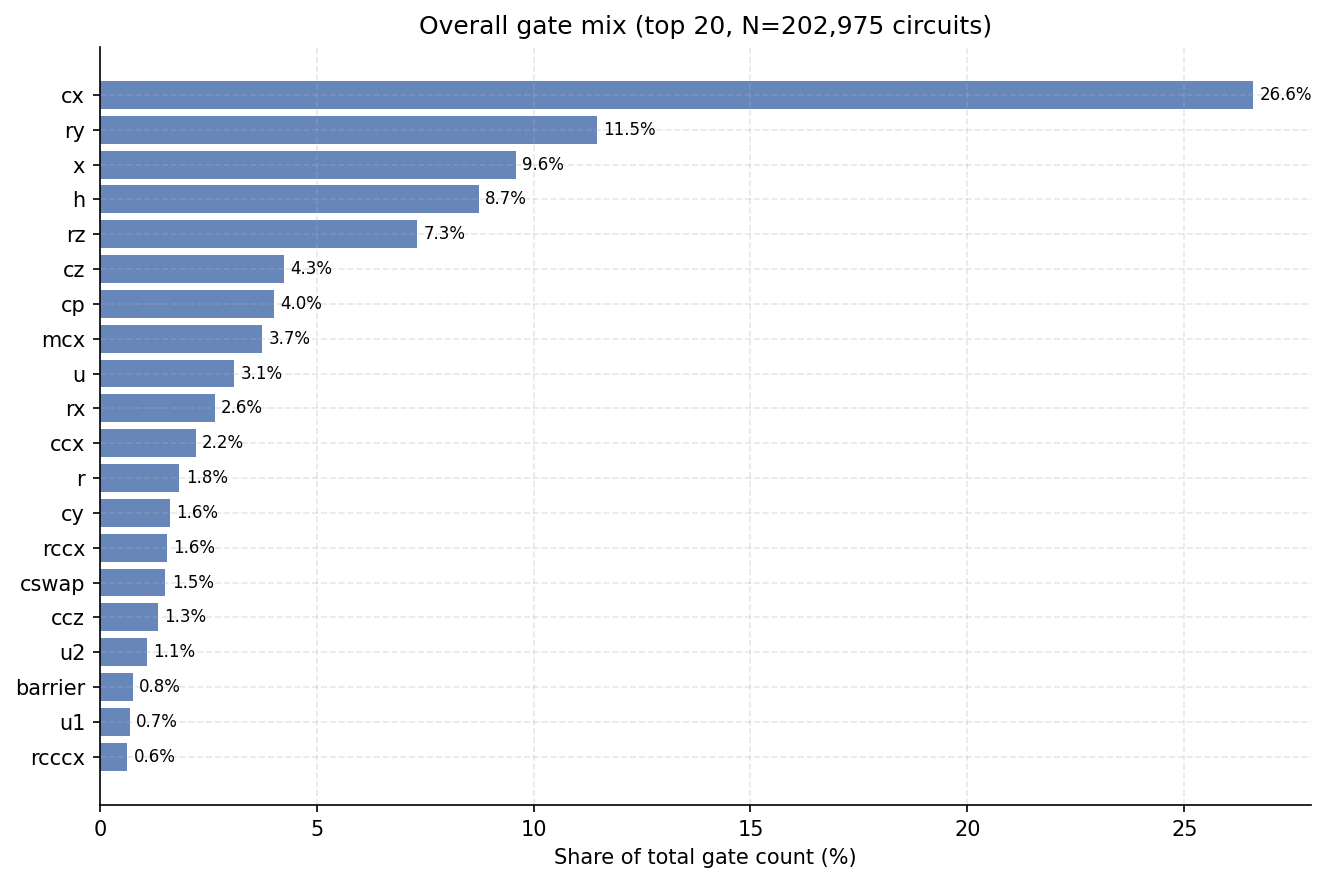}
        \caption{Top-20 gate-type distribution.}
        \label{fig:gate-mix}
    \end{subfigure}\hfill
    \begin{subfigure}[t]{0.48\textwidth}
        \centering
        \includegraphics[width=\linewidth]{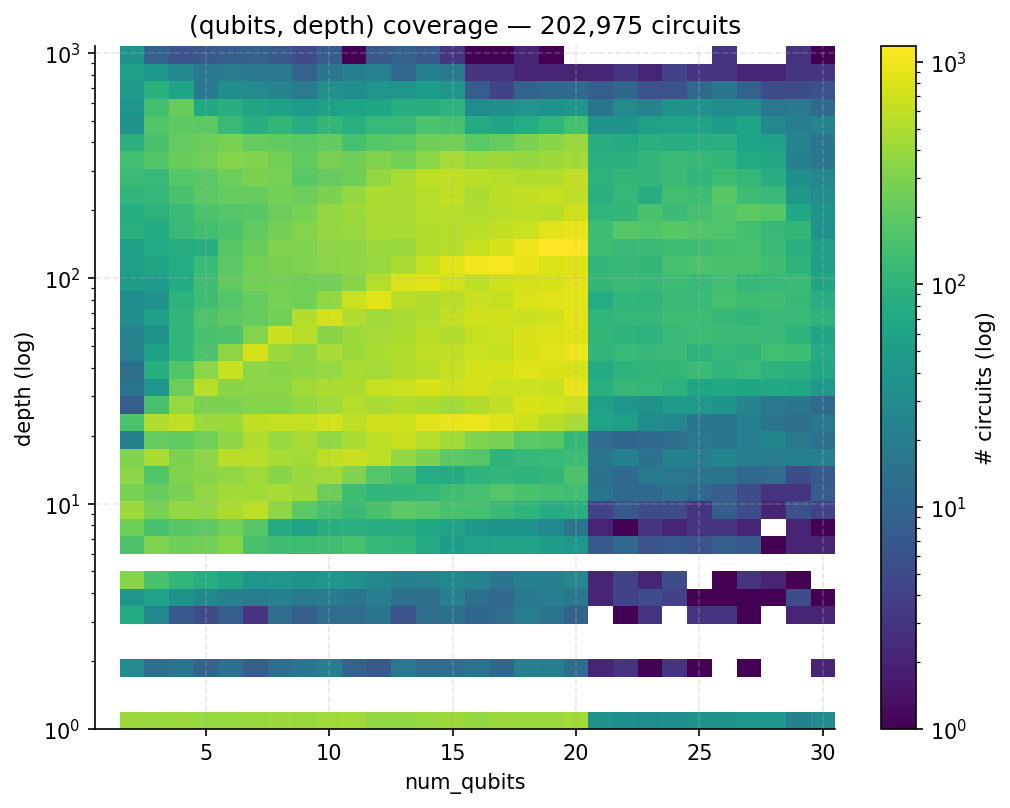}
        \caption{Depth/width coverage.}
        \label{fig:depth-width}
    \end{subfigure}
    \caption{Generated-corpus gate mix and depth/width coverage. Color in \subref{fig:depth-width} indicates the number of circuits in each bin on a logarithmic scale.}
    \label{fig:generated-characterization-overview}
\end{figure*}

\begin{figure*}[t]
    \centering
    \includegraphics[width=\textwidth]{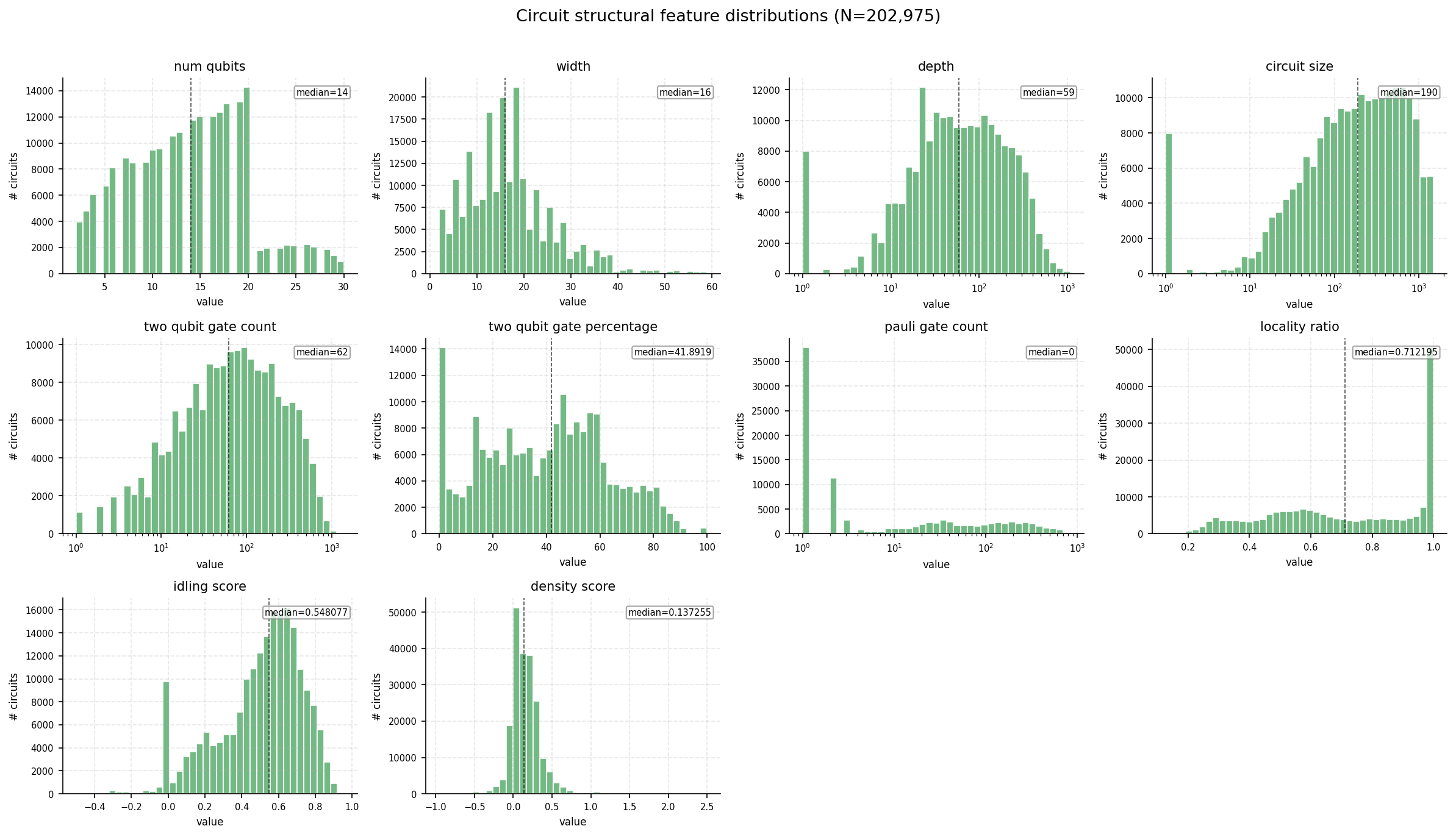}
    \caption{Generated-corpus static circuit-feature distributions. Dashed lines mark medians.}
    \label{fig:circuit-structure}
\end{figure*}

\begin{figure*}[t]
    \centering
    \includegraphics[width=\textwidth]{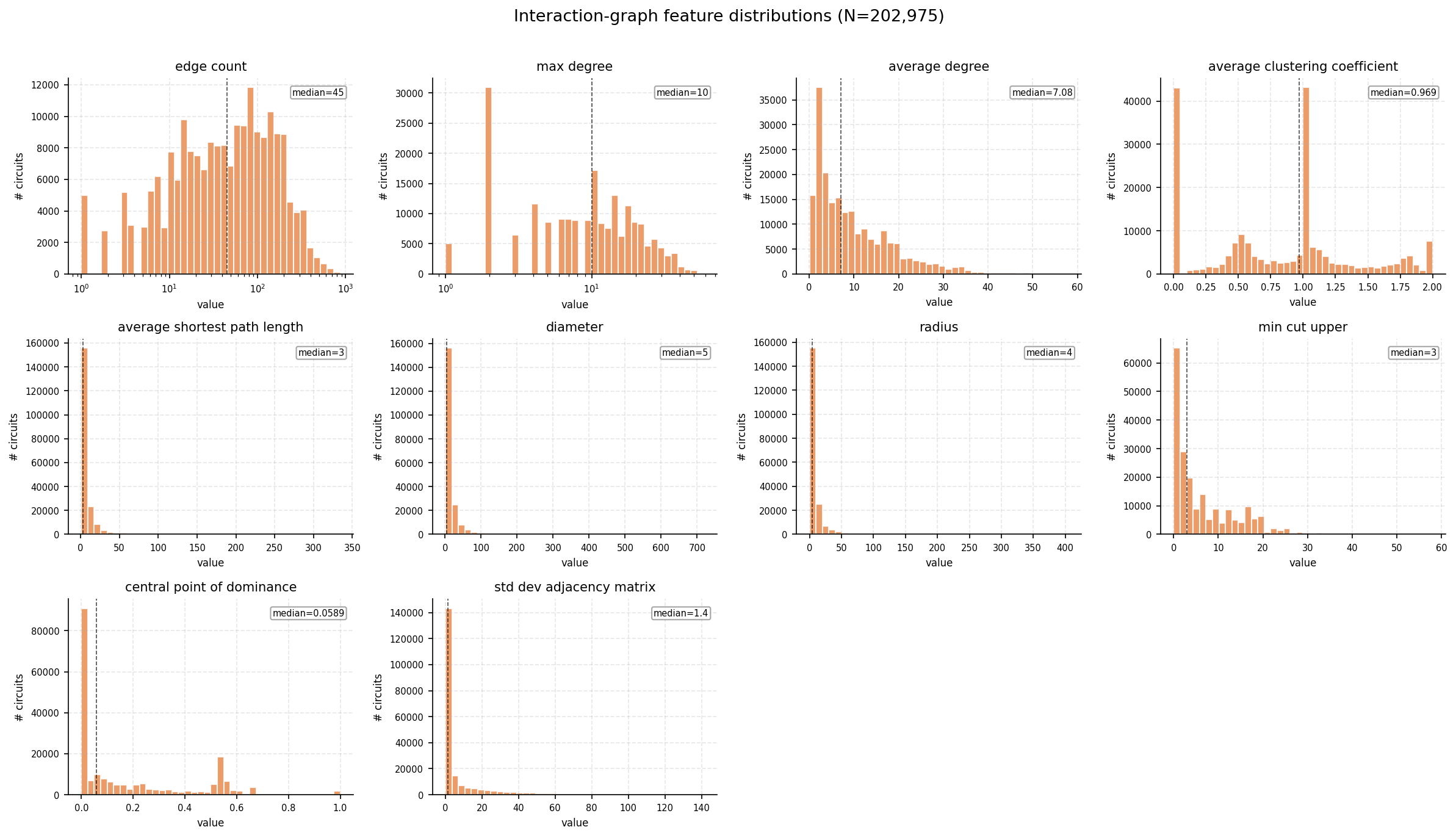}
    \caption{Generated-corpus interaction-graph feature distributions. Dashed lines mark medians.}
    \label{fig:interaction-graph}
\end{figure*}

To make the generated workload distribution transparent and reproducible, \sys records per-circuit metadata for each generated circuit and aggregates these fields over the full corpus. Figures~\ref{fig:generated-characterization-overview}, \ref{fig:circuit-structure}, and~\ref{fig:interaction-graph} summarize the resulting 202{,}975-circuit inventory from complementary perspectives: gate mix, static circuit structure, width/depth coverage, and interaction-graph structure. Together, these summaries show that the generator produces a broad workload rather than a collection of circuits that differ only in size.

\medskip
\para{Gate-mix distribution}
Figure~\ref{fig:gate-mix} reports the top-20 gate types by share of total gate count. The corpus has a substantial entangling component: \texttt{cx} alone accounts for 26.6\% of all gates, alongside common one-qubit rotations and Clifford-style operations such as \texttt{ry}, \texttt{x}, \texttt{h}, and \texttt{rz}. The remaining top gates include controlled-phase, controlled-Z, multi-controlled-X, swap, and higher-controlled variants, indicating that the generated workload includes both simple local transformations and larger multi-qubit control patterns.

\medskip
\para{Static circuit structure}
Figure~\ref{fig:circuit-structure} expands the view from gate identities to circuit-level structure. The generated corpus spans small and moderate-width circuits, with median values of 14 qubits, width 16, depth 59, and circuit size 190. It also varies substantially in two-qubit-gate count and two-qubit-gate percentage, which are important because they influence both tensor contraction structure and the amount of intermediate state growth during simulation. Additional static descriptors such as Pauli-gate count, locality ratio, idling score, and density score capture whether a circuit is mostly local or contains broader qubit interactions and idle regions. These distributions make the benchmark more informative than a width/depth grid alone.

\medskip
\para{Depth/width coverage}
Figure~\ref{fig:depth-width} shows the joint coverage over number of qubits and circuit depth. The heat map reports the number of circuits in each \emph{(number of qubits, depth)} bin using a logarithmic color scale, exposing both dense regions of the generated space and sparsely populated edge cases. The workload includes shallow circuits, deep circuits with depths above $10^3$, and circuit sizes extending to roughly 30 qubits in this corpus. This joint view is useful for interpreting performance results because simulator and DBMS behavior can change sharply when both width and depth increase, even when either feature alone appears moderate.

\medskip
\para{Interaction-graph structure}
Figure~\ref{fig:interaction-graph} characterizes the qubit interaction graphs induced by the generated circuits. We construct an interaction graph by treating qubits as vertices and adding an edge when two qubits participate in a multi-qubit operation. The resulting features include edge count, maximum and average degree, clustering coefficient, shortest-path statistics, diameter, radius, cut-related measures, central-point dominance, and adjacency-matrix variation. These graph features capture structure that is not visible from gate count, width, or depth alone: two circuits with the same number of qubits and a similar depth can still differ substantially in interaction density, path length, graph diameter, or degree distribution. Such differences affect SQL query shape, contraction opportunities, intermediate-result sizes, and simulator behavior.

Overall, these summaries make the generated workload transparent, reproducible, and analytically useful. With \sys, users can inspect the concrete properties that shape simulation performance, including gate composition, static circuit size, depth/width coverage, and qubit-interaction topology. This allows engine comparisons to be interpreted in terms of the circuits being simulated and helps identify which circuit families require particular simulation strategies.

%% file: chapters/profiling.tex
\section{Profiling of SQLite and Qiskit Aer}
\label{sec:deep-profile-sqlite-aer}

To understand why SQLite uses less peak memory than Qiskit Aer on many circuits,
we profile both systems on 162 generated circuits.\footnote{We have sampled these 162 circuits
  from the large \num{7705} circuits used in Section~6. This profiling comparison is separate from the all-backend tuned
winner count reported in Figure~7 and Appendix~G.}

For SQLite, each circuit is compiled into a tensor-contraction program represented
as a chain of Common Table Expressions (CTEs). Each contraction step is implemented
as a join followed by an aggregation. We record the number of total and contraction
CTEs, estimated intermediate-relation sizes, result cardinalities, SQLite
execution-plan operators, temporary B-tree usage, automatic-index usage,
page-cache overflow counters, peak resident set size (RSS), and wall-clock
runtime. For Qiskit Aer, we compare against the fastest   Aer method for
runtime, rather than fixing a single simulator representation across all circuits.
For memory, we compare against the lowest-RSS   Aer configuration for
the same circuit.

The profiling results\footnote{\url{https://github.com/InfiniData-Lab/InferQ/blob/main/analysis/memory_profiling/sqlite_vs_qiskit_summary.csv}} in Table~\ref{tab:sqlite-aer-profile} characterize when
each system is faster. Across 162 circuits, SQLite is faster in 46 cases, while
Qiskit Aer is faster in 116 cases. SQLite is faster when the relational workload
remains small: the median SQLite-winning circuit has 26.5 total CTEs, 22
contraction CTEs, the largest intermediate relation of 256\,B, and 16 final result
rows. In these cases, SQLite runs in a median of 2.03\,ms, compared with
2.82\,ms for the fastest   Aer method. In contrast, when Aer is faster,
the median workload is larger: 55 total CTEs, 49 contraction CTEs, a largest
intermediate relation of 1.0\,KB, and 128 final result rows. In these cases,
SQLite takes a median of 6.63\,ms, while Aer takes 3.09\,ms.

\begin{table}[t]
\centering
\small
\caption{Median properties of completed SQLite--Aer profiling runs, grouped by
the backend with lower runtime. CTE sizes are estimated from intermediate-relation
cardinalities and row widths.}
\label{tab:sqlite-aer-profile}
\begin{tabular}{lrr}
\toprule
 & SQLite faster & Aer faster \\
\midrule
Number of cases & 46 & 116 \\
Total CTEs & 26.5 & 55 \\
Contraction CTEs & 22 & 49 \\
Total estimated CTE size & 2.9\,KB & 10.5\,KB \\
Largest intermediate CTE & 256\,B & 1.0\,KB \\
Final result rows & 16 & 128 \\
SQLite runtime & 2.03\,ms & 6.63\,ms \\
Fastest Aer runtime & 2.82\,ms & 3.09\,ms \\
\bottomrule
\end{tabular}
\end{table}

These results show that SQLite's advantage is not determined by qubit count
alone. Some SQLite-winning circuits have more qubits than Aer-winning circuits,
but their relational intermediates remain small. The decisive factor is the size
and structure of the induced SQL workload. When the tensor-contraction sequence
preserves sparsity, SQLite materializes only a small number of nonzero amplitudes,
and the join-and-aggregate pipeline can complete before Aer's method setup and
simulation overheads dominate. For such circuits, the sparse SQL representation
provides a compact representation of the circuit state.

The same representation becomes less favorable once the contraction pipeline
grows. Longer CTE chains introduce repeated joins, aggregations, and intermediate
execution structures. SQLite query plans use temporary B-trees for grouping in
all completed runs and automatic indexes for joins in 154 out of 162 runs. These
are standard relational execution mechanisms, but in a long contraction sequence
their cumulative cost can dominate the arithmetic. Thus, when Aer is faster, it
is not because SQLite fails to represent the computation; rather, the relational
execution machinery becomes the bottleneck.

The memory measurements follow a different pattern from the runtime measurements.
Among the 162 circuits, SQLite uses less peak RSS than the lowest-RSS  
Aer method in 139 cases. Median peak RSS is 203\,MB for SQLite and 206\,MB for
Aer. This difference is small in absolute terms for these circuits, but it is
consistent with SQLite benefiting from storing sparse relational intermediates
rather than allocating the simulator representation used by Aer. At the same
time, this memory advantage does not automatically translate into a runtime
advantage, because SQLite still pays the cost of repeated SQL execution.

The SQLite-specific profiling counters support this interpretation. The
page-cache overflow counter is zero in all completed runs, and we do not observe
external-sort behavior. Thus, the slowdown is not explained by observed
out-of-core execution in these runs. Instead, the dominant cost is the repeated
construction and consumption of temporary B-trees and automatic indexes across
the CTE chain.

Unlike the RDBMS backends, Aer's performance depends strongly on which simulator
method is applicable and efficient for each circuit. In the profiling runs, the
fastest  Aer method varies across circuits: \texttt{unitary} is fastest
in 46 cases, \texttt{automatic} in 42 cases, \feat{statevector} in 37 cases,
\feat{extended\_stabilizer} in 26 cases, \feat{matrix\_product\_state} in
5 cases, \feat{density\_matrix} in 4 cases, and \feat{stabilizer} in 2
cases. This indicates that a single fixed Aer method is not a stable baseline for
heterogeneous circuits. Some circuits admit efficient specialized simulation,
while others require a more general representation. For peak memory, the
lowest-RSS Aer configuration is usually \feat{statevector} or
\feat{density\_matrix}:  \feat{statevector} has the lowest RSS in 88 cases,
\feat{density\_matrix} in 49 cases, \feat{automatic} in 14 cases,
\feat{matrix\_product\_state} in 9 cases, and \feat{unitary} in 2 cases.

We do not enable Aer cache blocking in this comparison. Our Aer baseline uses the
standard CPU simulator methods, whereas Aer's cache-blocking options (e.g.,
\texttt{blocking\_enable} and \texttt{blocking\_qubits}) are documented for
distributed GPU and/or multi-node simulation~\cite{QiskitAerMultiGPU}. They
partition the state into chunks to reduce data exchange across distributed memory
spaces, rather than providing a disk-backed CPU cache comparable to an RDBMS
buffer manager. Enabling these options would therefore change the simulator
configuration instead of providing a fair comparison to SQLite's
relational execution.  

\medskip
  \vspace{0.2cm}
\noindent
\setlength{\fboxsep}{6pt}
\fcolorbox{black!15}{black!4}{%
  \parbox{\dimexpr\linewidth-2\fboxsep-2\fboxrule\relax}{
\noindent\textbf{Takeaway.}
The relative performance of SQLite and Qiskit Aer is governed by both
representation and execution strategy. SQLite is memory-efficient when sparsity
keeps intermediate relations small, but Aer is faster once repeated joins,
aggregations, temporary B-trees, and automatic indexes make SQL execution
overhead dominate the contraction pipeline.
}}

%% file: chapters/feature.tex
\begin{figure}[t]
    \centering
    \includegraphics[width=\linewidth]{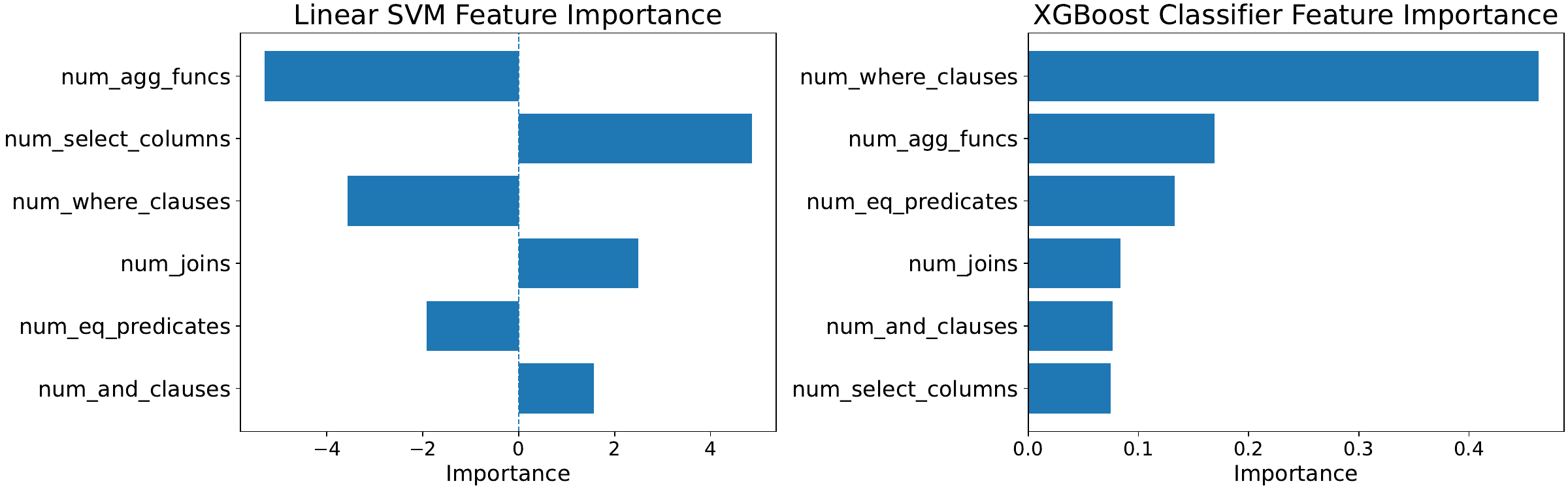}
    \caption{Importance of various SQL features for SVM and XGBoost trained for optimizing execution time.}
    \label{fig:rdbmsqiskitimportancetime}
\end{figure}
\begin{figure}[t]
    \centering
    \includegraphics[width=\linewidth]{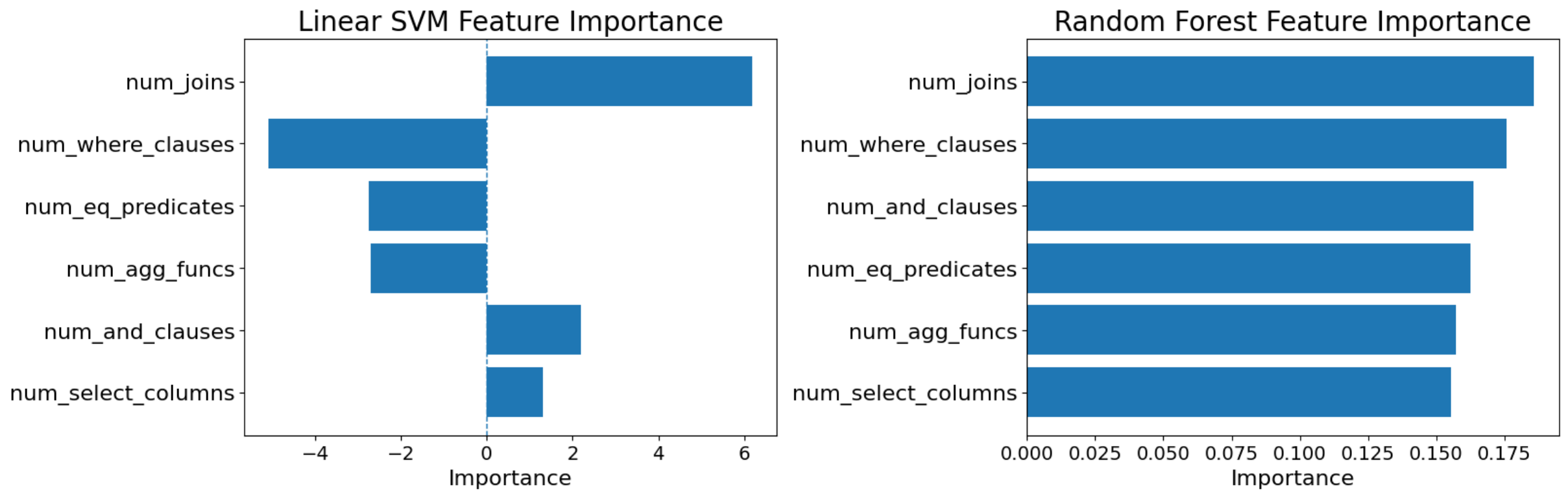}
    \caption{Importance of various SQL features for SVM and RF trained for optimizing memory.}
    \label{fig:rdbmsqiskitimportancemem}
\end{figure}
\section{Additional Results on SQL-only features.}
\label{ssec:sqlf}
Figures~\ref{fig:rdbmsqiskitimportancetime} and~\ref{fig:rdbmsqiskitimportancemem} show that the
dominant SQL signals are structural counts, in particular the number of \texttt{JOIN}s,
predicates, and aggregates. For runtime, the Linear SVM assigns negative coefficients to
\feat{num_where_clauses}, \feat{num_eq_predicates}, and \feat{num_agg_funcs}, suggesting that
queries with heavier filtering and aggregation tend to favor Qiskit Aer. In contrast,
\feat{num_joins} and \feat{num_and_clauses} shift the decision toward the RDBMS backend. For memory,
\feat{num_joins} is the dominant factor, while predicate and aggregation features retain
negative influence, reflecting the benefit of the optimized join algorithms and cost-based planning employed by RDBMS.

%% file: scalability.tex
\begin{figure}[t]
    \centering
    \includegraphics[width=\linewidth]{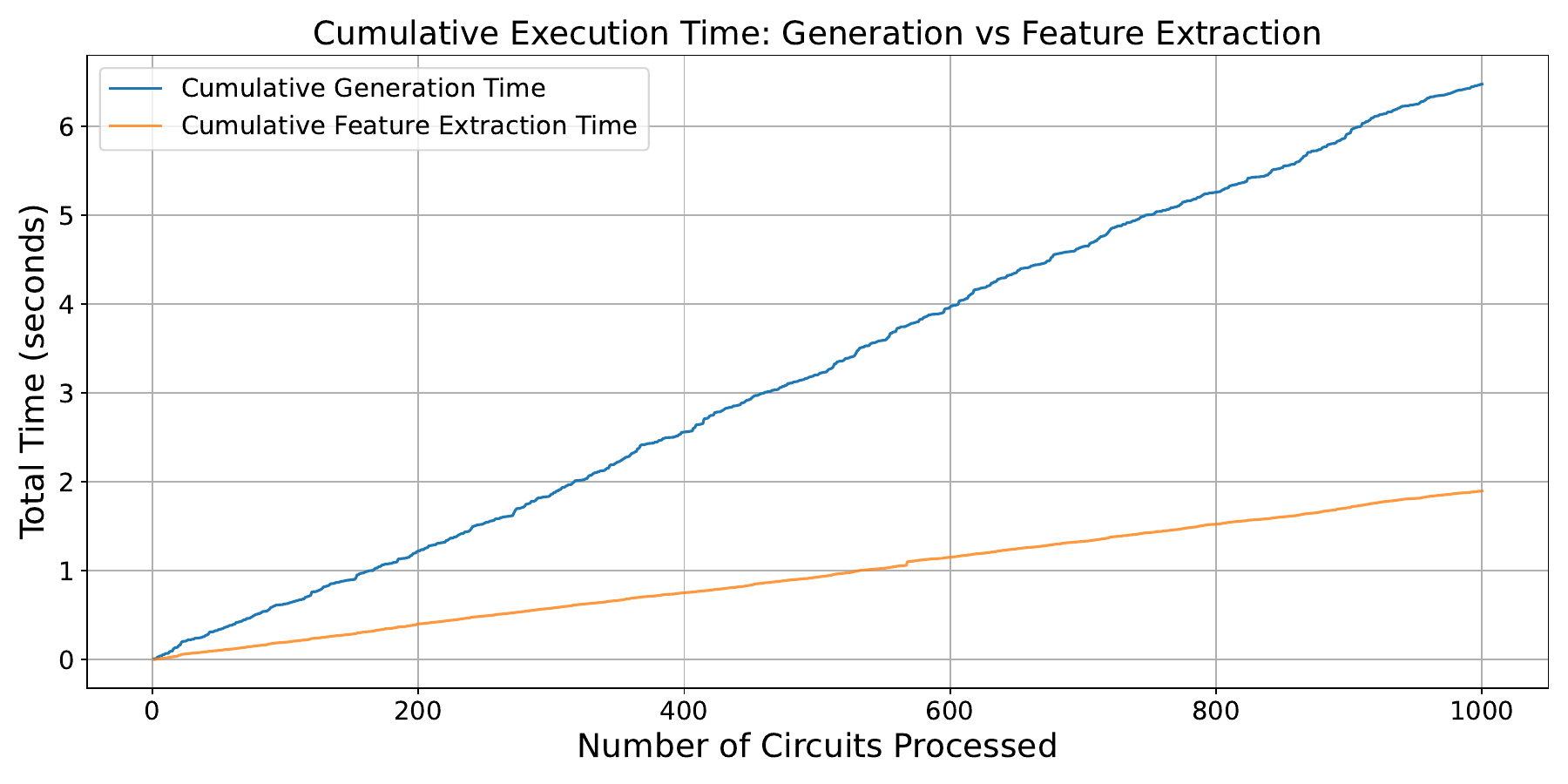}
    \caption{Cumulative execution time versus number of circuits processed. The figure shows cumulative generation time, and cumulative feature extraction time.}
    \label{fig:scalability_runtime}
\end{figure}




\section{Scalability}
\label{sec:scalability}

Figure~\ref{fig:scalability_runtime} reports how the \sys preprocessing cost grows as we increase the number of generated circuits.
We observe near-linear scaling for both \emph{circuit generation} and \emph{feature extraction} (static, graph, and SQL features): generating
1{,}000 circuits takes only 6.47~seconds, and extracting features takes 1.9~seconds.\footnote{We exclude dynamic features here, as extracting them requires running the simulation, while Section~\ref{ssec:uc2} suggests that they can be predicted using static, graph, and SQL features.}
This tackles Gap 3 in Section~\ref{ssec:limit}. That is, \sys can scale to large circuit corpora: generation and feature extraction contribute negligible overhead.

\medskip
  \vspace{0.2cm}
\noindent
\setlength{\fboxsep}{6pt}
\fcolorbox{black!15}{black!4}{%
  \parbox{\dimexpr\linewidth-2\fboxsep-2\fboxrule\relax}{%
\para{Take-away}
\sys scales well in preprocessing: circuit generation and (static/graph/SQL) feature extraction grow linearly and remain inexpensive. This supports building large circuit datasets for database- and learning-based analyses.
}}%